\documentclass[letterpaper,11pt]{article}
\usepackage[utf8]{inputenc}

\pdfoutput=1
\usepackage{jheppub}
\usepackage{graphicx,verbatim}
\usepackage{blindtext}
\usepackage[T1]{fontenc}
\usepackage{epsf}
\usepackage{mathtools}
\usepackage{lmodern}
\usepackage{dsfont}
\usepackage[mathscr]{euscript}
\usepackage{mathrsfs}

\usepackage{amsmath}
\usepackage{amsfonts}
\usepackage{amssymb}
\usepackage{mathrsfs}
\usepackage{slashed}
\usepackage{xcolor}
\usepackage{caption}
\usepackage{subcaption}
\definecolor{green}{RGB}{0,255,0}

\usepackage{ytableau,tikz,varwidth}
\usetikzlibrary{calc}
\usepackage{tikz-cd}
\usepackage{caption}
\usetikzlibrary{shapes.geometric}
\usetikzlibrary{decorations.markings}

\def\fg{\mathfrak{g}}
\def\hfg{\widehat{\mathfrak{g}}}
\def\qe{\mathfrak{q}}
\def\kq{\mathfrak{q}}

\def\bla{{\boldsymbol\lambda}}
\def\fR{\mathfrak{R}}

\def\fgl{\mathfrak{gl}}

\def\fsl{\mathfrak{sl}}

\def\ii{\mathrm{i}}

\def\be{\mathbf{e}}
\def\bx{\mathbf{x}}
\def\ba{\mathbf{a}}

\def\bd{\mathbf{d}}
\def\bs{\mathbf{s}}

\def\bu{\mathbf{u}}

\def\BC{\mathbb{C}}
\def\BR{\mathbb{R}}
\def\BE{\mathbb{E}}
\def\BZ{\mathbb{Z}}

\def\BP{\mathbb{P}}

\def\bT{\mathbf{T}}

\def\CalQ{\mathcal{Q}}
\def\CalB{\mathcal{B}}
\def\CalE{\mathcal{E}}

\def\CalA{\mathcal{A}}
\def\CalG{\mathcal{G}}
\def\CalV{\mathcal{V}}
\def\CalP{\mathcal{P}}
\def\CalR{\mathcal{R}}
\def\CalZ{\mathcal{Z}}
\def\CalH{\mathcal{H}}
\def\CalW{\mathcal{W}}
\def\CalU{\mathcal{U}}
\def\CalO{\mathcal{O}}
\def\CalM{\mathcal{M}}
\def\CalS{\mathcal{S}}

\def\EuT{\EuScript{T}}
\def\qe{\mathfrak{q}}

\def\sG{\mathscr{G}}
\def\sE{\mathscr{E}}
\def\sF{\mathscr{F}}

\def\ve{{\varepsilon}}

\def\sK{\mathsf{K}}

\def\sF{\mathsf{F}}

\def\EQ{\EuScript{Q}}

\def\EN{\EuScript{N}}
\def\EY{\EuScript{Y}}

\def\EZ{\EuScript{Z}}

\def\EB{\EuScript{B}}

\def\ET{\EuScript{T}}
\def\EM{\EuScript{M}}

\def\ED{\EuScript{D}}

 \def\p{\partial}
 
 \def\a{\alpha}
 \def\b{\beta}
 \def\g{\gamma}
 \def\d{\delta}

 \def\th{\theta}
 
 \def\k{\kappa}
 \def\l{\lambda}
 \def\m{\mu}
 \def\n{\nu}
 
 \def\r{\rho}

 \def\s{\sigma}
 
 \def\th{\theta}

 \def\G{\Gamma}
 \def\D{\Delta}

 \def\S{\Sigma}
 
 \def\O{\Omega}
 \def\o{\omega }

 \def\bn{\mathbf{n}}

 \def\bl{\boldsymbol{\lambda}}

 \makeatletter
\newsavebox{\@brx}
\newcommand{\llangle}[1][]{\savebox{\@brx}{\(\m@th{#1\langle}\)}%
  \mathopen{\copy\@brx\kern-0.5\wd\@brx\usebox{\@brx}}}
\newcommand{\rrangle}[1][]{\savebox{\@brx}{\(\m@th{#1\rangle}\)}%
  \mathclose{\copy\@brx\kern-0.5\wd\@brx\usebox{\@brx}}}
\makeatother

\def\beq{\begin{equation}}
\def\eeq{\end{equation}}

\makeatletter
\def\blfootnote{\xdef\@thefnmark{}\@footnotetext}
\makeatother

\title{{di-Langlands correspondence and extended observables}}
\author[a]{Saebyeok Jeong}
\author[b]{, Norton Lee}
\author[c]{, and Nikita Nekrasov}

\affiliation[a]{Department of Theoretical Physics, CERN, \\ 1211 Geneva 23, Switzerland}
\affiliation[b]{Center for Geometry and Physics, Institute for Basic Science (IBS),\\Pohang 37673, Republic of Korea}
\affiliation[c]{Simons Center for Geometry and Physics, \\ Stony Brook University, Stony Brook, NY 11794-3636, USA}
\emailAdd{saebyeok.jeong@cern.ch}
\emailAdd{nortonxs@gmail.com}
\emailAdd{nnekrasov@scgp.stonybrook.edu}

\preprint{CERN-TH-2023-220, CGP24002}

\vspace{2cm}
\abstract{We explore the \textit{difference Langlands correspondence} using the four dimensional ${\EN}=2$ super-QCD. Surface defects and surface observables play the crucial r{o}le. As an application, we give the first construction of the full set of quantum integrals, i.e. commuting differential operators, such that the partition function of the so-called regular monodromy surface defect is their joint eigenvectors in an evaluation module over the Yangian $Y(\fgl(2))$, making it the wavefunction of a $N$-site $\mathfrak{gl}(2)$ spin chain with bi-infinite spin modules. We construct the $\mathbf{Q}$- and $\tilde{\mathbf{Q}}$-surface observables which are believed to be the $Q$-operators on the bi-infinite module over the Yangian $Y(\fgl(2))$, and compute their eigenvalues, the $Q$-functions, as vevs of the surface observables.\blfootnote{
Dedicated to the 40th anniversary of the BPZ paper \cite{bpz}}}

\begin{document}

\maketitle

\section{Introduction}

This paper explores the BPS/CFT correspondence at the example of the ${\EN}=2$ asymptotically free super-QCD in four dimensions, i.e. gauge theory with $8$ supercharges,
with $SU(N)$ vector multiplet $(A_{\mu}, {\phi}, {\bar\phi})$, and $N_{f}= 2N$ hypermultiplets 
$(Q_{f}, {\tilde Q}^{f} )_{f=1}^{2N}$ in fundamental representation. The theory has been thoroughly explored, its two-derivatives low-energy effective action computed exactly, both indirectly using the constraints of electro-magnetic duality \cite{Seiberg:1994rs} and by the direct field theory computation employing equivariant localization \cite{Nekrasov:2002qd}. This approach works well for computations of supersymmetric partition functions and correlators of observables, commuting with a specific supercharge ${\mathfrak{Q}}$. 

This gauge theory has a string realization, making it possible to establish connections to
other gauge theories, e.g. maximally supersymmetric super-Yang-Mills theory on different spacetimes, and mathematical structures associated with them.  It would be very hard, if even possible, to envision these theoretical bridges without string theory framework. 
One such structure is the celebrated geometric Langlands (GL for short) program. Another related one is the mathematics of two dimensional conformal field theory, and that of quantum integrable systems. 

We explore the correlation functions of extended observables associated to two-dimensional surfaces
in Euclidean spacetime, and find their r{o}le in establishing the {\it difference Langlands correspondence}, or {\it di-Langlands} for short. Other names, recently appeared in the literature, are the $\hbar$-Langlands correspondence (for additive difference operators) and $q$-Langlands correspondence (for multiplicative difference operators). We prefer not to use these terms, in order not to confuse with the setting of quantum Langlands correspondence (which has to do with general $(q_1, q_2)$ or general $({\ve}_{1}, {\ve}_{2})$ $\Omega$-deformation reviewed below).

\paragraph{A few remarks on our terminology} From the four dimensional gauge theory we are extracting, via taking the vacuum expectation values, a set of functions of various parameters of the theory. These functions will be organized into a vector space ${\CalH}$. 
The gauge theory operators whose expectation values produce vectors in $\CalH$ will be called {\it observables}, and denoted using the boldface font, e.g. ${\bf O}, {\bf Q}, {\bf Y}, {\bf H}$-observables. 
The expectation values of some of these observables will be called {\it functions}, and denoted using the regular font, e.g. $O$, $Q$, $H$-functions. Some of observables correspond to defects, e.g. ${\bf\Psi}$, so they expectation values will be called the {\it vectors}, e.g. ${\Psi} \in {\CalH}$:
\beq
{\Psi} = \frac{\langle {\bf\Psi} \rangle}{\langle 1 \rangle}
\eeq
Some of our observables have the bonus property
that they can be expressed as polynomials in the topological charges, up to $\mathfrak{Q}$-exact terms. In this case the insertion of such an observable in the correlation function
computing $\Psi$
is equivalent to action on the $\Psi$ by differential operators in topological fugacities, e.g. complexified theta-angles, K\"{a}hler moduli, etc. In this case we shall call such an observable and {\it operator}, and denote its action as
\beq
{\hat O} {\Psi} : = \frac{\langle {\bf O} {\bf \Psi} \rangle}{\langle 1 \rangle}
\eeq
Another remark concerns the use of {\it gauge origami}. This is a non-commutative gauge theory setup inspired by IIB string theory D-brane configuration, which geometrizes the ${\EN}=4$ gauge theory together with codimension two and codimension four defects, as well as the truncated version of the theory, corresponding to orbifolding the closed string background by a subgroup of $SU(4)$, preserving the geometry of branes. 
It is in this frame of mind that we often approach the gauge theory of interest: the $A_{1}$-quiver gauge theory: $SU(N)$ theory with $2N$ fundamental flavors. It is the limit ${\qe}_{1}, {\qe}_{2} \to 0$
of the ${\hat A}_{2}$-quiver gauge theory, which is the theory with the
gauge group $SU(N)_{0} \times SU(N)_{1} \times SU(N)_{2}$, with bi-fundamental hypermultiplets in $({\bf N}_{1}, {\bar{\bf N}}_{0})$, $({\bf N}_{2}, {\bar{\bf N}}_{1})$, 
$({\bf N}_{0}, {\bar{\bf N}}_{2})$-representations. In our conventions, the ten-dimensional space-time is viewed as a product ${\BC}_{1} \times {\BC}_{2} \times {\BC}_{3} \times {\BC}_{4} \times {\BC}_{5}$, with the physical spacetime of gauge theory being the product ${\BC}^{2}_{12} = {\BC}_{1} \times {\BC}_{2}$, spanned by $N$ D3-branes. The defects are produced by 
adding D3-branes stretched along ${\BC}^{2}_{34} = {\BC}_{3} \times {\BC}_{4}$, giving rise to {\it $qq$-observables}, and/or ${\BC}^{2}_{13}$ or ${\BC}^{2}_{14}$, producing
{\it $Q$- or ${\tilde Q}$-observables}. To produce the ${\hat A}_{2}$-quiver theory
one performs the orbifold by the ${\BZ}_{3}$-group acting via $(z_3, z_4) \mapsto
( e^{\frac{2\pi\ii}{3}} z_3, e^{-\frac{2\pi\ii}{3}} z_4)$

Finally, one of the main applications of gauge theory is seen in the limit ${\ve}_{2} \to 0$. We shall call such a specialization of the general $\Omega$-background the $\Omega_1$-background, thereby stressing that $\BC_1 \times \BC_3 \times \BC_4$ planes are equivariant topological, and $\BC_2$ is topological. 

\subsection{The main results and outline}

The main results of the paper are: the identification of $\Psi$ with the state of quantum $N$-site integrable $\mathfrak{gl}(2)$ XXX spin chain, whose inhomogeneities, spins, and reference charges (the parameters of the HW-modules \cite{Nekrasov:2021tik} which are the local site representations of $\mathfrak{gl}(2)$) are determined by the masses and Coulomb parameters in the units of $\hbar = \ve_1$, the $\Omega_1$-background parameter, while the twist $\qe$ is determined by the gauge coupling and the theta angle of the theory. We construct the full set of quantum integrals of motion (QIM). The conventional way of organizing them uses the transfer matrix formalism and representation theory of the Yangian $Y(\mathfrak{gl}(2))$.  There are constructions in terms of supersymmetric interfaces in lower dimensional supersymmetric gauge theories \cite{Aganagic:2017gsx, Dedushenko:2021mds}, but these apply to finite dimensional representations of the Yangian (or bounded modules for affine quivers). We do not quite follow this route.

The outline of the paper is as follows. In section \ref{sec:gauge}, we recall the setup of the theory, both in field theory language, and through string and M-theory constructions. In section \ref{sec:Qopssrf}, we focus on specific surface observables, called ${\bf Q}(x)$ and ${\tilde{\bf Q}}(x)$-observables. In section \ref{sec:tq}, we review the argument showing 
the expectation values $Q(x)/{\tilde Q}(x)$ obey the Baxter's TQ equation. We then study the fractionalization of $\mathbf{Q}/\tilde {\mathbf{Q}}$-observables in the presence of the monodromy surface defect. In section \ref{sec:hbarlang}, we show their vacuum expectation values obey a set of linear equations (which we prove by an extension of the method employing the regularity of partition functions of the gauge origami models), which can be organized in the form of Lax operator formalism of the Leningrad school \cite{Faddeev:1996iy}. This allows us to identify the
action of the 
trace of the twisted transfer matrix on a state of the spin chain (realized by differential operators
in parameters of the monodromy surface defect) with the operator product expansion of the local bulk chiral operators with the said surface defect. Using supersymmetry and cluster decomposition, we demonstrate the surface defect is an eigenvector of the transfer matrix, with the eigenvalue given by the vacuum expectation value of these bulk operators. As a result, we establish the di-Langlands to infinite and bi-infinite dimensional setting (in the usual considerations of mathematicians only bounded or finite dimensional modules are considered). Finally, we conclude with discussions in section \ref{sec:discussion}. The appendices contain a brief review of the Yangian algebra and Manin matrices, and some computational details.

\paragraph{Acknowledgement}
The authors thank Mina Aganagic, Kevin Costello, Mykola Dedushenko, Chris Elliott, Alba Grassi, Nathan Haouzi, Nafiz Ishtiaque, Shota Komatsu, Jihwan Oh, Andrei Okounkov, Miroslav Rap\v{c}\'{a}k, and Yehao Zhou for discussions and collaboration on related subjects. SJ is grateful to Du Pei for helpful discussion and support during his visit to Center for Quantum Mathematics at University of Southern Denmark, where a part of the work was done. The work of SJ is supported by CERN and CKC fellowship. The work of NL is supported by IBS project IBS-R003-D1. Research of NN is partly supported by NSF grant 2310279. 

\section{The webs of gauge and string theories} \label{sec:gauge}

In this section we briefly review the interconnections between the quantum $4d$ ${\EN}=2$ supersymmetric gauge theory and two and three dimensional gauge theories whose moduli spaces of special classical solutions describe the moduli spaces of vacua of the four dimensional theories. We also briefly describe the surface defects and observables, and what is known or conjectured about their presentations in the framework of gauge and string theories in various dimensions. Finally, we review the geometric Langlands correspondence and its extension $-$ di-Langlands correspondence $-$ which we develop in this paper in the specific case. 

\subsection{Four dimensional quantum field theory}

The equivariant localization approach to exact computations in quantum field theory uses two deformations.

\subsubsection{Noncommutativity and stability}

The first, curing the ultraviolet singularities of the moduli spaces of instantons deforms the Euclidean space-time ${\BR}^{4}$ to its noncommutative version ${\BR}^{4}_{\Theta}$, with the flat coordinates
$X^{i}$ obeying
\beq
[ X^{i}, X^{j} ] = {\ii} {\Theta}^{ij}\, , \ 
\eeq
with constant antisymmetric tensor ${\Theta}$, 
while keeping the metric Euclidean
\beq
g = \sum_{i=1}^{4} dX^{i} dX^{i}
\eeq
By an $SO(4)$ rotation of $X^i$'s the tensor $\Theta$ can be brought to the normal form
\beq
{\Theta} = {\theta}_{1} \frac{\partial}{\partial x^{1}} \wedge \frac{\partial}{\partial x^{2}} + {\theta}_{2} \frac{\partial}{\partial x^{3}} \wedge \frac{\partial}{\partial x^{4}}
\label{eq:normfth}
\eeq
The important for the structure of our observables is the sign of the combination
\beq
{\zeta}_{\BR} = {\theta}_{1} + {\theta}_{2} 
\eeq
We shall call the theory at ${\zeta}_{\BR} > 0$ the {\it holomorphic phase}, and that at
${\zeta}_{\BR}< 0$ an {\it anti-holomorphic phase}.

\subsubsection{Equivariance with respect to rotations}

The second deformation cures the infrared divergencies, by placing the theory in an effectively rotating frame, creating a potential well for localized field configurations attempting to run away to infinity. This is achieved by the $\Omega$-deformation, i.e. a
supergravity background, whose effect on the bosonic part of the gauge multiplet kinetic term is to replace the covariant derivatives $D_{A}{\phi}$, $D_{A}{\bar\phi}$ and the commutators $[{\phi}, {\bar \phi}]$ of the adjoint scalars in the vector multiplet by
\beq
D_{A}{\phi} + \iota_{V_{\ve}} F_{A}\, , \ D_{A}{\bar\phi} + \iota_{V_{\bar\ve}} F_{A}\, , \ [ {\phi}, {\bar\phi}] + \iota_{V_{\ve}} D_{A}  {\bar\phi} - \iota_{V_{\bar\ve}} D_{A} {\phi} + \iota_{V_{\ve}}\iota_{V_{\bar\ve}} F_{A}\, , 
\eeq
respectively, with the vector fields $V_{\ve}, V_{\bar\ve}$ being the infinitesimal $SO(4)$ rotations, preserving $\theta$ \eqref{eq:normfth}:
\beq
\begin{aligned}
& V_{\ve} = {\ve}_{1} \left( X^{2} \frac{\partial}{\partial X^{1}}  - X^{1} \frac{\partial}{\partial X^{2}} \right) +  {\ve}_{2} \left( X^{4} \frac{\partial}{\partial X^{3}}  - X^{3} \frac{\partial}{\partial X^{4}} \right) \\
& V_{\bar\ve} = {\bar\ve}_{1} \left( X^{2} \frac{\partial}{\partial X^{1}}  - X^{1} \frac{\partial}{\partial X^{2}} \right) +  {\bar\ve}_{2} \left( X^{4} \frac{\partial}{\partial X^{3}}  - X^{3} \frac{\partial}{\partial X^{4}} \right) \end{aligned}
\label{eq:veve}
\eeq
Here ${\ve}_{1}, {\ve}_{2} \in {\BC}$ are the two complex parameters of the $\Omega$-deformation, and ${\bar\ve}_{1}, {\bar\ve}_{2}$ are their conjugates. 

In this paper, we shall be often working in the limit ${\ve}_{2} \to 0$, with ${\ve}_{1} \neq 0$. We call this particular case of $\Omega$-deformation the $\Omega_{1}$-{background}.

In this paper we are going to study the gauge theory observables, both local and extended, 
which supercommute with a supercharge ${\mathfrak{Q}}_{\ve}$ preserved by the $\Omega$-background. 

One such class of observables are the gauge invariant functions of $\phi (X)$, evaluated
at the fixed point locus $X = X_{*}$ of $V_{\ve}$, i.e. where $V_{\ve}(X_{*}) = 0$:
\beq
{\bf O}^{(0)}_{k}(X) = {\rm tr} {\phi}(X)^{k}\, , \ k = 2, 3, \ldots , N
\label{eq:locobs}
\eeq
Classically, all gauge invariant polynomial functions of ${\phi}(X)$ can be expressed polynomially in \eqref{eq:locobs}. In performing the instanton calculations it is convenient to operate with all single-trace operators at once, by defining the generating function, the $Y$-observable, 
\beq
{\bf Y}(x) = x^{N} \, {\exp} \, \sum_{k=1}^{\infty} - \frac{1}{k x^{k}} {\bf O}^{(0)}_{k}
\eeq
where the $X$-dependence is omitted, while the auxiliary (at this stage) variable $x \in E$
is introduced. Here $E$ is a valuation group for the equivariant parameters of global symmetries. For four dimensional theories $E \approx {\BC}$ (which we shall call, more specficially, ${\BC}_{5}$ in what follows), with additive structure. Our theory admits a lift to five dimensions, where $E \approx {\BC}^{\times}$, and related theories admit lifts to six dimensions, where $E$ becomes a compact elliptic curve.
This observable survives the limit ${\ve}_{1}, {\ve}_{2} \to 0$, although its expectation value exhibits a complicated analytic behavior. To describe it we need to specify a few more details about our gauge theory. Having $N_{f}= 2N$ means the ultraviolet theory is superconformal, and is characterized by the microscopic gauge coupling ${\rm e}^2$ and theta angle ${\vartheta}$, conveniently combined into the parameter 
\beq
{\qe} = {\exp} \, 2\pi \ii {\tau} :  = \ e^{{\ii} {\vartheta} - \frac{8\pi^2}{{\rm e}^{2}}} 
\label{eq:qpar}
\eeq
The $2N$ hypermultiplets are characterized by the set 
$\{ m_{1}, \ldots, m_{2N} \} \subset E$. We shall often use
\beq
P(x) = \prod_{f=1}^{2N} (x - m_{f})
\label{eq:masspol}
\eeq
to encode the masses. 

Finally, a vacuum state of the theory is characterized by the asymptotic eigenvalues
of ${\phi}(x)\,  \xrightarrow{x \to \infty} \, {\rm diag} ( a_{0}, \ldots, a_{N-1} )$.
The set ${\bf a} = \{ a_{0}, \ldots, a_{N-1} \} \subset E$ will be a parameter of our correlation functions. The normalized vacuum expectation value of an observable
${\bf O}$ will be denoted as:
\beq
O_{\bf a} = \frac{\langle {\bf O} \rangle_{\bf a}}{\langle 1 \rangle_{\bf a}}
\label{eq:oa}
\eeq
We can now recall the structure of the moduli space of vacua of the theory, i.e. the behavior of correlators in the limit ${\ve}_{1}, {\ve}_{2} \to 0$ \cite{Nekrasov:2003rj}. The expectation values of ${\CalO}_{k}^{(0)}$ can be recovered from two pieces of information:
first, define the hyperelliptic curve 
\beq
 y + {\qe} \frac{P(x)}{y} = t (x) : = t_{0} x^{N} + t_{1} x^{N-1} + u_{2} x^{N-2} + \ldots
 + u_{N}
\label{eq:swcurve}
\eeq
where 
\beq
t_{0} = 1 + {\qe}\, , \ t_{1} = - {\qe} \sum_{f} m_{f} - (1 - {\qe}) \sum_{\alpha} a_{\alpha}
\eeq
while
\beq
{\bf u} = (u_{2}, u_{3}, \ldots, u_{N}) \in {\CalU} \approx {\BC}^{N-1}
\eeq
implicitly determined from requiring $A$-periods of the differential $\frac{1}{2\pi\ii} x \frac{dy}{y}$ to be equal to $a_1, \ldots , a_{N-1}$;  
second 
$(x, Y_{\bf a} (x))$ is a branch of $(x,y)$ which asymptotes to $(x, x^{N})$ for $x \to \infty$. 

\subsection{Two and three dimensional classical gauge theories and integrability} \label{subsec:dualp}

The curve \eqref{eq:swcurve} is called the Seiberg-Witten curve. It has been long known to coincide with the spectral curves of several classical algebraic integrable systems: an XXX Heisenberg $\mathfrak{sl}(2)$ spin chain with $N$ spin sites, or a Gaudin-Garnier $\mathfrak{sl}(N)$ spin chain with $4$ spin sites. 

There are both physical and mathematical aspects of such identification. 
Mathematically it means that the total space of fibration of Jacobians of the curves \eqref{eq:swcurve} over the affine space ${\CalU}$ of polynomials $t(x)$ is a holomorphic symplectic manifold $({\CalP}, {\omega}^{\BC})$, with the Lagrangian projection ${\pi}: {\CalP} \to {\CalU}$ having polarized abelian varieties as fibers. The Coulomb moduli  $\bf a$ are the holomorphic action variables, they are supplemented with ${\bf a}_{D}$, a dual set of periods, such that
\beq
{\omega}^{\BC} = \sum_{i=1}^{N-1} da_{i} \wedge d{\varphi}^{i} = \sum_{i=1}^{N-1} da_{D}^{i} \wedge d{\varphi}^{D}_{i}
\eeq
where the corresponding angle variables ${\varphi}^{i}$ on the fiber ${\pi}^{-1}(t(x))$
defined up to a lattice
\beq
{\varphi}^{i} \sim {\varphi}^{i} + 2\pi n^{i} + 2\pi \sum_{j} {\tau}^{ij} m_{j} 
\, ,\ n^{i} , m_{j} \in {\BZ}
\eeq
with ${\varphi}^{D}_{i} = \sum_{j} {\tau}^{-1}_{ij} {\varphi}^{j}$, and ${\tau}^{ij}$
the period matrix, obeying
\beq
{\tau}^{ij} = \frac{{\partial}^{2} {\mathcal{F}}}{{\partial} a_{i} {\partial} a_{j}}\, , \
a_{D}^{i} = \frac{{\partial} {\mathcal{F}}}{{\partial} a_{i}}
\eeq
with the {\it prepotential} ${\mathcal{F}}({\bf a})$ governing the ${\ve}_{1}, {\ve}_{2} \to 0$
limit of the partition function
\beq
\langle 1 \rangle_{\bf a} \sim e^{\frac{{\mathcal{F}}({\bf a})}{{\ve}_{1}{\ve}_{2}}}
\label{eq:pfnprep}
\eeq
In addition to this action-angle picture of ${\CalP}$, identifying it with such and such integrable model usually means having another Lagrangian foliation ${\CalP} \to X$,  the coordinates on $X$ playing the role of coordinates (suitable coordinates on the fiber being momenta). For example, Hitchin system ${\CalP} \approx \EM_H$ is almost isomorphic to $T^{*}\text{Bun}_{G}(C,S)$, for some curve $C$ and colored divisor $S$
so that $X$ is a moduli space of parabolic $G$-bundles. Specifically for the subject of this paper, $G = SL(N)$, $C \approx {\BP}^{1}$, and $S = {\bf\mu}_{0} 0 + {\bf\mu}_{\qe} {\qe} + {\bf\mu}_{1} 1 + {\bf\mu}_{\infty} {\infty}$ are four points $\{ 0, {\qe}, 1, {\infty} \}$ 
where ${\bf\mu}_{\alpha}$ are conjugacy classes in $G$, with ${\bf\mu}_{0}, {\bf\mu}_{\infty}$ being diagonal matrices with distinct eigenvalues, while ${\bf\mu}_{\qe}, {\bf\mu}_{1}$ are the diagonal matrices with eigenvalues of multiplicity $(N-1,1)$. The special nature of a genus zero Hitchin system is that it can be described by a finite dimensional Hamiltonian reduction, specifically in our case (it is trivial to generalize this formalism to the case of multiple punctures \cite{Jeong:2018qpc}): 
\beq
{\CalP} = \left( {\CalO}_{{\bf\mu}_{0}} \times {\CalO}_{{\bf\mu}_{\qe}} \times {\CalO}_{{\bf\mu}_{1}} \times {\CalO}_{{\bf\mu}_{\infty}} \right)^{\rm stable}/\!\!/G
\eeq
where ${\CalO}_{\bf\nu} \subset \mathfrak{g} \approx \mathfrak{g}^{*}$ is the coadjoint orbit of $\bf\nu$. The Higgs field ${\phi}_{w} dw$ of Hitchin's equations is a meromorphic $1$-form on $C$ valued in $\mathfrak{g}$,
\beq
{\phi}_{w} dw = J_{0} \frac{dw}{w} + J_{\qe} \frac{dw}{w-{\qe}} + J_{1} \frac{dw}{w-1} 
\, , \ J_{\infty} = - J_{0} - J_{\qe} - J_{1}
\eeq
with $J_{\alpha} \in {\CalO}_{{\bf\mu}_{\alpha}}$. Now let us remember that $G = SL(N)$, so that each orbit, and all of $\CalP$, can be given a quiver variety description (the physical significance of this picture is justified in going to three dimensions and applying mirror symmetry), as a symplectic quotient $T^* {\sE}/\!\!/{\sG}$ of the cotangent bundle  $T^* {\sE}$ of the vector space ${\sE}$ of homomorphisms between the vector spaces attached to the end-points of oriented edges of star-shaped graph, by the action of the group ${\sG}$ of general linear transformations of these vector spaces. 
By keeping only one of the two homomorphisms per edge, and taking a quotient instead of the symplectic quotient, one arrives at ${\sE}/{\sG}$, a stand-in for $\text{Bun}_{G}(C,S)$. 

The beauty of Hitchin system is that ${\CalP}$ has not only  the complex structure $I$ and the symplectic structure ${\omega}_{\BC} =: {\Omega}_I$, but also a metric $g$, which is K\"{a}hler with respect to the complex structure, i.e. defines a $(1,1)$-type symplectic form ${\omega}_{\BR} = g(I \cdot , \cdot)$, and two more complex structures $J,K$, such that: $IJK = I^2 = J^2 = K^2 = -1$, and ${\Omega}_{I} = g (J \cdot, \cdot) + {\ii} g (K {\cdot}, {\cdot})$. This metric $g$ stems from the identification of ${\CalP}$ with the moduli space of solutions to Hitchin equations  with sources at $\{ 0, {\qe}, 1, {\infty} \}$ on a pair $({\bf A}, {\bf\Phi})$ of a gauge field and adjoint-valued one-form, living on $C \backslash \{ 0, {\qe}, 1, {\infty} \}$. The hyperk\"{a}hler metric $g$ depends on a conformal class $[h_{C}]$ of a metric on $C\backslash \{ 0, {\qe}, 1, {\infty} \}$. 
The $(I, {\omega}_{\BC})$-part of data does not depend on $g$ and $[h_{C}]$. However, for special choices of $[h_{C}]$ one finds yet another interpretation of ${\CalP}$. Namely, by making $C \backslash \{ 0, {\infty} \}$ a flat cylinder ${\BR} \times S^{1}_{R}$, and temporarily forgetting about ${\qe}, 1$ one can interpret Hitchin's equations as a loop 
space version of Nahm equations describing $SU(2)$-monopoles, i.e. solutions to Bogomolny equations
\beq
D_{\sl A} {\sigma} + \star F_{\sl A} = 0
\eeq
 on ${\BR}^{2} \times S^{1}_{1/R}$, with $N$ Dirac monopole-like singularities of $-1$ charge, and $N$ Dirac monopole-like singularities of $+1$ charge. 
Here ${\BR}^{2} \approx {\BC}$ is the \textit{target} of the Higgs field ${\bf\Phi}
\in {\BC} \otimes {\mathfrak{sl}({N})} $ of Hitchin's equations, while the ${\BR}$ in ${\BR} \times S^{1}$ of Hitchin's equation is the \textit{target} of the Higgs field $\sigma
\in {\BR} \otimes {\mathfrak{sl}({2})}$ of Bogomolny equations. The circle $S^{1}_{R}$ of the cylinder supporting Hitchin data is dual to the circle $S^{1}_{1/R}$ of ${\BR}^{2} \times S^{1}_{1/R}$ supporting Bogomolny data, i.e. one parametrizes flat $U(1)$-connections on the other. The positions of $\pm 1$ Dirac monopoles map to the ${\mu}_{\infty}$-(${\mu}_{0}$-)monodromy data on Hitchin's side, respectively. The cross ratio $\qe$ of the four points $\{ 0, {\qe}, 1, {\infty} \}$ determines the asymptotics of 
\beq
g(x) \sim \text{P}\exp \oint_{(x_{1}, x_{2})  \times S^{1}_{1/R}} ( A + {\sigma})
\eeq 
at $(x_{1}, x_{2}) \to \infty$ in ${\BR}^{2}$. 

The identification with the monopole moduli space (which goes by Nahm transform)
gives another interpretation of $\CalP$ as an integrable system. Perhaps the short way of describing it is as a Hamiltonian reduction of $T^{*}{\CalA}_{\widehat{L{\check{G}}}}({\check{C}}, {\check{D}})$, with ${\CalA}_{\widehat{L{\check{G}}}}$ the space of  $(0,1)$-$\widehat{L{\check{G}}}$-connections on a trivial bundle over ${\check{C}} = {\BC} \cup {\infty}$, with $\check{G} = SL(2)$. Since affine Lie algebra $Lie({\widehat{L{\check{G}}}}) = {\BC} \oplus L{\check{\mathfrak{g}}}$, i.e. central extension of the loop algebra $L{\check{\mathfrak{g}}}$, is not isomorphic
to its dual $Lie({\widehat{L{\check{G}}}})^{*} \approx$ the space of $\lambda$-connections on $S^{1}$, the Higgs field has different nature from the gauge field.

\subsection{Langlands duality for algebraically integrable systems}

The overused name {\it Langlands duality} borrowed from the original Langlands program in the theory of automorphic functions is now used in the geometric context mostly in reference to Hitchin system $\EM_H$. The reason is that the moduli space of principal $G$-bundles over an algebraic curve has a double coset presentation similar to the one underlying the automorphic side of the number-theoretic Langlands program, while the fundamental group of the curve is an analogue of the Galois group. The representations of the fundamental group in the $^{L}G$-group are the local systems, whose moduli space is the phase space of the Hitchin system for the $^{L}G$ gauge group, albeit in the $J$ complex structure. 

The physics approach to the geometric Langlands uses the mirror symmetry of the two dimensional ${\EN}=(4,4)$ supersymmetric sigma model with $\EM_H$ target. More specifically, using the algebraic integrable structure of $\EM_H$ in the $I$ complex structure, where it presents itself as a fibration of abelian varieties, one employs the fiber-wise T-duality, passing to the dual abelian varieties. Of course, this is an approximate picture, valid far away from the discriminant locus. The magic of ${\EN}=(4,4)$ theory is that this approximate  picture completes to the exact equivalence of superconformal field theories.

The consequences of this equivalence for the Fukaya/$D^{b}(\text{Coh}\,\EM_H)$-category (the hyperk\"{a}hler nature of $\EM_H$ implies the multi-faceted nature of D-branes in the theory) can be very interesting mathematically. For example, the quantization of Hitchin system \cite{BD1} viewed as  a commutative subalgebra of $K^{\frac 12}_{\EM_H}$-twisted differential operators on $\EM_H$ of the algebra of morphisms $\text{End}({\EB}_{cc})$ of the canonical coisotropic brane on $\EM_H$, maps, under the mirror symmetry, to the Lagrangian brane of $^{L}G$-opers \cite{BD2}-- special holomorphic flat $^{L}G$-connections.

\subsection{$\bf Q$- and ${\tilde{\bf Q}}$-observables, surface defects, and spin chains}

Gauge theories have lots of extended observables. Wilson and 't Hooft operators are associated to one-dimensional and codimension-three chains. There are codimension-two defects defined by prescribing the singularity of the gauge field near the support of the defect. In four dimensional theory of our interest these are called monodromy surface defects. 

The main object of study in this paper are the $\mathbf{Q}$- and ${\tilde {\mathbf{Q}}}$-observables, which we define as peculiar surface observables in gauge theory. 

Recall that Donaldson theory (a twisted version of pure ${\EN}=2$ super-Yang-Mills theory) 
has the surface observables
\beq
{\bf O}^{(2)}_{k, \Sigma} = \int_{\Sigma} \, k \, \text{Tr}\,  {\phi}^{k-1} F_{A} + {\rm fermions}
\label{eq:otwo}\eeq
which are defined by the slant product of the universal characteristic class associated to 
${\bf O}_{k}$ with the two dimensional homology cycle $\Sigma$ in spacetime. In the theory
on ${\BR}^{4}$ there are no nontrivial two-cycles, moreover equivariantly every ${\mathfrak{Q}}_{\ve}$-descendant
of ${\bf O}_{k}$ such as \eqref{eq:otwo} is equivalent to a local observable inserted at
the fixed point of $V_{\ve}$. However, one can define something nontrivial using infinite-dimensional bundles. Specifically, given a $V_{\ve}$-invariant two dimensional submanifold
$\Sigma \approx {\BR}^{2}$, one can take the space ${\CalE}_{\Sigma}$ of solutions of Dirac equation $\slashed{D}_{B} {\Psi} = 0$ on $\Sigma$ with the gauge field $B = A |_{\Sigma}$ restricted from the bulk. The characteristic classes of ${\CalE}_{\Sigma}$ viewed as a bundle over the gauge equivalence classes of gauge fields on ${\BR}^{4}$ are the observables we are interested in. Were ${\CalE}_{\Sigma}$ finite dimensional, we would organize them into the Chern polynomial
\beq
{\CalQ}_{\Sigma} (x) = \sum_{k=0}^{{\rm rk}{\CalE}_{\Sigma}} (-)^{k} x^{{\rm rk}{\CalE}_{\Sigma} - k} c_{k}({\CalE}_{\Sigma})
\label{eq:naiveq}
\eeq
But ${\CalE}_{\Sigma}$ is infinite-dimensional (like a the first Landau level on a infinite plane) so \eqref{eq:naiveq} does not make literally sense. 
Nevertheless, with the help of $\Omega$-deformation \eqref{eq:naiveq} can be defined
by using $\zeta$-regularization. For example, for $\Sigma = {\BC}_{1} \subset {\BC}^{2}$ with $A=0$ the space ${\CalE}_{\Sigma} \approx {\BC}[z]$ is the space of all holomorphic functions on $\BC$, viewed equivariantly with respect to the group $U(1)$ of rotations 
$z \mapsto z e^{\ii {\alpha}}$. Then
\beq
{\CalQ}_{\Sigma}(x) \sim x (x - {\ve}_{1}) (x - 2 {\ve}_{1}) \ldots \sim
{\exp} \, \frac{d}{ds} \Biggr\vert_{s=0} \, \frac{1}{{\Gamma}(s)} \int_{0}^{\infty} \frac{dt}{t} t^{s} \frac{e^{tx}}{1-e^{-t{\ve}_{1}}} : = {\BE} \left[ \frac{e^{x}}{1-q_{1}^{-1}} \right]
\eeq
where we introduced the notation ${\BE} [ \,{\cdots} \, ]$ for plethystic exponent.  The general instance of $\bf Q$-observable associated with surfaces, invariant under the rotations of $\Omega$-background is defined in the main body of the paper.

\subsection{Topological sigma model with Hitchin target space}
In this paper, the $\EN=2$ theory in consideration is the class $\CalS$ theory \cite{gai1} of type $A_{N-1}$ associated to the cylinder $\BC^\times = \BP^1 \setminus \{0,\infty\}$ with $n$ marked points $S = \{p_1,p_2,\cdots, p_n \}\subset \BC^\times$, which is the linear $A_{n-1}$-quiver $SU(N)$ gauge theory in an appropriate S-duality frame. The corresponding Coulomb branch is the moduli space $\EM_H (SU(N),\BP^1;S)$ of Hitchin's equations on $\BP^1$ with (regular) ramification data at $S \cup \{0,\infty\}$ \cite{Gaiotto:2009hg}. 
The main example will be the case $n=2$. Some details for the general $A_{n-1}$-quiver theory are discussed in \cite{GNtoappear}

To see the Hitchin moduli space at hand, we consider the $\EN=2$ theory subject to the Donaldson-Witten twist \cite{cmp/1104161738} and further to the $\Omega_{\ve_1,\ve_2}$-background associated to the $U(1)^2$ isometry of the worldvolume $\BC^2$ \cite{Nekrasov:2002qd}. Then we compactify this theory along the torus $T^2$, where $\BC^2$ is viewed as a $T^2$ fibration over $\S = \BR^+ \times \BR^+$, to obtain a two-dimensional $\EN=(2,2)$ sigma model of maps from the worldsheet $\S$ to the Hitchin moduli space target $\EM_H (SU(N),\BP^1;S)$ \cite{Bershadsky:1995qy,Harvey:1995tg}. As a cohomological field theory in the preserved supercharge, it is a topological sigma model; either an $A$-model associated with one of the symplectic structures (with a B-field in general) or a $B$-model associated with one of the complex structures. This association is determined by the ratio $\k = -\frac{\ve_2}{\ve_1}$ of the $\O$-background parameters in the original four-dimensional $\EN=2$ theory \cite{Nekrasov:2010ka,Jeong:2023qdr}. 

In our recent study \cite{Jeong:2023qdr}, it was demonstrated that certain types of branes and functors acting on them descend from half-BPS surface defects in the $\EN=2$ gauge theory supported on a complex plane in $\BC^2$. Specifically, a regular monodromy surface defect descends to a brane of $\l$-connections, while a canonical surface defect leads to a Hecke operator. The magnetic eigenbrane corresponding to an oper, on which the Hecke operators act diagonally, was revealed to have its origin in the boundary condition at infinity, where the opers are parametrized by the Coulomb moduli \cite{Jeong:2018qpc,Jeong:2023qdr}. We showed that the vacuum expectation values and the correlation functions of surface defects satisfy the constraints for the sections of corresponding twisted $\ED$-modules. Moreover, it was suggested that this $\EN=2$ gauge theoretical formulation of the geometric Langlands correspondence (with ramifications) \cite{BD1,BD2,Frenkel:2005ef,Frenkel:2006nm} is related to more conventional approach of a topologically twisted (GL-twisted) $\EN=4$ gauge theory \cite{Kapustin:2006pk,gukwit2,Gaiotto:2016hvd,Frenkel:2018dej} by a string duality. See also \cite{Frenkel:2005pa,Teschner:2010je,Balasubramanian:2017gxc,Teschner:2017djr} for vertex algebra approach of the geometric Langlands correspondence.

\subsection{di-Langlands correspondence from $\EN=2$ gauge theory}
The goal of the present work is to view the geometry of the same moduli space from a dual perspective (see section \ref{subsec:dualp}). The Hitchin moduli space $\EM_H (SU(N),\BP^1;S)$ is known to be isomorphic as a hyper-K\"{a}hler space to the moduli space of periodic $U(n)$-monopoles on $\BP^1 \times S^1$ with a framing at $\infty \in \BP^1$ and Dirac singularities at $D\times \{0\} $, where $D = \{m_0^+ ,m_1 ^+ ,\cdots, m_{N-1} ^+ \} \subset \BP^1 \setminus \{\infty\} = \BC$ \cite{Cherkis:2000cj,Cherkis:2000ft,Nekrasov:2012xe,Nikita-Pestun-Shatashvili,Elliott:2018yqm}. In particular, these two moduli spaces are related to each other by a Nahm transform. 
From this presentation, it also follows that it is isomorphic to the moduli space $\EM_{\text{mHiggs}} (GL(n),\BP^1;D)$ of \textit{multiplicative} $GL(n)$-Higgs bundles on $\BP^1$ with a framing at $\infty \in \BP^1$ and regular singularities at $D$ \cite{Cherkis:2000cj,charbonneau2011,Elliott:2018yqm}.

Due to the isomorphism, the target space of the aforementioned topological sigma model can now be viewed as $\EM_{\text{mHiggs}} (GL(n),\BP^1;D)$, and it is tempting to reevaluate our findings from the $\EN=2$ gauge theory within this perspective. This shift leads to an exploration of relationships between geometric structures on $\EM_{\text{mHiggs}} (GL(n),\BP^1;D)$. The $\EN=2$ gauge theory of class $\CalS$ thus offers a methodology to investigate the \textit{$\hbar$-deformation} of the geometric Langlands correspondence.

The geometric Langlands correspondence has two sides $-$ the \textit{automorphic} side and the \textit{Galois} side $-$ both of which have a respective $\hbar$-difference analogs defining the di-Langlands (``di'' for difference) correspondence. On the Galois side, (quasi-coherent sheaves on) the moduli space of local systems in the ordinary case gets replaced by the moduli space of \textit{$\hbar$-difference connections} on $\BP^1$ with a framing at $\infty \in \BP^1$ and regular singularities at $D$. We focus on a special subspace spanned by $\hbar$-opers, for which the $\hbar$-difference connections can be expressed by scalar $\hbar$-difference operators. In our $\EN=2$ gauge theory setup, we construct the $\hbar$-opers as the quantized chiral ring relation for  $\bf Q$-observable \cite{Jeong:2018qpc,Jeong:2023qdr}.\footnote{See \cite{Jeong:2019fgx} for the derivation of the chiral ring relation in the absence of surface observables.} The vacuum expectation value $Q(\ba;x)$ of the $\bf Q$-observables in the limit $\ve_2 \to 0$ is shown \cite{Jeong:2017pai,Jeong:2018qpc,Jeong:2023qdr} 
to be solutions to the $\hbar$-oper 
\begin{align}
    0 = \left[1 - t(\ba;x) e^{-\hbar \p_x} + \qe P(x) e^{-2\hbar \p_x} \right]Q(\ba;x), \qquad \text{ for } n=2.
\end{align}

On the automorphic side, we expect twisted $\ED$-modules appearing in the ordinary case to be replaced by \textit{$\hbar$-difference modules} on the moduli space $\text{Bun}_{GL(n)} (\BP^1;D)$ of parabolic $GL(n)$-bundles over $\BP^1$ with a framing at $\infty \in \BP^1$ and parabolic structures at $D$. Our aim will not be formulate this object geometrically as a sheaf of modules, as our methodology lies in the $\EN=2$ gauge theory. Instead, we only note that the $\hbar$-difference modules on $\text{Bun}_{GL(n)} (\BP^1;D)$ should arise from the $(\EB_{cc},\EB')$-strings stretched between a canonical coisotropic brane $\EB_{cc}$ and another $A$-brane $\EB'$, on which the $(\EB_{cc},\EB_{cc})$-strings act by joining the strings. The latter gives the quantized algebra of holomorphic functions on the moduli space $\EM_{\text{mHiggs}} (GL(n),\BP^1;D)$, which is known to be a module over the Yangian $Y(\fgl(n))$ \cite{Nekrasov:2012xe,Nikita-Pestun-Shatashvili,Elliott:2018yqm}. 

Our proposal is to realize the Yangian module in the vector space spanned by vacuum expectation values of monodromy surface defects in the $\EN=2$ gauge theory, 
defined over a $\BZ$-lattice of Coulomb vacua, specified below. In particular, we demonstrate that the regular monodromy surface defect (also supported on $\BC_{1} \times \{0\} \subset \BC_{1} \times \BC_{2}$) gives rise to a distinguished basis of a module $\CalH$ over the Yangian $Y(\fgl(n))$ enumerated by the Coulomb moduli $\ba$, by its vacuum expectation values: 
\begin{align}
    \psi(\ba)  \in \CalH,\qquad Y(\fgl(n)) \to \text{End}(\CalH).
\end{align}
We show this by constructing the Yangian R-matrices from the correlation function of the monodromy surface defect, the $\bf Q$-observable, and the $qq$-characters. This Yangian module $\CalH$ is evaluation module realized on a tensor product of $N$ bi-infinite (i.e., neither highest- nor lowest-weight) $\fgl(n)$-modules. The $\fgl(n)$ generators are represented by $\hbar$-difference operators, where the Yangian deformation parameter $\hbar$ is identified with the $\Omega_1$-background parameter $\ve_1 = \hbar$. 

Had we realized the $\hbar$-difference modules as a sheaf over $\text{Bun}_{GL(n)} (\BP^1;D)$, we could have defined the $\hbar$-analogue of the Hecke operator by employing the Hecke correspondence and the projections to $\text{Bun}_{GL(n)} (\BP^1;D)$. This is not the approach we take in this work. Instead, we define a $Q$-operator, the $\hbar$-analogue of the Hecke operator, within our $\EN=2$ gauge theory context. More specifically, we define the $Q$-operator on the aforementioned Yangian module by 
inserting the $\bf Q$-observable on top of the regular monodromy surface defect in the correlation function. Using the cluster decomposition of the two surface defects in the limit $\ve_2 \to 0$, we prove that the $Q$-operators act diagonally on the distinguished basis elements $\psi(\ba) \in \CalH$ in the limit $\ve_2 \to 0$, where the \textit{eigenvalues} are given by the $Q$-functions $Q(\ba;x)$ satisfying the $\hbar$-oper equation. Namely,
\begin{align}
    \mathbf{Q}_i(x) \cdot \psi(\ba) = Q_i(\ba;x) \psi(\ba),\qquad i=1,2,\cdots, n.
\end{align}
Further, we demonstrate that the $Q$-operator satisfies the universal $\hbar$-oper equation for Yangian $Y(\fgl(n))$ represented on the module $\CalH$, viewed as a quantized chiral ring relation of the coupled system. From this, we show that the $Q$-eigenstate property of the distinguished basis implies that its elements are also common eigenstates of the transfer matrices $\hat{t}(x)$ of the Yangian $Y(\fgl(n))$ represented on $\CalH$, 
\begin{align}
    0= (\hat{t}(x)-t(\ba;x))\psi(\ba),
\end{align}
with the eigenvalues (coefficients of $t(\ba;x)$) parametrizing the space of $\hbar$-opers. 

All in all, we conclude the following statement: for a $\hbar$-oper associated to the Coulomb moduli $\ba$, there is a corresponding $Q$-eigenstate $\psi(\ba)$ in the Yangian module $\CalH$ realized by the vacuum expectation value of the regular monodromy surface defect in the limit $\ve_2 \to 0$ at the vacuum $\ba$. The $Q$-eigenstates also represent the spectral $\hbar$-difference equations for the associated XXX spin chain. The quantum spectra are precisely determined by the Coulomb moduli $\ba$ that specify the $\hbar$-oper. This can be regarded as a $\hbar$-deformation of the ordinary geometric Langlands correspondence associated to twisted $\ED$-modules and opers \cite{BD1,BD2}.

\section{Four dimensional gauge theory, surface defects and surface observables} \label{sec:Qopssrf}

In this section, we begin our $\EN=2$ gauge theory formulation. 

The path integral of gauge theory is a sum over $k \in {\BZ}$ of
integrals over the space ${\CalA}_{k}/{\CalG}_{k}$ of gauge equivalence classes of connections on a principal $G$-bundle over $S^{4} = {\BR}^{4} \cup {\infty}$. Although the moduli space ${\CalM}_{k}$ of instantons (i.e. solutions to $F^{+}_{A} = 0$ equation) depends only on the conformal class of the metric on the four dimensional spacetime, the metric on ${\CalM}_{k}$ depends on the metric itself. The proper setup for computations of the low-energy effective theory is the moduli space ${\CalM}_{k}^{\rm framed}$ of framed instantons on ${\BR}^{4}$. It is hyperk\"ahler, yet not complete. Path integral of ${\EN}=2$ theory has a mathematical interpretation as Mathai-Quillen representative of the Euler class of certain infinite-dimensional vector bundle over ${\CalA}_{k}/{\CalG}_{k}$. The bundle depends on the field content of the theory. In this paper we are mostly dealing with asymptotically conformal super-QCD $G=SU(N)$ gauge theory. The corresponding
Euler class can be modeled, for each $k$, on a finite-dimensional model of ${\CalM}_{k}^{\rm framed}$, which admits smooth partial $G \times G_{\rm rot}$-equivariant compactification. Here $G_{\rm rot} = U(2)$ is the spin cover of the group of rotations of ${\BR}^{4}$, compatible with the structure used in the compactification.

We present the construction of the base four-dimensional $\EN=2$ supersymmetric gauge theory and its half-BPS surface defects in the gauge origami formulation \cite{Nikita:I,Nikita:II,Nikita:III}. 

The gauge origami is the configuration of intersecting D3-branes in the IIB theory on the ten-dimensional spacetime $X \times \BC$, where $X$ is a local Calabi-Yau four-fold. We will consider the cases $X= \BC_1\times \BC_2 \times (\BC_3 \times \BC_4)/\G_{34}$ and $X= \BC_1 \times (\BC_2 \times \BC_3 \times \BC_4)/(\G_{34} \times \G_{24})$, where $\G_{34} = \BZ_{n+1}$ and $\G_{24}=\BZ_l$. Here, $\G_{ab}$ gives an orbifolding action rotating $\BC_a$ and $\BC_b$ in the opposite direction. We implement the $\O$-background for the $U(1)^3 \subset SU(4)$ rotations of $X$. The three independent $\O$-background parameters are written as $\ve_a$, $a=1,2,3,4$, $\sum_{a=1} ^4 \ve_a = 0$; $\ve_a$ associated to the rotation of the $\BC_a$-plane. In order to preserve the supersymmetry, the D3-branes must wrap two complex planes among four in $X$ and be located at the origin of the remaining two planes, while they can be positioned at any points on the transverse $\BC$. We abbreviate the notation so that, for instance, $\BC^2 _{12}$ indicates $\BC_1 \times \BC_2 \times \{0\} \subset X$ and certain position on $\BC$.

\begin{table}[h!] 
    \centering
    \begin{tabular}{ c||c|c|c|c|c|c|c|c|c|c } 
        \text{IIB branes} & 0 & 1 & 2 & 3 & 4 & 5 & 6 & 7 & 8 & 9  \\ \hline\hline
         D3 & \rm{x} & \rm{x}& \rm{x}& \rm{x} &  & & & &  &  \\  KK5${}_{n+1}$ & \rm{x} & \rm{x} & \rm{x} & \rm{x} & &  &  & & \rm{x} & \rm{x}   \\ \hline KK5${}_l$ & \rm{x} & \rm{x} & &  &\rm{x}& \rm{x} & & & \rm{x} &  \rm{x} \\ D3 & \rm{x} & \rm{x} & & & \rm{x}& \rm{x}  &&&
    \end{tabular}
    \caption{IIB brane configuration for gauge origami}
    \label{table:IIBori}
\end{table}

\begin{table}[h!]
    \centering
    \begin{tabular}{c|c|c|c|c}
        $\BC_{1}$ & $\BC_{2}$ & $\BC_{3}$ & $\BC _{4}$ & $\BC_{5}$  \\ \hline
        $x^0,x^1$ & $x^2,x^3$ & $x^4, x^5$ & $x^6,x^7$ & $x^8,x^9$ 
    \end{tabular}
    \caption{Spacetime for gauge origami}
    \label{table:spacetimeori}
\end{table}

On top of stacks of D3-branes, we have D($-1$)-instantons. The path integral of the gauge origami configuration is exactly computed by equivariant localization \cite{Nikita:III}. The framing equivariant weights of the Chan-Paton bundles are given by the positions of the D3-branes on the transverse plane $\BC_{5}$.

After introducing the gauge origami construction of the base $\EN=2$ gauge theory, we will define $Q$-operators as surface defects realized by coupling two-dimensional $\EN=(2,2)$ chiral multiplets. Also, we will define the monodromy defect as assigning singular behavior of the gauge field at the chosen surface, modelled by an orbifold. Their vacuum expectation values are computed by the gauge origami partition functions. See also \cite{Kimura:2015rgi,Bourgine:2017jsi,Awata:2017lqa,Bourgine:2019phm} for an algebraic approach of engineering the gauge theory and its monodromy surface defects.

\subsection{Four-dimensional $\EN=2$ supersymmetric gauge theory}

By a well-described construction, the $A_1$-type quiver theory is a limit $\qe_1 ,\qe_2 \to 0$
of the theory of $N$ regular branes on ${\BZ}_{3}$-orbifold. In the gauge origami setup it is 
a stack of $3N$ D3-branes on $\BC_{12} ^2 \times \{ 0 \} \subset \BC^2 _{12} \times \BC^2 _{34} / \BZ_3$ (see the first two rows of Table \ref{table:IIBori} with $n=2$). Unlike the fully dynamical ${\hat A}_{2}$-type quiver theory, the $A_{1}$-limit does not depend on ${\ve}_{3}$, so we can simplify our formulas below by setting it to zero. The Chan-Paton bundle has the character decomposition
\begin{align} \label{def:GO-setup}
    {\bf N}_{12} = N_{0} {\CalR}_{0} + N_{1} {\CalR}_{1} + N_{2} {\CalR}_{2} = \sum_{\alpha=0}^{N-1} e^{a_\alpha} \cdot \CalR_0 +  e^{m_\alpha^- + {\ve}_{1} + {\ve}_{2}} \cdot \CalR_1 + e^{m_\alpha^+} \cdot \CalR_2,
\end{align}
where $\CalR_i$, $i=0,1,2$, denote the  three irreducible one-dimensional representations of $\BZ_3$, ${\CalR}_{0}$ being the trivial one, and ${\CalR}_{1} = {\CalR}_{2}^{*}$.\footnote{so that
\beq
{\BC}[{\BZ}_{3}] = {\CalR}_{0} \oplus {\CalR}_{1} \oplus {\CalR}_{2}
\eeq}
The gauge origami partition function, upon dropping a $1$-loop contribution of non-dynamical $(N_{1}, {\bar N}_{2})$-hypermultiplets, reduces to  
\begin{align} \label{eq:n2part}
\begin{split}
    \CalZ & = \sum_{\boldsymbol\lambda} \kq^{|{\boldsymbol\lambda}|} \, \BE \left[ \frac{-SS^* + M^+ S^* + q_{12}^{-1} (M^- )^* S }{P_{12}^*} \right]  =: \sum_{\bl} \qe ^{\vert \bl \vert} \m_{\ba, \bl}  ,
\end{split}
\end{align}
where $S := S_{0} = N_{0} - P_{12} K$, $M^{+} =  S_{2} = N_{2}$, $M^{-} = S_{1} = N_{1}$, and $\bl$ now denotes (slightly abusing the notation) an $N$-tuple of partitions. This is exactly the partition function of the $A_1$-quiver gauge theory, i.e., $U(N)$ gauge theory with $N$ fundamental and $N$ anti-fundamental hypermultiplets. The explicit formulas for the $1$-loop contribution $ \m_{\ba, \bl}$ of the perturbative fluctuations around the instanton configuration $\bl$ in the Coulomb vacuum $\ba$ is well-known \cite{Nekrasov:2002qd}.

In $\Omega_1$-background the four-dimensional theory is described by an effective two-dimensional $\EN=(2,2)$ theory on the $\BC_2$-plane, with the effective twisted superpotential $\widetilde{\EuScript{W}}(\ba;\qe)$. The partition function \eqref{eq:n2part} in the ${\ve}_{2} \to 0$ limit behaves as \cite{Nekrasov:2009rc}\footnote{This limiting behavior of the partition function can also be obtained from the limit shape \cite{Nekrasov:2003rj, Poghossian:2010pn}.}
\begin{align} \label{eq:partns}
    \lim_{\ve_2\to 0} \CalZ (\ba, \mathbf{m},\ve_1,\ve_2 ;\qe) = e^{ \frac{\widetilde{\EuScript{W}}(\ba,\mathbf{m},\ve_1;\qe)}{\ve_2}}.
\end{align}
We also note that the $\BC_2$-plane becomes topological in the limit $\ve_2 \to 0$ (as the supercharge $\mathfrak{Q}_{\ve}$ becomes the twisted supercharge of a hybrid $A/B$-model with worldsheet ${\BC}_{2}$).  In particular, the local $\mathfrak{Q}_{\ve}$-closed observables in the effective two-dimensional theory can be translated modulo $\mathfrak{Q}_{\ve}$-exact corrections.

The path integral with insertions of local or non-local BPS observables can be  localized to a sum over colored partitions  $\bl$. For $SU(N)$ or $U(N)$ gauge theory $\bl$ stands for a collection 
\beq
{\bl} = ({\lambda}^{(0)}, \ldots , {\lambda}^{(N-1)})
\label{eq:vecyd}
\eeq
of partitions
\beq
{\lambda}^{({\alpha})} = \left( {\lambda}^{({\alpha})}_{1} \geq {\lambda}^{({\alpha})}_{2} \geq \ldots \geq {\lambda}^{({\alpha})}_{{\ell}({\lambda}^{({\alpha})})} > 0 \right)
\eeq
of sizes
\beq
| {\lambda}^{({\alpha})} | = {\lambda}^{({\alpha})}_{1} + {\lambda}^{({\alpha})}_{2} + \ldots + {\lambda}^{({\alpha})}_{{\ell}({\lambda}^{({\alpha})})}
\eeq
so that
\beq
k = |{\lambda}^{(0)}| + \ldots + |{\lambda}^{(N-1)}|
\label{eq:ic}
\eeq
is the instanton charge.

For observable $\mathcal{O}$ we denote by $\mathcal{O}[{\ba}, \bl]$ its evaluation at the instanton configuration corresponding to the $N$-tuple of partitions $\bl$, in the vacuum characterized by the Coulomb moduli ${\ba}$. Therefore the
normalized vacuum expectation value of $\mathcal{O}$ is computed by
\begin{align}
    \frac{\langle \mathcal{O}\rangle_{\ba}}{ \langle 1\rangle_{\ba}} = \frac{\sum_{\bl} \, \qe^{\vert \bl \vert}  \m_{\ba, \bl} \, \mathcal{O}[{\ba}, \bl]}{\sum_{\bl} \, \qe^{\vert \bl \vert}  \m_{\ba, \bl}} \ .
    \label{eq:normvevo} 
\end{align}

\subsubsection{Local observables in four dimensions}
The local observables in the four-dimensional $\EN=2$ gauge theory are polynomials in the single trace invariants of the complex scalar $\phi$ in the vector multiplet. In our main example of $\EN=2$ theory with $SU(N)$ gauge group, there are $N-1$ independent local observables,
\begin{align}
    \text{Tr}\, \phi^k,\qquad k=2,3,\cdots, N.
\end{align}
We organize them into a generating function, called the $\bf\EY$-observable,
\begin{align}
    {\bf\EY} (x) = \exp \text{Tr}\, \log (x-\phi) = x^N \exp \left[ - \sum_{l=1} ^\infty \frac{1}{l x^l} \text{Tr}\,\phi ^l \right],
\end{align}
whose  evaluation at the instanton configuration $\bl$ is the virtual Chern polynomial of the universal sheaf,
\begin{align} \label{eq:yobs}
    {\bf\EY}(x)[{\ba} , \bl] = \BE \left[ -e^x S [{\ba}, \bl] ^* \right].
\end{align}
with, for $\zeta_\BR > 0$, 
\beq
S[{\ba}, {\bl} ] = \sum_{\alpha = 0}^{N-1} e^{a_{\alpha}} \left( 1 - (1-q_{1})(1-q_{2})  \sum_{(i,j) \in {\lambda}^{({\alpha})}} q_{1}^{i-1} q_{2}^{j-1} \right)
\label{eq:sevalpos}
\eeq
and, for $\zeta_\BR <0$,
\beq
S[{\ba}, {\bl} ] = \sum_{\alpha = 0}^{N-1} e^{a_{\alpha}} \left( 1 - (1-q_{1})(1-q_{2})  \sum_{(i,j) \in {\lambda}^{({\alpha})}} q_{1}^{-i} q_{2}^{-j} \right).
\label{eq:sevalneg}
\eeq

\subsection{${\bf Q}/{\tilde{\bf Q}}$-surface observables} \label{subsec:Qobser}
Having fixed our $\EN=2$ gauge theory, we will turn to the construction of various surface defects and observables. We introduce the $\bf Q$- and $\tilde{\bf Q}$-observables which are built by inserting an additional D3-brane that intersects the worldvolume of the original stack of D3-branes of the gauge theory along a surface.

\subsubsection{$\mathbf{Q}/\tilde{\mathbf{Q}}$-observables} \label{subsubsec:Qobs}
On top of the stack of D3-branes wrapping $\BC^2 _{12} $ which engineers the four-dimensional $\EN=2$ theory, we add a single D3-brane wrapping $\BC^2 _{13}$ carrying the $\BZ_3$-charge $1$ for $\bf Q$ or $0$ for ${\tilde {\bf Q}}$, respectively (cf. the fourth row of Table \ref{table:IIBori}),
\begin{align}
\begin{split}
    N_{12} & = \sum_{\alpha=0}^{N-1} e^{a_\alpha} \cdot \CalR_0 + \sum_{\alpha=0}^{N-1} e^{m_\alpha^{-} + {\ve}_{1}+{\ve}_{2} } \cdot \CalR_1 + \sum_{\alpha=0}^{N-1} e^{m_\alpha^+} \cdot \CalR_2 \\
    N_{13} & = e^{x+\ve_1} \cdot \CalR_1 \, , \ {\rm or}\ = e^{x+\ve_1} \cdot \CalR_0
\end{split}
\end{align}
The gauge origami partition function, in the limit ${\qe}_{1} = {\qe}_{2} = 0$, becomes, by simple computation:
\beq 
\label{eq:Qobswithout}
    \left( \begin{matrix} \CalZ_{\bf Q} \\
    \CalZ_{\tilde {\bf Q}} \end{matrix} \right) = \sum_{{\boldsymbol\lambda}} \kq^{|{\boldsymbol\lambda}|} \m_{\ba, \bla} \; 
    \left( \begin{matrix} {\bf Q}(x) [\bla] \\
    {\tilde {\bf Q}}(x) [\bla] \end{matrix} \right)\ .
\eeq
with
\begin{align}\label{eq:dualQ}
\begin{split}
&{\tilde{\bf Q}}(x) = {\qe}^{\frac{x}{{\ve}_{1}}}\, \sum_{d=0}^{\infty}  {\qe}^d  \, {\phi}_{d} \,  {\tilde{\bf Q}}_d (x),  \\
&{\phi}_{d} = \prod_{j=1}^d \left( 1  + \frac{{\ve}_{2}}{j {\ve}_{1}} \right)  ,\qquad  {\tilde{\bf Q}}_d (x) :=\frac{{\bf Q} (x)\, M(x  +d \ve_1)}{{\bf Q} (x+ d  \ve_1)  {\bf Q}(x +\ve_{2} +(d+1) \ve_1) }
\end{split}
\end{align}
where $M(x)$ is entire function of $x$ solving
\beq \label{eq:dualqpart}
\frac{M(x)}{M(x-\ve_1)} =  P(x).
\eeq
We call these surface observables $\bf Q$- and ${\tilde{\bf Q}}$-\textit{observable}, respectively. 

Finally, the evaluation of ${\bf Q}(x)[{\bla}]$ at the instanton configuration $\bla$ is entire function of $x$, with the zeroes, at $\zeta_{\BR} > 0$, at
\beq
x = a_{\alpha} + {\ve}_{1}(i-1) + {\ve}_{2} {\lambda}^{({\alpha})}_{i}\, , \ {\alpha} = 1, \ldots, N\, , \ i = 1, 2, 
\eeq
Explicitly, we have
\beq
{\bf Q}(x) [{\bla}] = \prod_{{\alpha} = 0}^{N-1} \left( \frac{{\ve}_{1}^{\frac{x-a_{\alpha}}{\ve_1}}}{{\Gamma} \left( - \frac{x-a_{\alpha}}{{\ve}_1} \right)} 
\prod_{i=1}^{\infty} \frac{x - a_{\alpha} - {\ve}_{1}(i-1) - {\ve}_{2} {\lambda}^{({\alpha})}_{i}}{x - a_{\alpha} - {\ve}_{1}(i-1)} \right)
\label{eq:qeval}
\eeq
It is easy to show by the methods of \cite{Nikita:I} that $\langle {\tilde{\bf Q}}(x) \rangle$ is also entire in $x$, for any $\ve_1, \ve_2$. 

For later purpose, we denote the normalized vacuum expectation value of the $\bf{Q}/\tilde{\bf Q}$-observables in the limit $\ve_2 \to 0$ by
\begin{align} \label{eq:qvevlim}
\begin{split}
    &\lim_{\ve_2 \to 0} \frac{\left\langle {\bf Q}(x) \right\rangle_\ba}{\left\langle 1 \right\rangle_\ba} = Q(\ba;x),\qquad 
    \lim_{\ve_2 \to 0} \frac{\left\langle \tilde{{\bf Q}}(x) \right\rangle_\ba}{\left\langle 1 \right\rangle_\ba} = \tilde{Q}(\ba;x),
\end{split}
\end{align}
where we specified the vacuum by the Coulomb parameter $\ba$.

\subsection{Monodromy surface defect} \label{subsec:monodef}
The monodromy surface defect ${\bf\Psi}_{\Sigma}$ in the four-dimensional $\EN=2$ gauge theory is engineered by assigning a singular behavior of the gauge field along a surface $\Sigma$ \cite{NN2004ii, gukwit}, effectively coupling the four dimensional gauge theory to a two dimensional sigma model \cite{Nikita:IV}.

\subsubsection{Curvature singularity}

The connection between the orbifold structure and the monodromy surface defect is the following. Requiring the singular behavior of the form
\beq
F_{z_{2} {\bar z}_{2}} (z_1, {\bar z}_1, z_2, {\bar z}_{2}) \sim J (z_1, {\bar z}_{1}) {\delta}^{(2)}(z_2, {\bar z}_{2})
\label{eq:curvf}
\eeq
with $J(z_1, {\bar z}_{1}) \in {\CalO}_{\bf m} \subset \mathfrak{g}$ being a $z_1$-dependent element of a conjugacy class in the Lie algebra of the gauge group, 
\beq
{\CalO}_{\bf m} = G/L_{gauge}
\label{eq:levi}
\eeq 
characterized by a choice 
of the Levi subgroup of the gauge group. Additionally, the monodromy surface defect depends on the choice of a Levi subgroup of the flavor group it preserves \cite{gukwit}. The choice of the Levi subgroup is conveniently encoded in \textit{coloring functions}. In our main example of the $SU(N)$ gauge theory with $N$ fundamental and $N$ anti-fundamental hypermultiplets, the coloring functions 
$(c,c_f,c_{af})$ assign $\BZ_l$-charges to the framing and the flavor bundles,
\beq \label{eq:colormono}
(c, c_{f}, c_{af}) : \{0, \ldots, N-1 \} \longrightarrow \{0,1,\cdots, l-1\} \\
\eeq
from which the preserved Levi subgroups are read off as 
\beq
\begin{aligned}
& L_{gauge} = S\left(\bigtimes_{\o=0} ^{l-1} U( {\#} c^{-1}(\o) )\right) ,  \\
& L_{matter} = S\left(\bigtimes_{\o=0} ^{l-1} U( {\#} c_{matter}^{-1}(\o) )\right) \\
\end{aligned}
\eeq
In the case of $l =N$ and the coloring functions $c$, $c_{f}$, and $c_{af}$ are chosen to be one-to-one (without loss of generality, $c(\a) = c_f (\a) = c_{af} (\a) = \a$), the associated monodromy surface defect is said to be \textit{regular}. For the purpose of studying the di-Langlands correspondence (as well as the ordinary geometric Langlands correspondence, as studied in \cite{Jeong:2023qdr}), we will mainly consider the regular monodromy surface defect throughout the work.

\subsubsection{Parabolic structure}

The curvature singularity \eqref{eq:curvf} translates to 
the language of algebraic geometry as a parabolic structure of the gauge bundle $E$, as a flag of subbundles $0 = F_0 \subset F_{1} \subset \ldots \subset F_{l-1} \subset E \vert_{\Sigma}$,  
of the restriction of gauge bundle on $\Sigma$. Working modulo the supercharge 
$\mathfrak{Q}$ means we can assume all bundles and subbundles holomorphic. Also, a $\mathfrak{Q}$-exact deformation leads to a partial compactification of the moduli space of parabolic bundles by the moduli space of parabolic torsion free sheaves:
\beq
 {\sF}_{0} = z_2 {\sE} \subset {\sF}_{1} \subset {\sF}_{2} \subset \ldots \subset {\sF}_{l-1} \subset {\sE}
 \label{eq:parabsh}
 \eeq
 where $z_{2} = 0$ is the equation defining $\Sigma$ (it is clear that \eqref{eq:parabsh} can be generalized to $\Sigma$ being a normal crossing divisor).

\subsubsection{Orbifold construction}
The singular behavior of the gauge field can be modelled by placing the $\EN=2$ gauge theory on an orbifold \cite{K-T}, $\BC_1 \times (\BC_2 /\BZ_l)$. The supersymmetry preserving construction is best formulated in the gauge origami language, where the ${\BZ}_{l}$-orbifold group acts via $(z_2, z_4) \mapsto (e^{\frac{2\pi\ii}{l}} z_2, e^{-\frac{2\pi\ii}{l}} z_4)$.

In orbifold computations we shall often use the following notation: 
\beq
{\tilde q}_{2} = q_{2}^{\frac 1l}\, , \quad {\tilde q}_{2}^{l} = q_{2}\ , 
\label{eq:qtilde}
\eeq
also, 
for ${\omega} \in {\BZ}$ we denote by 
\beq
[{\o}] := \o \text{ mod } l \in \{0,1,\cdots, l-1\}
\label{eq:om}
\eeq
its projection to the fundamental domain for ${\BZ}/ l {\BZ}$. 

The map $z_2 \mapsto z_2 ^l$ sends the spacetime of gauge theory on the $\BZ_l$-orbifold to the ordinary spacetime $\BC_1 \times \BC_2$. The branching locus represents the singularity of gauge fields along the locus $\{z_2 = 0\}$. This map is holomorphic but not isometric, of course. The actual field redefinition requires complex gauge transformations, which, in turn, entail complicated functional Jacobians. Fortunately, in computations involving $\mathfrak{Q}$-closed observables all these complications are irrelevant, up to $\mathfrak{Q}$-exact terms.

To summarize, the monodromy surface defect in ${\hat A}_{n}$-theory is implemented by the gauge origami on the orbifold $\BC_1 \times (\BC^3 _{234} /\G_{13} \times \G_{24} )$, where $\G_{13} = \BZ_{n+1}$ and $\G_{24} = \BZ_l$ (see the third row of Table \ref{table:IIBori}), with Chan-Paton spaces now being representations of both $\BZ_l$ and $\BZ_{n+1}$. They are encoded in the coloring functions,
\begin{align}
    c_{A,i} : \{ 1, \ldots , N_{A,i} \} \longrightarrow \{0,1,\cdots, l-1\},\qquad A \in \{ 12, 13, 14, 23, 24, 34 \}, \; i\in \{0,1,\cdots, n\}.
\end{align}
In the $A_1$ limit, with $n=2$, and $\qe_1,\qe_2 \to 0$, where we can also set ${\ve}_{3} = 0$ without any loss of generality, we arrive at ${\EN}=2$ gauge theory with a specific monodromy surface defect.

The path integral of the $\EN=2$ gauge theory on the $\BZ_l$-orbifold localizes to an equivariant integral over the moduli space $\mathfrak{M}^{\BZ_l}$ of instantons on the $\BZ_l$-orbifold \cite{Nikita:III}. As explained in \cite{Jeong:2023qdr}, one associates a projection $\r:\mathfrak{M}^{\BZ_l} \to \mathfrak{M}$ to the map $z_2 \mapsto z_2^l$ of the underlying spaces.  The pushforward ${\r}_{*}1$, summed over all fractional topological charges but $k_{N-1}$ is an observable in the $\EN=2$ gauge theory, the monodromy surface defect representative ${\bf \Psi}(u)$ in ${\mathfrak{Q}}_{\ve}$-cohomology.

Now we present the evaluation ${\bf \Psi}(u)[{\bl}]$ of the monodromy surface defect associated to the coloring function $\mathbf{c}$ \eqref{eq:colormono}. First, we can write the equivariant Chern character of the framing bundle as
\begin{align}
    \hat{N}_{12} = \sum_{\a  = 1}^{N} \left( e^{a_{\a}} {\tilde q}_{2}^{c({\a})} {\CalR}_{0} \otimes {\fR}_{c({\a})} +  e^{m^{+}_{\a}} {\tilde q}_2^{c_{f}({\a})} \CalR_2 \otimes {\fR}_{c_{f}({\alpha})}  + 
     e^{m_{\a}^{-} +{\ve}_{1}} q_2 ^{\d_{{c_{af}({\a})},0}}  \CalR_1   \otimes {\tilde q}_2^{c_{af}({\a})}  \fR_{c_{af}({\a})}\right)\ ,
\end{align}
Here, $\fR_\o$ denotes the one-dimensional representation of $\BZ_l$ with charge $\o$. Accordingly, the universal sheaf of the instantons on the $\BZ_l$-orbifold, carrying a representation of $\BZ_l$, can be written as
\begin{align}
    \hat{S}_{12} = \sum_{\o=0}^{l-1} \left[ S_\o  \CalR_0  + 
  q_{13} q_2 ^{ \d_{\o,0}} M_{[{\o-1}]} ^-  \CalR_1  +  q_3^{-1} M_\o ^+  \CalR_2  \right] \otimes {\tilde q}_2^\o \fR_{\o},
\end{align}
where we defined $S_\o = \sum_{\a \in c^{-1} (\o) } (N_\a - P_1K_\a) + q_2^{\d_{\o,0}} 
\sum_{\a\in c^{-1} ([{\o-1}])}  P_1K_{\a}$, $M_\o^\pm = \sum_{\a \in c^{-1} _{f,af} (\o)} e^{m^\pm_\a}$. The projection map $\r:\mathfrak{M}^{\BZ_l} \to \mathfrak{M}$ gives the universal
sheaf for the instantons and the flavor bundles in the absence of the orbifold by
\begin{align}
    S = \sum_{\o=0}^{l-1} S_\o = N - P_{12} K_{l-1}, \quad M^\pm = \sum_{\o=0}^{l-1} M_\o^\pm. 
\end{align}
In particular, the projection $\r$ induces a map of fixed point sets $\r : \left(\mathfrak{M}^{\BZ_l} \right)^{\mathsf{T}_\mathsf{H}} \to \mathfrak{M}^{\mathsf{T}_{\mathsf{H}}}$, where the fixed points of $\mathfrak{M}^{\BZ_l}$ enumerated by colored partitions $\{\hat{\bl}\}$ map to the sub-partitions $\{\bl\}$ formed by the columns carrying $\BZ_l$-charge $l-1$.
It is convenient to relate the fractionalized universal sheaf to the parabolic sheaf by working with
\begin{align}
    \Sigma_\o = S_{\o+1} + \cdots + S_{l-1},\quad\quad \o=0,1,\cdots, l-2.
\end{align}

Then, the gauge origami partition function gives, by a standard computation
\beq
{\CalZ} =  \langle \Psi_{\mathbf{c}}(\mathbf{u}) \rangle_\ba
\eeq
with the {\it surface defect observable} 
\begin{align} \label{eq:monoobser}
\begin{split}
    &\Psi _{\mathbf{c}} (\mathbf{u}) [\bl] =\sum_{\hat{\bl} \in \r^{-1} (\bl)} \prod_{\o=0} ^{l-2} \qe_\o ^{k_\o - k_{l-1}}  \BE\left[ \frac{S \S_0 ^*}{P_1 ^*} + \sum_{\o=0} ^{l-2} \frac{- \S_\o (\S_\o - \S_{\o+1})^* - M^+ _\o \S_\o ^* + q_1 ^{-1} (M^- _{\o}) ^* \S_\o}{P_1 ^*} \right],
\end{split}
\end{align}
labelled by the coloring function $\mathbf{c}$. There are $l-1$ defect parameters $(\qe_\o)_{\o=0} ^{l-2}$ encoding the singularity of the gauge field and the magnetic flux along the support $\BC_1$. 

We parameterize these defect parameters using $\mathbf{u} = (u_\o)_{\o=0} ^{l-1}$,
\begin{align}
    \qe_\o = \frac{u_{\o+1}}{u_\o},\qquad \o=0,1,\cdots, l-2,
\end{align}
with a redundancy of overall scaling. Also, we introduce the tree-level normalization of the surface defect by multiplying \eqref{eq:monoobser} by 
\begin{align} \label{eq:classcal}
    \Psi_{\mathbf{c}} (\bu) ^{\text{classical}} = \prod_{\a=0} ^{N-1} u_{c({\a})}^{\frac{m^+ _\a - a_\a }{\ve_1}}.
\end{align}

\subsubsection{Basis of Yangian module from monodromy surface defect} \label{subsubsec:basis}
Here, we will restrict our consideration to the regular monodromy surface defect. We define a vector space $\CalH$ as a suitably completed vector space of multi-valued functions
${\bf\Psi}(u_{0}, \ldots , u_{N-1})$ on $(\BC^\times)^{N}$, which are the eigenfunctions of the monodromy
around the coordinate axes:
\beq
{\bf\Psi}( e^{2\pi\ii n_{0}} u_{0}, \ldots , e^{2\pi \ii n_{N-1}} u_{N-1})\, = \, \prod_{\o = 0}^{N-1} t_{\o}^{n_{\o}} 
{\bf\Psi} (   u_{0}, \ldots ,   u_{N-1}),
\label{eq:monpsi}
\eeq
with the \emph{monodromy eigenvalues}
\beq
t_{\a} = e^{\frac{2\pi \ii (a_{\a}-m_{\a}^{+})}{\ve_1}}\, , \ {\a} = 0, \ldots , N-1.
\label{eq:monta}
\eeq
We denote collectively
\beq
{\bf t} = \left( t_{\a} \right)_{\a = 0}^{N-1} \in \left({\BC}^{\times} \right)^{N}.
\eeq
A typical element of $\CalH$ has the form
\beq
\prod_{\o=0} ^{N-1} u_\o ^{  \frac{m^+ _\o - a_\o }{\ve_1} } \times \text{(a homogeneous Laurent polynomial in $u_{\o}$)} .
\eeq
We are completing the space of Laurent polynomials by allowing infinite (convergent) series in $u_{\o}/u_{\o'}$, with ${\o} > {\o}'$. We also allow formal paths in $\CalH$, parametrized by $\qe$, such that at $\qe^k$ order one admits finite polynomials in $u_{\o'}/u_{\o}$ of degree bounded by $k$.\footnote{As we will see in section \ref{subsubsec:rmatgauge}, we introduce a twist parameter $\qe$ that parametrizes a twist of the periodic boundary condition of the $\fgl(2)$ XXX spin chain. Then, we construct a $\qe$-parametric family of basis elements of $\CalH$ by the construction above.} 
Thus the vacuum expectation values of the monodromy surface defect becomes a vector in $\CalH $,
\begin{align} \label{eq:psiele}
   \Big\langle \Psi (\mathbf{u}) \Big\rangle_\ba \in \CalH .
\end{align}
Importantly, so do vevs of the monodromy surface defects at other vacua of the $\EN=2$ theory enumerated by the $\BZ$-lattice of Coulomb parameters:
\beq\label{eq:psielesh}
   \Big\langle \Psi (\mathbf{u}) \Big\rangle_{\ba + {\ve}_{1} {\bn}} \in \CalH,
\eeq
for any vector ${\bn} = (n_{0}, \ldots , n_{N-1}) \in {\BZ}^{N}$ satisfying $n_{0} + \ldots + n_{N-1} = 0$. 

As we will clarify in section \ref{subsubsec:rmatgauge}, the space $\CalH $ is the weight-zero subspace in a $U(\fgl(2))$-module $\tilde{\CalH}$, from which we construct the evaluation module over the Yangian $Y(\fgl(2))$ (recall section \ref{subsec:rmat}). When considering the Bethe subalgebra, we may restrict to $\CalH$ since the degree-zero condition only amounts to setting one of the Hamiltonians (the total momentum) to a number.

In the $\Omega_1$-background, the monodromy surface defect becomes a local observable on the $\BC_2$-plane. Its normalized vacuum expectation value becomes
\begin{align} \label{eq:monodefns}
    \lim_{\ve_2 \to 0} \frac{\Big\langle \Psi(\mathbf{u}) \Big\rangle_\ba}{\langle 1 \rangle_{\ba}} = \psi (\ba;\mathbf{u};\qe),
\end{align} which
we can view as above-mentioned $\qe$-parametric family of basis vectors in $\CalH$. 

\subsection{Parallel surface defects}
Finally, we consider the configuration where a regular monodromy surface defect and $\bf Q$-
or $\tilde{\bf Q}$-observable are extended along the same $\BC_1$-plane in parallel. The parallel surface defect configuration here should not be confused with the intersecting surface defect configuration studied in \cite{Jeong:2020uxz,Jeong:2021bbh} in the regard of the associated isomonodromy problem. Rather, the configuration here is a $\hbar$-analogue of the parallel configuration of the monodromy surface defect and the canonical surface defect studied in \cite{Jeong:2023qdr}, utilized for the $\EN=2$ gauge theoretical account of the ordinary geometric Langlands correspondence.

In the gauge origami setup, the parallel surface defect configuration descends from adding a stack of $N$ D3-branes to the stack of D3-branes engineering the $\EN=2$ gauge theory with the monodromy surface defect, each of which carries a $\BZ_N$-charge $\o=0,1,\cdots, N-1$, as we now explain.

\subsubsection{$\bf Q$-observable and monodromy surface defect}

First, we assign the coloring functions for the framing Chan-Paton bundles. 
We keep the coloring function \eqref{eq:colormono}, where $(c,c_f,c_{af})$ are 
one-to-one functions, for the stack of $3N$ D3-branes supported on $\BC_1\times \BC_2 \times \{0\}$, while also giving an one-to-one coloring function for the new stack of $N$ D3-branes supported on $\BC_1 \times \{0\} \times \BC_3 \times \{0\} \times {\bx}$, where $\bx=  \{ x_0,x_1,\cdots, x_{N-1} \} \subset {\BC}_{5}$. The configuration is summarized by the following equivariant Chern characters of Chan-Paton spaces,
\begin{align}
\begin{split}
    &\hat{N}_{12} = \sum_{\o=0}^{N-1} \left[  e^{a_{\o+1}}  \CalR_0  + e^{m_{[{\o-1}]} ^- +\ve_1} q_2 ^{\d_{\o,0}}  \CalR_1  +   e^{m^+ _\o} \CalR_2 \right] \otimes {\tilde q}_2^\o  \fR_\o ,\\
    &\hat{N}_{13} = \sum_{\o=0} ^{N-1} e^{x_\o +\ve_1 +\ve_2} {\tilde q}_2 ^\o \cdot \CalR_1 \otimes \mathfrak{R}_\o \ . 
\end{split}
\end{align}
The corresponding gauge origami partition function reads
\beq
{\CalZ}_{12, 13} = \llangle \EQ (\bx) \rrangle
\label{eq:oriparall}
\eeq
with the \textit{generalized} $\bf Q$-observable $\EQ (\bx)$ defined by its evaluation at the instanton configuration:
\begin{align}
    \EQ (\bx) [\hat{\bl}] = \prod_{\o=0} ^{N-1} Q_\o (x_\o) [\hat{\bl}] : = \prod_{\o=0} ^{N-1} \BE\left[ \frac{e^{x_\o} (M^- _\o - S_\o[\hat{\bl}])^*}{P_1 ^*}  \right]
\end{align}
where we use the double bracket $\llangle \cdots \rrangle$ to indicate the ensemble average over the colored partitions $\{\hat{\bl}\}$.

Instead of considering arbitrary positions $\bx$, let us set a reference value $x$ and $\ve_1$-integral shifts, namely, $x_\o = x-n_\o \ve_1$ with $n_\o \in \BZ$. Then the generalized $\bf Q$-observable factorizes as a product of the $\bf Q$-observable, and a fractional $0$-observable inserted at the intersection of the surface defect and the surface observable:
\begin{align}
    \EQ_\mathbf{n} (x)[\hat{\bl}] = {\bf Q}(x) [\bl] \, \BE \left[  \sum_{\o=0} ^{N-1} \sum_{i=0} ^{n_\o-1} e^{x-i\ve_1} S^* _\o [\hat{\bl}]  \right],
\end{align}
Therefore, the gauge origami partition function \eqref{eq:oriparall} computes the correlation function of the two surface defects and the fractional $0$-observable at their intersection. 

Let us consider the case where $\bx = \bx_\o := (\underbrace{x,\cdots, x}_{\o+1 \text{\;times}}, \underbrace{x-\ve_1, \cdots, x-\ve_1}_{N-\o-1 \;\text{times}})$.  The corresponding generalized $\bf Q$-observable factorizes as  
\begin{align} \label{eq:genQspec}
\begin{split}
   &\EQ_\o (x)  := \EQ(\bx_\o) = \prod_{\o'=0} ^\o Q_{\o'} (x) \prod_{\o'=\o+1} ^{N-1} Q_{\o'}(x-\ve_1) \ ,  \\
   & \EQ_\o (x) [\hat{\bl}] = {\bf Q}(x) [\bl]\, \BE \left[ \sum_{\o'=\o+1} ^{N-1} e^x S^* _\o [\hat{\bl}] \right]
\end{split},\qquad \o=0,1,\cdots, N-1,
\end{align}
relating the $0$-observables to the parabolic structure \eqref{eq:parabsh}. The $0$-observable is absent in the case of $\omega = N-1$, $\EQ_{N-1} (x) = {\bf Q}(x)$, where the gauge origami becomes the OPE of the regular monodromy surface defect and the bulk $\bf Q$-observable,
\begin{align} \label{eq:correl}
    \llangle \EQ_{N-1} (x) \rrangle = \Big\langle {\bf Q}(x) \Psi(\mathbf{u}) \Big\rangle_\ba = \sum_{\bl} \qe^{\vert \bl \vert} \m_{\ba, \bl} \, {\bf Q}(x)[ \bl] \Psi(\mathbf{u})[ \bl] ,
\end{align}
with ${\bf Q}(x)[\bl]$ and $\Psi(\mathbf{u})[ \bl]$ given by \eqref{eq:Qobswithout} and \eqref{eq:monoobser}.

\subsubsection{${\tilde{\bf Q}}$-observable and monodromy surface defect}
Next, we bring together ${\tilde {\bf Q}}$-observable and monodromy surface defect, again with both surfaces being parallel. The gauge origami configuration is represented by the equivariant Chern characters of Chan-Paton bundles, 
\begin{align}
\begin{split}
    &\hat{N}_{12} = \sum_{\o=0}^{N-1} \left[  e^{a_\o}  \CalR_0  + e^{m_{[\o-1]} ^- +\ve_1 +  \ve_3} q_2 ^{\d_{\o,0}}  \CalR_1  +   e^{m^+ _\o -\ve_3} \CalR_2 \right] \otimes \hat{q}_2^\o  \fR_\o , \\
    & \hat{N}_{13} = \sum_{\o=0}^{N-1} e^{x_{\o} + \ve_1} \hat{q}_2^{\o} \CalR_0 \otimes \fR_{\o},
\end{split}
\end{align}
where we placed $N$ D3-branes supported on $\hat{\BC}^2 _{13}$ at at generic positions $\bx = (x_0,x_1,\cdots,x_{N-1}) \in \BC^{N-1}$. In $A_1$-quiver gauge theory the instantons on $\hat{\BC}^2 _{13}$ only grow along $\hat{\BC}_1$. The corresponding partitions ${\lambda}_{13}$ classifying the fixed points on the instanton moduli space are single-columns:
\begin{align}
\begin{split}
    & \hat{K}_{13} = \sum_{\o=0}^{N-1} e^{x_{\o}}q_1 \frac{1-q_1^{d_{\o}} }{1-q_1} {\tilde q}_2^{\o} \CalR_0 \otimes \fR_{\o}, \\ 
    & \hat{S}_{13} 
    = \sum_{\o=0}^{N-1} e^{x_{\o}+\ve_1} {\tilde q}_2^{\o} \left( q_1^{d_{\o}} \CalR_0 + q_3 (1-q_1^{d_{\o}}) \CalR_1 \right) \otimes \fR_\o,
\end{split}
\end{align}
where we set the lengths of the columns as ${\bd}\, = \, (d_{\o})_{\o=0}^{N-1}$.

The gauge origami partition function is given by
\beq
    {\tilde\CalZ}_S ({\bx}) = \sum_{\hat{\boldsymbol\lambda}} \, \prod_{\o=0}^{N-1} \kq_\o^{k_\o} \ \m_{\ba, \hat{\bl}} ^{\BZ_N} \,  \tilde{\EQ}(\bx)[\ba, \hat{\boldsymbol\lambda}] 
    \label{eq:gentildq}
\eeq
where we defined, after including the classical part, the \textit{generalized}  $\tilde{\bf Q}$-observable by
\beq
    {\tilde\EQ}(\bx) 
       =  \sum_{\bd} \prod_{\o=0}^{N-1} \kq_{\o}^{\frac{x_{\o}}{\ve_1}+d_{\o}} {\Phi}_\bd(\bx) \tilde\EQ_\bd(\bx),
\label{eq:genftildq}
\eeq
where 
\begin{align} \label{eq:dualnot}
\begin{split}
     {\Phi}_\bd(\bx) & \, = \, \prod_{\o=0}^{N-1} \, \prod_{j=1}^{d_{\o}} \, \frac{ x_{[{\o-1}]} - x_{\o}+ (d_{[{\o-1}]}+1-j)\ve_1 + \ve_2\d_{\o,0} }{j\ve_1}, \\
    \tilde\EQ_\bd(\bx) & \, = \, \prod_{\o=0}^{N-1} \, \frac{Q_{\o}(x_{\o}) \, M_{\o}(x_{\o}+d_{\o}\ve_1) }{Q_{\o}(x_{\o}+d_{\o}\ve_1) Q_{[{\o+1}]}(x_{\o} + (d_{\o}+1)\ve_1+\ve_2\d_{\o,N-1}) }.
\end{split}
\end{align}
They satisfy the following recursion relations:
\begin{align}\label{eq:EZ-ER conditions}
\begin{split}
    &  {\Phi}_{\bd-e_\o}(\bx+\ve_1 {\be}_\o) \, = \, \frac{d_\o\ve_1}{x_{[{\o-1}]}-x_{\o} + d_{[{\o-1}]}\ve_1 + \ve_2 \d_{\o,0} } \, {\Phi}_{\bd}(\bx)  \\
    &  {\Phi}_{\bd+e_\o}(\bx-\ve_1 {\be}_\o)\,  = \, \frac{x_{[{\o-1}]}-x_{\o} + (d_{[{\o-1}]}+1)\ve_1 +\ve_2 \d_{\o,0} }{(d_{\o}+1)\ve_1} \, {\Phi}_{\bd}(\bx)  \\
    & \tilde\EQ_{\bd-e_\o}(\bx+\ve_1 {\be}_1) \, = \, \EY_{\o}(x_\o+\ve_1) \, \tilde\EQ_\bd(\bx)  \\
    & \tilde\EQ_{\bd+e_\o}(\bx-\ve_1 {\be}_1) \, = \, \frac{1}{\EY_\o(x_\o)} \, \tilde\EQ_{\bd}(\bx) .
\end{split}
\end{align}
In the  $\ve_2 \to 0$ limit with all $x_{\o}=x$, $\o=0,1,\cdots,N-1$, identical,  
all the non-negative integers $d_\o $ in the sum \eqref{eq:genftildq} would have to be the same (otherwise the numerator of $ {\Phi}_{\bd}$ vanishes, cf. \eqref{eq:dualnot}): 
\begin{align}
    \lim_{\ve_2 \to 0} \tilde\EQ(x, x, \ldots, x)[\hat{\bl}] = \lim_{\ve_2 \to 0} \tilde{\bf Q}(x) [\bl]. 
\end{align} In this sense  
$\tilde{\EQ}_\o ({\bx})$ is a generalization of the $\tilde{\bf Q}$-observable.

\subsubsection{$\bf Q$-observables as $Q$-operators} \label{subsubsec:paralleldefs}

Even with the additional insertion of the $\bf Q$-observable, the correlation function \eqref{eq:correl} of the $\bf Q$-observable and the monodromy surface defect is valued in the space of degree-zero Laurent polynomials in the monodromy defect parameters $\mathbf{u}=(u_\o)_{\o=0} ^{N-1}$,
\begin{align}
     \mathbf{Q}(x)\cdot \Big\langle \Psi(\bu) \Big\rangle_\ba := \Big \langle Q(x) \Psi(\mathbf{u}) \Big\rangle_\ba  \in \CalH  = \left( \bigotimes_{\o=0}^{N-1} \CalH_\o \right)^{\BC^\times},
\end{align}
just as the vacuum expectation value of the monodromy surface defect itself. In other words, we can view the $\bf Q$-observable also as an operator
\begin{align} \label{eq:Qoper}
\mathbf{Q}(x) \in \text{End}\left(\CalH \right),
\end{align}
whose action on the basis element $\langle \Psi(\mathbf{u}) \rangle_\ba $ is given by the above correlation function. Under this context, we will refer to $\mathbf{Q}(x)$ as the \textit{$Q$-operator}. Since the contribution at each partition $\bl$ is modified by the insertion of the $\bf Q$-observable, the action of $Q$-operator is highly non-trivial.

In the limit $\ve_2 \to 0$, the surface defect and the surface observable can be moved around the plane $\BC_2$ topologically. Therefore the vacuum expectation value \eqref{eq:correl} factorizes:
\begin{align}
  \lim_{\ve_2 \to 0} \frac{\Big \langle {\bf Q}(x) \Psi(\mathbf{u}) \Big\rangle_\ba}{\langle 1 \rangle_{\ba}} =Q(\ba;x) \psi(\ba;\mathbf{u}),
\end{align}
by cluster decomposition. The normalized vacuum expectation values $Q(\ba;x)$ and $\psi(\ba;\mathbf{u})$ are precisely the ones we already obtained in \eqref{eq:qvevlim}and \eqref{eq:monodefns}. In the point of view of the action \eqref{eq:Qoper}, the factorization implies that the $Q$-operator acts diagonally on the basis $\psi(\ba;\mathbf{u})$ of the space $\CalH $ enumerated by the Coulomb moduli $\ba$, with the eigenvalue $Q(\ba;x)$. Under this context, we will refer to the eigenvalue $Q(\ba;x)$ also as the \textit{$Q$-function}.

Note that such a factorization does not occur for other $\bf Q$-observables \eqref{eq:genQspec}, thus they do not act diagonally in the basis $\psi(\ba;\mathbf{u})$.

The $\tilde{\bf Q}$-observable also defines an action on the space $\CalH$ by its insertion in the correlation function,
\begin{align}
\begin{split}
  \tilde{\mathbf{Q}}(x) \cdot \Big \langle  \Psi(\mathbf{u}) \Big\rangle_\ba :=  \Big \langle \tilde{\bf Q}(x) \Psi(\mathbf{u}) \Big\rangle_\ba  \in \CalH ,\qquad  \tilde{\mathbf{Q}}(x) \in \text{End}\left(\CalH \right).
\end{split}
\end{align}
In the limit $\ve_2 \to 0$, the action is diagonal on the basis $\psi(\ba;\mathbf{u})$ of the space $\CalH$ enumerated by the Coulomb moduli $\ba$, with the eigenvalue $\tilde{Q}(\ba;x)$,
\begin{align}
    \lim_{\ve_2 \to 0} \frac{\Big\langle \tilde{\bf Q}(x) \Psi(\mathbf{u}) \Big\rangle_\ba}{\langle 1 \rangle_{\ba}} = {\tilde{Q}}(\ba;x) \psi(\ba;\mathbf{u}).
\end{align}

To sum up, the $\bf Q$-observables can be viewed as $Q$-operators on the space $\CalH $. In the limit $\ve_2 \to 0$, the action of $Q$-operators are simultaneously diagonalized on the basis $\psi(\ba;\mathbf{u}) \in \CalH $, the normalized vacuum expectation value of the monodromy surface defect. The eigenvalues are the normalized vacuum expectation values ($Q(\ba;x)$ and $\tilde{Q}(\ba;x)$) of the $\bf Q$-observables, which we call the $Q$-functions.

\section{Non-perturbative Dyson-Schwinger equations in the presence of surfaces} \label{sec:tq}

Having introduced the surface defects and observables, we turn to the modification of the  chiral ring of the four-dimensional gauge theory in their presence. We derive
the so-called TQ equations and show they provide the quantization of the chiral ring in the form of $\hbar$-opers.

First, we give geometric constructions of the $\hbar$-difference connections and $\hbar$-difference opers, or $\hbar$-opers, for short. Then we move on to their $\EN=2 $ gauge theoretical realization. Within the gauge origami configuration, we add D3-branes wrapping $\BC^2_{34}$, thereby engineering the $qq$-characters \cite{Nikita:I}. By the regularity property of the vacuum expectation value of the $qq$-character \cite{Nikita:II}, we show that the vacuum expectation values of $\bf Q$- and $\tilde {\bf Q}$-observables obey quantum $\hbar$-oper equation. Further, the correlation functions of the $\bf Q$-, ${\tilde{\bf Q}}$-observables and the regular monodromy surface defect are shown to satisfy \textit{fractional TQ equations}.

\subsection{Moduli space of $\hbar$-connections and affine space of $\hbar$-opers} \label{subsec:hopergeom}
We introduce the moduli space of $\hbar$-connections on $\BP^1$ with a framing at $\infty \in \BP^1$ and regular singularities at $D \subset \BP^1 \setminus \{\infty \} = \BC$, and the affine space of $\hbar$-opers as a subspace. The presentation in this section resembles the one in \cite{Koroteev:2018jht,Frenkel:2020iqq,Koroteev:2022tzv}, but is different in detail.

Fix $\hbar \in \BC$. Let us consider the shift automorphism $e^{\hbar \p_x} : \BP^1 \longrightarrow \BP^1$ defined by $x \mapsto x+\hbar$. Note that it restricts to a well-defined automorphism on $\BP^1\setminus \{\infty\} =\BC$, $e^{\hbar \p_x} : \BC \longrightarrow \BC$. Given a rank $n$ vector bundle $E\longrightarrow \BP^1$, let $E^\hbar$ be the pullback of $E$ under the map $e^{\hbar \p_x}$.

Consider a map $A : E \longrightarrow E^\hbar$. Upon trivialization of $E$ on an open dense subset $U \subset \BP^1$, the map $A$ is determined by the matrix $A(x) \in \fgl(n) \otimes \BC(x)$ representing linear map $E_x \to E_{x+\hbar}$ with respect to a chosen basis on $U \cap e^{-\hbar \p_x} (U)$. A change of trivialization by $g(x)$ modifies the matrix $A(x)$ by
\begin{align}
    A(x) \mapsto g(x+\hbar) A(x) g(x) ^{-1}.
\end{align}
The shift automorphism induces an operator $e^{\hbar \p_x} : E \longrightarrow E^\hbar$ that sends a section $s(x)$ to $s(x+\hbar)$. The above gauge transformation can also be expressed as
\begin{align}
    \mathds{1}_n - A(x) e^{-\hbar \p_x} \mapsto g(x+\hbar) \left( \mathds{1}_n - A(x) e^{-\hbar \p_x} \right)g(x+\hbar) ^{-1}.
\end{align}
The map $A:E \longrightarrow E^\hbar$ is called \textit{meromorphic $\hbar$-connection} on $E$ with a framing $K \in GL(n)$ and regular singularities at $D=\{m_0 ^+  , m_0 ^+ , \cdots, m^+ _{N-1}  \} \subset \BP^1 \setminus \{\infty\} = \BC$, if $A(x)$ has simple poles only at $D$ and $A(x) \stackrel{x\to \infty}{\sim} K$. In particular, $A(x)$ is regular at $\infty \in \BP^1$ and $\det A (x) = \det K \frac{P^- (x)}{P^+ (x)} = \det K \prod_{a=0} ^{N-1} \frac{x-m_\a ^-}{x - m_\a ^+}$ for some $\{ m_\a ^- \,\vert \, \a=0,1,\cdots, N-1 \} \subset \BP^1 \setminus \{\infty\} = \BC$. In other words, $\{m^- _\a  \, \vert \, \a=0,1,\cdots, N-1\}$ is the locus where $A(x)$ is not invertible, and we regard it also as an input data for the $\hbar$-connection. We will consider the generic case where the $\hbar$-lattices originating from the zeros and the simple poles do not overlap anywhere, i.e., $\{m^\pm _\a+  \hbar \BZ_{\geq 0} \} \cap \{m^\pm _\b+  \hbar \BZ_{\geq 0} \} = \varnothing$ for any pair of $\pm$ and $\a,\b \in \{0,1,\cdots, N-1\}$.

Now let us restrict to our main case $n=2$. We start from a $\hbar$-connection in the form of
\begin{align}
    A(x) = \begin{pmatrix}
        \a(x) & \b(x) \\ \g(x) & \d(x) 
    \end{pmatrix}, \quad A(x) \stackrel{x\to \infty}{\sim}  \begin{pmatrix}
        \qe & 0 \\ 0 & 1
    \end{pmatrix}  ,\quad \det A(x) = \qe \frac{P^-(x)}{P^+ (x)}.
\end{align}
We can always make a gauge transformation to set the lower-left component to be $\frac{1}{P^+ (x)}$, at the price of creating singularities of the $\hbar$-connection at the zeros of $\g(x)$. The $\hbar$-opers are defined to be the $\hbar$-connections which do not contain any additional singularity in this gauge, i.e., $\g(x) = \frac{1}{P^+ (x)}$. Then, by making a triangular gauge transformation by
\begin{align}
        g(x) = \begin{pmatrix}
       1 &  P^+ (x) \d(x) \\ 0 & 1 
    \end{pmatrix} ,
\end{align}
we get
\begin{align} \label{eq:hgauge}
    A'(x) = \begin{pmatrix}
      \frac{t(x)}{P^+ (x)} &  -\qe P^- (x)  \\ \frac{1}{P^+ (x)} & 0
    \end{pmatrix},
\end{align}
where we defined $t(x) : = P^+ (x) \a(x) + P^+ (x+\hbar) \d (x+\hbar) = (1+\qe)x^N 
 + \cdots$, which is a degree $N$ polynomial in $x$. The coefficients of the polynomial $t(x)$ span the space of $\hbar$-opers. The leading coefficient is fixed to be $1+\qe$ by the framing as noted above. We will also soon restrict to the subspace where the next-to-leading order coefficient is fixed to a number. After doing so, the space of $\hbar$-opers is a $(N-1)$-dimensional affine space.
 
In the gauge \eqref{eq:hgauge}, we can write the $\hbar$-difference equation for a horizontal section as
\begin{align} \label{eq:hopermat}
   0 = \left(  \mathds{1}_2 -\begin{pmatrix}
      \frac{t(x)}{P^+ (x)} &  -\qe P^- (x)  \\ \frac{1}{P^+ (x)} & 0
    \end{pmatrix} e^{-\hbar \p_x} \right) \frac{1}{M^+ (x)} \begin{pmatrix}
         Q(x+\hbar) \\   Q(x)  
    \end{pmatrix},
\end{align} 
where, as a convention, we normalized the horizontal section by a product of $\G$-functions $M^+ (x)$, defined by $M^\pm (x) = \prod_{\a=0} ^{N-1} \frac{(-\hbar)^{\frac{x-m^\pm _\a}{\hbar}} }{ \G\left(-\frac{x-m^\pm _\a}{\hbar} \right)} $ satisfying $\frac{M^\pm (x)}{M^\pm (x-\hbar)} = {P^\pm (x)}$. It follows that the lower component $Q(x)$ of the horizontal section satisfies a second-order scalar $\hbar$-difference equation,
\begin{align} \label{eq:hopergeom}
   0= \left(1 - t(x) e^{-\hbar \p_x} + \qe P(x) e^{-2\hbar \p_x} \right)Q(x+\hbar).
\end{align}
We call this $\hbar$-difference equation the \textit{$\hbar$-oper equation}, and call $Q(x)$ the \textit{$\hbar$-oper solution}. Note that the $\hbar$-oper equation is nothing but the scalar Baxter TQ equation satisfied by $Q$-functions for the $\fgl(2)$ XXX spin chain with $N$ sites. We will give a gauge theoretical construction of the $\hbar$-opers in the next subsection, and come back to the spectral problem of the XXX spin chain as a consequence of our $\EN=2$ theoretical formulation of the $\hbar$-Langlands correspondence in section \ref{subsubsec:hlanglands}.

Let us be given with two independent $\hbar$-oper solutions $Q(x)$ and $\tilde{Q}(x)$. The $\hbar$-Wronskian of the corresponding horizontal sections is defined by
\begin{align}
\begin{split}
    W_\hbar (x) &:= \frac{1}{M^+ (x)}\begin{pmatrix}
        Q(x+\hbar) \\ Q(x)
    \end{pmatrix} \wedge \frac{1}{M^+ (x)} \begin{pmatrix}
        \tilde{Q}(x+\hbar) \\ \tilde{Q}(x)
    \end{pmatrix} \\
    &= 
\frac{1}{M^+ (x) ^2} \left(Q(x+\hbar)\tilde{Q}(x) - \tilde{Q}(x+\hbar)Q(x)\right).
\end{split}
\end{align}
From \eqref{eq:hopermat}, it follows that
\begin{align} \label{eq:hopercond}
    \frac{W_\hbar (x)}{W_\hbar (x-\hbar)} = \qe \frac{P^-(x)}{P^+ (x)},\qquad W_\hbar (x) = \qe^{\frac{x}{\hbar}} \frac{M^-(x)}{M^+ (x)}.
\end{align}
Note that the $\hbar$-Wronskian $W_\hbar (x)$ of the $\hbar$-oper has $N$ semi-infinite $\hbar$-lattices of zeros at $x \in m^- _\a +  \hbar \BZ_{\geq 0}$, as well as $N$ semi-infinite $\hbar$-lattices of simple poles at $x \in m^+ _\a + \hbar \BZ_{\geq 0}$, $\a \in \{0,1,\cdots, N-1\}$. This is in contrast with the $\hbar$-(or $q$-)opers studied in \cite{Koroteev:2018jht,Frenkel:2020iqq,Koroteev:2022tzv} whose $\hbar$-(or $q$-)Wronskian have finite lattices of zeros (without any pole). As we will show in section \ref{sec:hbarlang}, this extension to semi-infinite $\hbar$-lattices of zeros and simple poles is necessary to incorporate bi-infinite Yangian modules in the XXX spin chain. When our formulation is restricted to the special loci where the positions of zeros $m^- _\a$ and poles $m^+ _\a$ are related by an integer multiple of $\hbar$, i.e., $m^+ _\a - m^- _\a  \in \hbar \BZ_{> 0}$, the bi-infinite modules precisely contain finite-dimensional submodules and the $\hbar$-Wronskian with finite $\hbar$-lattices of zeros is recovered by cancellation between $\G$-functions.\footnote{To be precise, the bi-infinite module $\CalH_\a$ in consideration contains a lowest-weight sub-Verma module at the discrete loci $m^+ _\a -a_\a \in \hbar \BZ$. If we impose further $m_\a ^+ - m^- _\a \in \hbar\BZ>0$, then we obtain the described finite-dimensional submodule. These specializations of the parameters can be understood in the view of higgsing. See \cite{Jeong:2021bbh}.}

The holomorphic symplectic structure of the moduli space of $\hbar$-connections was studied in \cite{Elliott:2018yqm}, from the twistor rotation for the hyper-K\"{a}hler structure of the moduli space $\EM_{\text{mHiggs}} (GL(n),\BP^1;D)$ of multiplicative Higgs bundles (the space of $\hbar$-opers is called the Hitchin section there in the view of the twistor rotation. See also \cite{Frassek:2018try}). We expect the space of $\hbar$-opers is a holomorphic Lagrangian submanifold of the moduli space of $\hbar$-connections. We postpone the description of the space of $\hbar$-opers in
Darboux coordinates on the moduli space of $\hbar$-connections to future work. See also \cite{Mukhin2005} for a different realization of $\hbar$-opers.

\subsection{TQ equation for $\bf{Q} / \tilde{\bf{Q}}$-observables and $\hbar$-opers} \label{subsec:tqoper}
\subsubsection{$\bf Q$-observable}
To the gauge origami setup for the $\bf Q$-observable we add one more D3-brane supported on $\{0\}\times \BC^2 _{34}$. The effect of that insertion is to engineer the $qq$-character. Consequently, the equivariant Chern characters of the Chan-Paton bundles are given by
\begin{align}
\begin{split}
    N_{12} & = \sum_{\alpha} e^{a_\alpha} \cdot \CalR_0 + \sum_{\alpha} e^{m_\alpha^- - \ve_4} \cdot \CalR_1 + \sum_{\alpha} e^{m_\alpha^+ - \ve_3} \cdot \CalR_{2} \\
    N_{13} & = e^{x'+\ve_1+\ve_3} \cdot \CalR_1 \\
    N_{34} & = e^x \CalR_0.
\end{split}
\end{align}
Flowing to the $A_1$ theory by setting $\qe_1,\qe_2 = 0$ we arrive at the gauge origami partition function 
\begin{align}
    Z (x, x') = \sum_{{\boldsymbol\lambda}_{12},{\boldsymbol\lambda}_{34}} \kq^{|{\boldsymbol\lambda}_{12}|+|{\boldsymbol\lambda}_{34}|} \BE & \left[ -\frac{\hat{P}_3\hat{S}_{12}\hat{S}_{12}^*}{\hat{P}_{12}^*} - \frac{\hat{P}_{2}\hat{S}_{13}\hat{S}_{13}^*}{\hat{P}_{13}^*} - \frac{\hat{P}_{1}\hat{S}_{34}\hat{S}_{34}^*}{\hat{P}_{34}} \right. \nonumber\\ 
    & \left. -{\tilde q}_{12}\hat{S}_{12}^*\hat{S}_{34} + {\tilde q}_{2} \hat{P}_4 \hat{S}_{13}\frac{\hat{S}_{12}^*}{\hat{P}_{1}^*} -{\tilde q}_{13}\frac{\hat{P}_{2}}{\hat{P}_{3}}\hat{S}_{34}\hat{S}_{13}^* \right]^{\BZ_3}.
\end{align}
It was shown in \cite{Jeong:2023qdr} that the contribution
\begin{align}
    -{\tilde q}_{13}\frac{\hat{P}_{2}}{\hat{P}_{3}}\hat{N}_{34}\hat{N}_{13}^*
\end{align}
can be replaced by 
\begin{align} \label{eq:simple}
    -{\tilde q}_{13} \hat{N}_{34} \hat{N}_{13}^* - {\tilde q}_{34} \hat{N}_{13} \hat{N}_{34}^*.
\end{align}
For completeness, we will repeat the derivation here. This term originates from the product of weights of an infinite set of equations
\begin{align}
    J_{13}\, B_3^k \, I_{34}=0, \quad J_{34}\, B_3^k \, I_{13} = 0, \quad k\geq 0.
\end{align}
on the origami ADHM data. 
We now show that in our setup the matrix $B_3$ vanishes, hence the corresponding weights are never zero, and can be dropped from the numerator without introducing poles in the relevant variable $x$. First, $B_{3}I_{12}=0$ by the standard stability condition of gauge origami. Secondly, the choice of $\hat{N}_{13}$ in $\CalR_2$ representation of $\BZ_3$ orbifolding ensures that instanton cannot move onto the ${\BC}_{13}^2$ subspace in the freezing, $K_{13}=0$. Thus, we get
$$
    B_3\, I_{13}=0.
$$
Finally, the would-be vectors $B_{3}I_{34}(N_{34})$ belong to the $\CalR_2$-component of $K_{34}$, which is empty in the ungauging limit. This implies
$$
    B_3 I_{34}=0.
$$
Therefore, all constraints are automatically satisfied for all $k>0$. The only remaining constraints imposed to the gauge origami data are
\begin{align}
    J_{13}I_{34}=0, \quad J_{34}I_{13}=0,
\end{align}
whose contributions to the partition function read exactly
$$
    \BE \left[-{\tilde q}_{13}\hat{N}_{34}\hat{N}_{13}^* - {\tilde q}_{34}\hat{N}_{13}\hat{N}_{34}^*+\hat{P}_{2}\hat{P}_4 \hat{N}_{13} \hat{K}_{34}^* \right]. 
$$
With the simplification \eqref{eq:simple}, the gauge origami partition function becomes 
\begin{align}
   Z(x, x') = \langle {\EuT}(x) \star {\bf Q}(x') \rangle_{\ba} \, = \, \sum_{{\boldsymbol\lambda}} \kq^{|{\boldsymbol\lambda}|}  \m_{{\ba}, \bl} \, \left( \EuScript{T}(x) \star {\bf Q}(x') \right) [ {\ba}, {\bla}] , 
   \label{eq:tqvev}
\end{align}
i.e. the vev of the OPE of the $qq$-character ${\bf\EuT}(x)$ and the $\bf Q$-observable ${\bf Q}(x')$, which evaluates to
\begin{align}
     {\bf\EuT}(x) \star {\bf Q}(x') [{\ba}, {\bla}] = \left( (x-x') {\EY}(x + \ve_1+\ve_2) [{\ba}, {\bla}] + \kq (x-x'+\ve_2) \frac{P(x)}{{\EY}(x)[{\ba}, {\bla}]} \right) {\bf Q}(x')[{\ba}, {\bla}]
     \label{eq:tqope}
\end{align}
with
\begin{align}
P(x) = P^+ (x) P^- (x),\qquad P^\pm (x) = \prod_{f=1}^{N} (x-m^{\pm}_f). 
\end{align}
on an instanton configuration $\bla$ in the vacuum $\ba$. Now, the absence of poles in $x$ implies that the left hand side of \eqref{eq:tqvev} is a degree $N+1$ polynomial in $x$, whose coefficients are entire functions of $x'$. 
We can write:
\beq
Z(x,x') = (x-x') \langle {\bf T}_{0}(x) \cdot {\bf Q}(x') \rangle + {\qe} {\ve}_{2} \langle {\bf T}_{1}(x) \cdot {\bf Q}(x') \rangle + \langle {\bf T}_{-1}\cdot {\bf Q}(x') \rangle 
\label{eq:zxx}
\eeq
where the observables 
\beq
\begin{aligned}
 & {\bf T}_{0}(x)  = \left( {\EY}(x + \ve_1+\ve_2)  + \kq  \frac{P(x)}{{\EY}(x)} \right)_{x^{\geq 0}} \\
 & {\bf T}_{1}(x)  = \left( \frac{P(x)}{{\EY}(x)}\right)_{x^{\geq 0}} \\
 & {\bf T}_{-1}  = {\rm Coeff}_{x^{-1}} \left( {\EY}(x + \ve_1+\ve_2)  + \kq  \frac{P(x)}{{\EY}(x)} \right)
 \end{aligned}
 \label{eq:t0t1}
\eeq
are, explicitly, polynomials\footnote{The notation $\left( \ldots \right)_{x^{\geq 0}}$ means taking the polynomial part in expansion near $x = \infty$, similarly the notation ${\rm Coeff}_{x^{-1}} ( \ldots )$ means taking the residue at $x = \infty$} in $x$, whose coefficients are some polynomials in ${\rm Tr}\, {\phi}^{k}$ and the masses, and ${\bf T}_{-1}$ is also a  polynomial in ${\rm Tr}\, {\phi}^{k}$'s and the masses. The product $\cdot$ in \eqref{eq:zxx} means that the evaluations at $\bla$ of the factors simply multiply, i.e. 
\beq
\left( {\bf T}_{0}(x) \cdot {\bf Q}(x') \right) [{\ba}, {\bla}] = 
 {\bf T}_{0}(x) [{\ba}, {\bla}]  \times {\bf Q}(x') [{\ba}, {\bla}] 
\eeq
At special values $x= x'$ and $x = x'-\ve_2$, the Eq. \eqref{eq:tqvev} gives
\begin{align} \label{eq:TQder}
\begin{split}
    & Z(x',x') = \ve_2 \kq \left\langle \frac{P(x')}{{\bf\EY}(x')} {\bf Q}(x') \right\rangle = \ve_2 \kq P(x') \langle {\bf Q}(x'-\ve_1) \rangle \\
    & Z (x=x'-\ve_2,x')  = -\ve_2 \langle {\bf\EY}(x'+\ve_1) {\bf Q}(x') \rangle = -\ve_2 \langle {\bf Q}(x'+\ve_1) \rangle
\end{split}
\end{align}
Let us now evaluate
\beq
\frac{Z(x,x) - Z(x-{\ve}_{2}, x)}{\ve_2}
\eeq
in two ways: first, from \eqref{eq:TQder}, second, from \eqref{eq:zxx}. We arrive at the equation:
\beq
\langle {\bf Q}(x+\ve_1)  + {\qe} P(x) {\bf Q}(x- {\ve}_{1} ) \rangle = 
\langle {\bf t}(x) \cdot {\bf Q}(x) \rangle
\label{eq:qTQ}
\eeq
where we
defined an observable-valued degree $N$ polynomial in $x$,
\beq \label{eq:tx}
{\bf t}(x) = {\bf T}_{0}(x-{\ve}_{2}) + {\qe} \left( 
{\bf T}_{1}(x)  -  {\bf T}_{1}(x-{\ve}_{2}) \right)
\eeq
whose coefficients are again the polynomials in the local observables ${\rm Tr}\, {\phi}^k$ and the masses. We call \eqref{eq:qTQ} the \textit{quantum TQ equation}. 

In $\Omega_1$-background the quantum TQ equation reduces to the Baxter's TQ equation on the $Q$-function:
\begin{align} \label{eq:qhopereq}
    0 = \left[ 1 - t(\ba;x)e^{-\ve_1 \p_x} + \qe P (x) e^{-2 \ve_1 \p_x} \right] Q(\ba;x+\ve_1),
\end{align}
which we shall also call the \textit{$\hbar$-oper equation}, 
where 
\beq \label{eq:tvev}
t(\ba;x) : = \lim_{\ve_2 \to 0} \frac{\langle {\bf t}(x) \rangle_\ba}{\langle 1 \rangle_\ba}
\eeq
is the normalized expectation value of the $qq$-character $\EuT(x)$ in the limit $\ve_2 \to 0$. 

The second-order scalar $\hbar$-difference equation \eqref{eq:qhopereq} is precisely the $\hbar$-oper equation \eqref{eq:hopergeom} that we derived in section \ref{subsec:hopergeom}. Thus, we gave a $\EN=2$ gauge theoretical realization of the $GL(2)$ $\hbar$-opers on $\BC$ (more precisely, $\BP^1$ with a framing at $\infty$) with regular singularities at $D = \{m^\pm _\a \, \vert \, \a=0,1,\cdots, N-1\} \subset \BC$ by the $\bf Q$-observable surface defect. 

Note that the positions of the singularities are given by the mass parameters $\mathbf{m} ^\pm$ and the twist parameter is identified with the gauge coupling $\qe$. Finally, $t(\ba;x)$ is a degree $N$ polynomial whose coefficients are parametrized by the Coulomb moduli $\ba$. The first two coefficients are respectively $1+\qe$ and a simple combination of $\bar{m}^\pm$ and $\ve_1$. The rest of $N-1$ coefficients are combinations of normalized vacuum expectation values of the local observables in the limit $\ve_2 \to 0$, namely, $\lim_{\ve_2 \to 0} \Big\langle \text{Tr}\, \phi^k \Big\rangle_\ba$, $k=2,3,\cdots, N$. We will collectively denote these coefficients as $E_k (\ba)$, $k=2,3,\cdots, N$. These coefficients $(E_k (\ba))_{k=2} ^N$, and therefore the Coulomb moduli $\ba$, parameterize the affine space of $\hbar$-opers. 

At this point, recall that we split the Coulomb moduli $(a_\a)_{\a=0} ^{N-1}$ into the values $\frac{a_\a}{\ve_1} \; \text{mod} \; \BZ $ and $\ve_1$-integral shifts. The former is considered to be fixed, and the $\ve_1$-integral shifts parameterize the $\hbar$-opers for the fixed values of $\frac{a_\a}{\ve_1} \; \text{mod} \; \BZ $. In turn, we regard the space of $\hbar$-opers as a $(N-1)$-dimensional $\ve_1$-lattice for each choice of $\frac{a_\a}{\ve_1} \; \text{mod} \; \BZ $ rather than an affine space over $\BC$.

Even though we do not describe the space of $\hbar$-opers in Darboux coordinates on the moduli space of $\hbar$-connections in this work, it would be desirable to do so to find an explicit relation between them and the Coulomb moduli $\ba$, and to express the generating function for the space $\hbar$-opers in terms of the Coulomb moduli. See \cite{Jeong:2018qpc} for the gauge theoretical derivation of the equivalence of the generating function for the space of opers, under a higher-rank generalization of the NRS (sometimes  called a higher-rank generalization of the complexified Fenchel-Nielsen, cf \cite{Hollands:2013qza,Hollands:2017ahy}) coordinates on the moduli space of flat connections, and the effective twisted superpotential of the $\Omega_1$-deformed $\EN=2$ theory, as conjectured in \cite{NRS2011}. 

\subsubsection{${\tilde{\bf Q}}$-observable}
We now verify that the identical TQ equation is satisfied by the $\tilde {\bf Q}$-observable in the $\Omega_1$-background. As in  the case of $\bf Q$-observable, we insert an additional D3-brane supported on $\{0\} \times \BC^2_{34}$. The gauge origami setup is represented by the following 
\begin{align}
\begin{split}
    N_{12} & = \sum_{\alpha} e^{a_\alpha} \cdot \CalR_0 + \sum_{\alpha} e^{m_\alpha^- - \ve_4} \cdot \CalR_1 + \sum_{\alpha} e^{m_\alpha^+ - \ve_3} \cdot \CalR_{1} \\
    N_{13} & = e^{x'+\ve_1} \cdot \CalR_0 \\
    N_{34} & = e^x \CalR_0.
\end{split}
\end{align}
The gauge origami partition function gives the correlation function of the $\tilde{\bf Q}$-observable and the $qq$-character,  
\begin{align} \label{eq:qqdualQ}
\begin{split}
    & {\tilde Z}(x, x')  =  \Bigg\langle {\EuT} (x) \star \tilde{\bf Q} (x') \Bigg\rangle  =  \, \qe^{\frac{x'}{\ve_1}}\sum_{d=0} ^{\infty} \qe^d  {\phi}_{d} \,  \times \\
    & \left[ \left( x'-x - {\ve}_{2} + \frac{d {\ve}_{1}\ve_{2}}{x'-x+d\ve_1} \right) \Bigg\langle \tilde{\bf Q}_d (x') {\bf\EY}(x+\ve_1+\ve_2) \Bigg\rangle \ + \right. \\
    & \qquad \left. + \qe P(x)  \left( x'-x   -{\ve}_{2}  \frac{\ve_{2} + (d+1) {\ve}_{1}}{x' -x +(d+1)\ve_1} \right)  \,  \Bigg\langle  \frac{\tilde{\bf Q}_d(x')}{{\bf\EY}(x)} \Bigg\rangle \right] ,
\end{split}
\end{align}
We stress that in \eqref{eq:qqdualQ} the potential poles at $x=x'+d\ve_1$ are absent even before taking the expectation value. 

Thus,  ${\tilde Z}(x, x')$ is a polynomial of degree $N+1$ in $x$. 
We can express ${\tilde Z}(x,x')$ in a form, somewhat similar to \eqref{eq:zxx} (cf. \eqref{eq:t0t1}):
\beq
{\tilde Z}(x,x') =  (x'-x - {\ve}_{2}) \langle {\bf T}_{0}(x) \cdot {\tilde{\bf Q}}(x') \rangle + {\qe} {\ve}_{2} \langle {\bf T}_{1}(x) \cdot {\tilde{\bf Q}}(x') \rangle - 
    \langle {\bf T}_{-1} \cdot {\tilde{\bf Q}}(x') \rangle \ + {\ve}_{1}{\ve}_{2}\,
    {\tilde z}(x,x')
\eeq
with the `error' term
\begin{align}
\begin{split}
&{\tilde z}(x,x') = 
 -  {\qe}^{\frac{x'}{{\ve}_{1}}}\, \sum_{d=1}^{\infty} \, d {\qe}^{d} \, {\phi}_{d} \, 
 \Biggl\langle \frac{{\tilde{\bf Q}}_{d-1}(x'){\bf G}(x, x'+ d {\ve}_{1} )}{{\EY}(x'+(d+1){\ve}_{1}+{\ve}_{2})}  \Biggr\rangle \, , \\
& {\bf G}(x, x') = 
 \frac{{\EY}(x+{\ve}_{1}+{\ve}_{2}) \frac{P(x')}{{\EY}(x')}-  {{\EY}(x'+{\ve}_{1}+{\ve}_{2})} \frac{P(x)}{{\EY}(x)}}{x'-x}  \\
 &\quad\quad\quad\quad =\frac{\left( {\bf T}_{0}(x) - {\qe} {\bf T}_{1}(x) \right) \frac{P(x')}{{\EY}(x')} -  {{\EY}(x'+{\ve}_{1}+{\ve}_{2})} {\bf T}_{1}(x)}{x'-x} + o(1/x) 
\end{split}
\end{align}
At the special values $x'=x+\ve_{2}$, or $x' = x$  the Eq. \eqref{eq:qqdualQ} simplifies to 
\beq \label{eq:dualTQ1}
{\tilde Z} ( x-\ve_2, x )   =  \ve_{2} \Bigg\langle \tilde{\bf Q} (x+\ve_1 ) \Bigg\rangle ,
\eeq
as there is no contribution from $d=0$ in the first line due to the factor
$x'-x-\ve_{2} +d \ve_1$ in the numerator, and
\beq \label{eq:dualTQ2}
{\tilde Z} (x, x)  =  -\ve_{2}  \qe P(x) \Bigg\langle {\tilde{\bf Q}} (x-\ve_1) \Bigg\rangle ,
\eeq
where we again used the definition \eqref{eq:dualQ}. 
Again, by computing $\frac{ {\tilde Z}  (x-\ve_{2}, x)  - {\tilde Z} (x,x)}{\ve_{2}}$ in two ways, we arrive at the equation
\begin{align} \label{eq:qtqdualq}
\begin{split}
 \Biggl\langle
{\bf {\tilde Q}}(x+{\ve}_{1}) + {\qe} P(x) {\tilde{\bf Q}}(x- {\ve}_{1}) \Biggr\rangle\ &= \Biggl\langle \left( {\bf t}(x+\ve_2)  +\qe \left(  {\bf T}_{1}(x-{\ve}_{2}) - {\bf T}_{1}(x+\ve_2) \right) \right) \cdot {\tilde{\bf Q}}(x)\Biggr\rangle \\
& \quad +  \sum_{d=1}^{\infty} {\qe}^{\frac{x}{\ve_1} + d}\, 
 \left( {\phi}_{d}-{\phi}_{d-1} \right) \,  \Biggl\langle {\tilde{\bf Q}}_{d-1}(x) {\bf T}_{1}(x) - {\tilde{\bf Q}_{d}}(x) \left( {\bf T}_{0}(x) - {\qe} {\bf T}_{1}(x) \right) \Biggr\rangle 
\end{split}
\end{align}
We call this quantum TQ equation for the $\tilde{\bf Q}$-observable.

In the limit ${\ve}_{2} \to 0$, the quantum TQ equation reduces to the Baxter's TQ equation, or equivalently the  $\hbar$-oper difference equation, on the ${\tilde Q}$-function:
\begin{align} \label{eq:dualTQ}
\begin{split}
   & \tilde{Q} (\ba;x+\ve_1)  +\qe  P(x)  \tilde{Q} (\ba;x-\ve_1)   =  {t}  (\ba;x) \tilde{Q} (\ba;x) , \\
   &0 = \left[ 1 - {t}(\ba;x) e^{-\ve_1 \p_x} + \qe P(x) e^{-2\ve_1 \p_x} \right] \tilde{Q}(\ba;x+\ve_1).
\end{split}
\end{align}

It is clear from \eqref{eq:qtqdualq} that $t(\ba;x)$, the degree $N$ polynomial whose coefficients are normalized vevs of the local observables, in the above equation is identical to the one appearing in the $\hbar$-oper equation \eqref{eq:qhopereq} for the $Q$-function. It can also be directly checked as follows. Multiplying the normalized vevs $t(\mathbf{a};x)$ and $\tilde{Q}(\mathbf{a};x)$ \eqref{eq:dualQ} in the limit $\ve_2 \to 0$, we get
\begin{align}
\begin{split}
    &t(\mathbf{a};x)\tilde{Q}(\mathbf{a};x) \\
    &= \qe^{\frac{x}{\ve_1}} \sum_{d=0} ^{\infty} \qe^d   \frac{ M(x  +d \ve_1)}{Q (\ba; x+ d  \ve_1) Q(\ba;x   +(d+1) \ve_1)} t(\ba;x) Q (\mathbf{a};x)\\
    & =  \qe^{\frac{x}{\ve_1}} \sum_{d=0} ^{\infty} \qe^d   \frac{ M(x  +d \ve_1)}{Q (\ba; x+ d  \ve_1) Q(\ba;x   +(d+1) \ve_1)}  (Q(\ba;x+\ve_1) + \qe P(x) Q(\ba;x-\ve_1)) \\
   &=\tilde{Q}(\ba;x+\ve_1)+\kq P(x) \tilde{Q}(\ba;x-\ve_1),
\end{split}
\end{align}
where we used the the TQ equation for the $\bf Q$-observable in the second line and the definition \eqref{eq:dualQ} of the $\tilde{\bf Q}$-observable in the fourth line. Accordingly, we conclude that the normalized vacuum expectation value $\tilde{Q}(\ba;x)$ of the dual $\bf Q$-observable in the limit $\ve_2 \to 0$ satisfies the same $\hbar$-oper equation \eqref{eq:dualTQ},
\begin{align} \label{eq:btqdualQ}
 0 = \left[ 1 - {t}(\ba;x) e^{-\ve_1 \p_x} + \qe P(x) e^{-2\ve_1 \p_x} \right] \begin{pmatrix}
     {Q}(\ba;x+\ve_1) & \tilde{Q}(\ba;x+\ve_1)
 \end{pmatrix} .
\end{align}

\subsubsection{Remark: Fourier transformation between $\hbar$-opers and opers}
Here, we make a short digression and remark about the relation between the quantum TQ equation for the ${\bf Q}/{\tilde{\bf Q}}$-observables and the quantum oper equation for the canonical surface defect \cite{Jeong:2018qpc}. The aim is only to motivate the quantum TQ equations with $\ve_2\neq 0$. The detail of the relation and its implication on the bispectral duality can be found in the companion paper \cite{Jeong:2024mxr}.

The canonical surface defect, for which we call the corresponding observable $\bf{H}$-observable, is related to the $\bf{Q}$-observable by the Fourier transformation \cite{Jeong:2018qpc}
\begin{align}
    \mathbf{H} ^{(\a)} (y) = \sum_{x \in L_\a} y^{-\frac{x}{\ve_1}} \mathbf{Q}(x), \qquad \a=0,1,\cdots, N-1,
\end{align}
where $L_\a = a_\a + \ve_1 \BZ $ is the $\ve_1$-lattice centered at the Coulomb parameter $a_\a$. The summation converges in the domain $\vert \qe \vert < \vert y \vert < 1$ due to the zeros of the $\G$-functions in ${\bf Q} (x)$. The description of the canonical surface defect as coupling a two-dimensional $\EN=(2,2)$ gauged linear sigma model and its observable expression are explained in \cite{Jeong:2023qdr, Jeong:2024mxr}.

 Under the Fourier transformation, the quantum TQ equation for ${\bf Q}$-observable passes to a $N$-th order differential equation for ${\bf H}$-observable,
 \begin{align}
     0 = \left[ \p_y ^N + \mathrm{t}_2 (y) \p_y ^{N-2} + \cdots + \mathrm{t}_N (y) \right] \widehat{\mathbf{H}}(y),
 \end{align}
 where we introduced a proper normalization factor to bring the differential into a canonical form. This is called the quantum oper equation \cite{Jeong:2018qpc, Jeong:2023qdr}. 
 
 Now, we consider the $\tilde{\mathbf{Q}}$-observable. We define the $\tilde{\mathbf{H}}$-observable by the Fourier transformation
 \begin{align}
     \tilde{\mathbf{H}} ^{(\a)} (y) = \sum_{x \in \tilde{L}_\a} y^{-\frac{x}{\ve_1}} \tilde{\mathbf{Q}}(x),\qquad \a=0,1,\cdots, N-1, 
 \end{align}
 where $\tilde{L}_\a = m^- _\a + \ve_1 \BZ$ is the $\ve_1$-lattice centered at the mass parameter $m^- _\a$. It is not difficult to see that the series converges in the domain $0 < \vert y \vert < \vert \qe \vert$ due to the $\G$-functions in $\tilde{\mathbf{Q}}(x)$.

 Since the quantum TQ equation \eqref{eq:qtqdualq} for the $\tilde{\mathbf{Q}}$-observable involves more terms compared to the one for the $\mathbf{Q}$-observable, it is not immediate that the two quantum TQ equations turn into the same kind of $N$-th order differential equation through the Fourier transformation. Here, we will show the two equations are actually the same, explicitly at the example of $N=2$.

 At $N=2$, it is straightforward to compute
 \begin{align}
\begin{split}
    \bT_0(x) & = \left( \EY(x+\ve_1+\ve_2) + \kq \frac{P(x)}{\EY(x)} \right)_{x^{\geq 0}} \\
    & = (x-a_1+\ve_+)(x-a_2+\ve_+) + \ve_1\ve_2 \langle |\lambda| \rangle + \kq (x^2 + (a-m)x + a_1^2+a_2^2+a_1a_2-am -\ve_1\ve_2\langle |\lambda| \rangle ) \\
    \bT_1(x) & = \left( \frac{P(x)}{\EY(x)} \right)_{x^{\geq 0}} \\
    & = x^2 + (a-m)x + a_1^2+a_2^2+a_1a_2-am -\ve_1\ve_2\langle |\lambda| \rangle,
 \end{split}
 \end{align}
 where we use the notation $\ve_+ = \ve_1 + \ve_2$, $a= a_1+ a_2$, and $m= m_1 ^+ + m_1 ^- + m_2 ^+ + m_2 ^-$. Also, we compute
 \begin{align}
 \begin{split}
     \tilde{z}(x,x')     &  = \left \langle  (x+\ve_1\nabla^\kq) \left[ (\qe-1)\left(\nabla^\kq- \frac{x'}{\ve_1} \right) +\kq \frac{\ve_+}{\ve_1} \right] \tilde{Q}(x') \right \rangle  \\
    & \quad -  \left \langle \left[ (2\ve_+-a) \left( \nabla^\kq - \frac{x'}{\ve_1} \right) + (m-a) \kq \left( \nabla^\kq - \frac{x'}{\ve_1} + \frac{\ve_1+\ve_2}{\ve_1}  \right) \right] \tilde{Q}(x') \right \rangle,
\end{split}
 \end{align}
 where $\nabla^\qe = \qe \p_{\qe}$. By a direct substitution into the quantum TQ equations, we get
\begin{align}
\begin{split}
    &\Big\langle P^+ (x+\ve_1) \EQ (x+\ve_1) + \qe P^- (x) \EQ (x-\ve_1) \Big\rangle \\
    &= \left\langle \left[ (1+\qe) x^2 + (2\ve_1 - \qe m -(1-\qe) a )x +(a_1-\ve_1)(a_2-\ve_1) +\qe (a(a-m)-a_1 a_2)   \right] \EQ(x) \right\rangle \\
    &\quad + \ve_1\ve_2 (1-\qe) \nabla^\qe \left\langle \EQ (x) \right\rangle,
\end{split}
\end{align}
and
\begin{align}
\begin{split}
    &\Big\langle P^+ (x+\ve_1) \tilde\EQ (x+\ve_1) + \qe P^- (x) \tilde\EQ (x-\ve_1) \Big\rangle \\
    & = \left\langle \left[ (1+\qe)x^2 + \left( 2\ve_1 -(1-\qe)(a-\ve_2) -\qe m  \right) x + (a_1 -\ve)(a_2-\ve) + \qe\left( (a-m)(a-\ve_2) -a_1 a_2 -\ve_1\ve_2  \right) \right]\tilde{\EQ}(x) \right\rangle \\
    & \quad  +\ve_1\ve_2 (1-\qe) \nabla^\qe \left\langle \tilde{\EQ}(x) \right\rangle,
\end{split}
\end{align}
where we normalized the $\mathbf{Q}$- and $\tilde{\mathbf{Q}}$-observables using some $\G$-functions. In particular, it is crucial to have $\tilde{z}(x,x')$-terms in the quantum TQ equation for $\tilde{\mathbf{Q}}(x)$ to pull out the $\qe$-derivative outside the bracket. We notice that the above two equations are identical, after shifting $a_1 \to a_1 + \ve_2$ in the latter and multiplying $\qe^{-\frac{a_1 - \ve_1}{\ve_1}}$ to $\tilde{\EQ}(x)$. Therefore, under such a redefinition taken into account, the Fourier transform of $\mathbf{Q}$- and $\tilde{\mathbf{Q}}$-observables give solutions to the same quantum oper equation, converging in different domains. For more detail of the Fourier transformation between the quantum TQ equation and the quantum oper equation and its implication on the bispectral duality, see \cite{Jeong:2018qpc,Jeong:2023qdr, Jeong:2024mxr}. For the quantum oper equation associated to the Fourier transform of the next-to-simplest $\mathbf{Q}$-observable, see \cite{Jeong:2017mfh}.

\subsubsection{$\hbar$-Wronskian and $Q \tilde{Q}$-relation}
So far, we achieved two solutions to the $\hbar$-oper difference equation by the normalized vacuum expectation values of the $\bf Q$-observable and the dual $\bf Q$-observable in the limit $\ve_2\to 0$. Here, we will show that these two solutions are indeed generically independent by computing the $\hbar$-Wronskian.

The $\hbar$-Wronskian\footnote{It is called quantum Wronskian in some literature. The term \textit{quantum} sometimes indicates the $\ve_2 \neq 0$ deviation (i.e. non-critical level $\k \neq 0$), so that we avoid using this terminology and call it $\hbar$-Wronskian not to cause any confusion.} is defined to be the determinant of the matrix formed by the $\hbar$-jets of solutions to $\hbar$-difference equation,
\begin{align}
    \det \begin{pmatrix}
        Q(\ba;x+\ve_1) & \tilde{Q}(\ba;x+\ve_1) \\ Q(\ba;x) & \tilde{Q} (\ba;x)
    \end{pmatrix} = Q(\ba;x+\ve_1) \tilde{Q} (\ba;x) - \tilde{Q}(\ba;x+\ve_1) Q(\ba;x).
\end{align}
By simply plugging the expression for the normalized vacuum expectation value $\tilde{Q}(\ba;x)$ of the dual $\bf Q$-observable in the limit $\ve_2 \to 0$ into the above definition, the $\hbar$-Wronskian is computed to be
\begin{align} \label{eq:hwrons}
     Q(\ba;x+\ve_1) \tilde{Q} (\ba;x) - \tilde{Q}(\ba;x+\ve_1) Q(\ba;x) = \qe^{\frac{x}{\ve_1}} M(x),
\end{align}
where $M(x)$ is a product of inverse $\G$-functions defined in \eqref{eq:dualqpart}. Thus, the $\hbar$-Wronskian is an entire function in $x \in \BC$ with simple zeros only at discrete loci, confirming independence of the two solutions to the $\hbar$-oper equation at generic $x \in \BC$. Note that the $\hbar$-Wronskian is precisely what appeared in the $\hbar$-oper condition \eqref{eq:hopercond}, up to a unimportant entire function.

We remind that the $\hbar$-Wronskian relation is a special of the $Q\tilde{Q}$-relation (see \cite{Bazhanov:2010ts} for instance), which reads
\begin{align}
    Q\left(\ba;x+ m \ve_1 \right) \tilde{Q}(\ba;x ) - \tilde{Q}(\ba;x+ m \ve_1) Q(\ba;x ) = \qe^{\frac{x}{\ve_1}} M(x) \, \mathtt{t} ^{(m)} (\ba;x+m\ve_1),
\end{align}
where $m \in \BZ_{>0}$ and $\mathtt{t} ^{(m)} (\ba;x)$ is the eigenvalue of the transfer matrix $\mathtt{t} ^{(m)} (x) = \text{Tr}_{\BC^m } K T(x) $ defined by using the auxiliary space $\BC^{m}$, at the eigenstate specified by $\ba$ (see appendix \ref{sec:appA} and section \ref{sec:yangian}). At $m=1$, we recover the $\hbar$-Wronskian relation \eqref{eq:hwrons}. We show here that the $Q\tilde{Q}$-relation holds in the next-to-simplest case $m= 2$, involving the standard transfer matrix for $Y(\fgl(2))$.

Using the expression for the normalized vacuum expectation value $\tilde{Q}(\ba;x)$ of the dual $\mathbf{Q}$-observable \eqref{eq:dualQ}, we compute
\begin{align}
\begin{split}
    &Q(\ba;x+2\ve_1) \tilde{Q}(\ba;x) - \tilde{Q}(\ba;x+2\ve_1) Q(\ba;x) \\
    &= \qe^{\frac{x}{\ve_1} } M(x) \left( \EY(\ba;x+2\ve_1) +\qe \frac{P(x+\ve_1)}{\EY(\ba;x+\ve_1)} \right) \\
    & =  \qe^{\frac{x}{\ve_1} } M(x) \frac{Q(\ba;x+2\ve_1) + \qe P(x+\ve_1) Q(\ba; x)}{Q(\ba;x+\ve_1)}  \\ 
    &= \qe^{\frac{x}{\ve_1}  }M(x)\, t (\ba;x+\ve_1).
\end{split}
\end{align}
We confirm the $Q\tilde{Q}$-relation is satisfied at each basis elements specified by $\ba$, with the transfer matrix $t(\ba;x) = \mathtt{t}^{(2)} (\ba;x+\ve_1)$ given by the vev of the $qq$-character of the $\EN=2$ gauge theory in the limit $\ve_2 \to 0$ (see \eqref{eq:hopereqyang} and footnote \ref{fn:shift}). This is indeed anticipated from our gauge theoretical formulation of the $\hbar$-oper equation \eqref{eq:qhopereq} and \eqref{eq:btqdualQ}. 

For higher $m>2$, the $Q\tilde{Q}$-relation expresses the transfer matrices defined by the higher-dimensional auxiliary space $\BC^m$ in terms of the higher $qq$-characters. In the present case of $Y(\fgl(2))$, the only non-trivial transfer matrix is at $m=2$ from which all the other higher transfer matrices can be expressed. This is precisely in accordance with that the higher $qq$-characters of the $A_1$-quiver gauge theory can be expressed in terms of the fundamental $qq$-character in the limit $\ve_2 \to 0$.

\subsection{Non-perturbative Dyson-Schwinger equations for surface defects and observables}
Having established the $\hbar$-oper equations as chiral ring equations for the $\bf Q$-observables, we turn to the configuration of the parallel surface defects: the (dual) $\bf Q$-observable and the regular monodromy surface defect inserted on top of each other.

Recalling from section \ref{subsubsec:paralleldefs} that the $\bf Q$-observables act as $Q$-operators when inserted on top of the monodromy surface defect, our goal is to construct the universal $\hbar$-oper equation (equivalently, the operator Baxter TQ equation) as chiral ring equations. In fact, the $0$-observables inserted at the interface of the two defects make the construction even richer, leading to the R-matrices of the Yangian which are building blocks of the monodromy matrix. Therefore, we investigate the chiral ring equations for the generalized $\bf Q$-observables \eqref{eq:genQspec}, which we refer to as the \textit{fractional quantum TQ equations}.

\subsubsection{$\bf Q$-observable in the presence of the surface defect}
For this, we insert an additional D3-brane supported on $\{0\} \times \BC^2 _{34}$ , which leads to the gauge origami configuration represented by
\begin{align}
\begin{split}
    & \hat{N}_{12} = \sum_{{\o'}=0}^{N-1} \left[ e^{a_{\o'}} \CalR_0  +  q_{13}  q_2 ^{\d_{\o',0}} e^{m_{[{\o'-1}]} ^- } \CalR_1  + q_3^{-1} e^{m_{\o'}^+}  \CalR_2  \right]  \otimes  {\tilde q}_2^{\o'} \fR_{\o'}, \\
    & \hat{N}_{13} = \sum_{\o'=0}^{N-1} e^{x'_{\o'} + \ve_1 + \ve_3} {\tilde q}_2^{\o'} \CalR_1 \otimes \fR_{\o'} , \\
    & \hat{N}_{34} = e^x {\tilde q}_2^\o \CalR_0 \otimes \fR_\o.
\end{split}
\end{align}
The gauge origami partition function\footnote{\begin{multline}  {\CalZ}_{\omega}(x, {\bx}') = \sum_{\hat{\boldsymbol\lambda}_{12},\hat{\boldsymbol\lambda}_{34}} \prod_{\o=0}^{N-1} \kq_\o^{k_{12,\o}+k_{34,\o}} \BE \left[ -\frac{\hat{P}_3\hat{S}_{12}\hat{S}_{12}^*}{\hat{P}_{12}^*} - \frac{\hat{P}_{2}\hat{S}_{13}\hat{S}_{13}^*}{\hat{P}_{13}^*} - \frac{\hat{P}_{1}\hat{S}_{34}\hat{S}_{34}^*}{\hat{P}_{34}^*} -{\tilde q}_{12}\hat{S}_{12}^*\hat{S}_{34}  \right. \nonumber\\ 
     \left. + {\tilde q}_{2} \hat{P}_4 \hat{S}_{13}\frac{\hat{S}_{12}^*}{\hat{P}_{1}^*} - {\tilde q}_{13} \hat{N}_{34}\hat{N}_{13}^* - {\tilde q}_{34} \hat{N}_{13} \hat{N}_{34}^* + \hat{P}_2 \hat{P}_{4} \hat{N}_{13} \hat{K}_{34}^*  \right]^{\BZ_3 \times \BZ_N} \end{multline}}
\begin{align}
\begin{split}
    \CalZ_{\o} (x, {\bx}') = \sum_{\hat{\boldsymbol\lambda}} \, \prod_{\o=0}^{N-1} \kq_\o^{k_\o} \,  \m_{\ba, \hat{\bl}}^{c} \ \ET_\o(x) \star \EQ(\bx') [ {\ba}, \hat{\boldsymbol\lambda} ],
\end{split}
\end{align}
becomes the expectation value of the OPE of the generalized $\bf Q$-observable and the $qq$-character,
\begin{align} \label{def:qq}
\begin{split}
    \ET_\o(x) \star \EQ(\bx')
    = & (x-x'_{\omega})\EY_{[{\omega+1}]}(x+\ve_1+\d_{\o,N-1} \ve_2) \EQ(\bx')  +  {\kq}_\omega (x-x'_{[{\omega+1}]} +\d_{\o,N-1} \ve_2) \frac{P_{\omega}(x)}{\EY_{\omega}(x)} \EQ(\bx'),
\end{split}
\end{align}
where $P_\o (x) = P^+ _\o (x) P^- _\o(x)$ and $P^\pm _\o (x) = x- m^\pm _{\o+1}$, and we use the notation \eqref{eq:om} for $l=N$.  
Recall the large $x$ expansions \footnote{$p_{[\omega+1]} = a_{[{\o+1}]}-\ve_1\nu_\o$, $\nu_\o = k_{\o} - k_{\o+1}$}:
\begin{multline} \EY_{[{\omega+1}]}(x) = x-p_{[\omega+1]} + x^{-1} \left( \ve_1 D^{(1)}_\o  +\frac{p_{[\o +1]}^2}{2} - \frac{a_{[\o +1]}^2}{2}  \right) + o (x^{-2}) \, , \\ 
 \frac{P_{\omega}(x)}{\EY_{\omega}(x)} =  x-m_{\o}^+-m_{{\o}}^-+p_{\o} +  \\ x^{-1} \left( - \ve_1 D^{(1)} _{\o-1} + \frac{\ve_1^2}{2} \n_{\o-1}^2 +(m_{\o}^+ - m_{{\o}}^- -a_\o - \ve_1 )\ve_1 \n_{\o-1} +P_{\o}^+ (a_\o) P_{{\o}}^-(a_\o) \right) + o(x^{-2}) \label{eq:largexy}
 \end{multline} 
 Since ${\CalZ}_{\omega}(x, {\bx}')$ has no singularities in $x$,  it is a polynomial of degree 2 in $x$, which can be computed by expanding the right hand side \eqref{def:qq} at large $x$ dropping all negative powers of $x$. 
Using \eqref{eq:largexy} we get:
\begin{align}
\label{eq:czoxx}
\begin{split}
    {\CalZ}_{\o}(x,\bx') = & 
    (x-x'_\o) \llangle (x-a_{[{\o+1}]}+\ve_1 + \ve_2\delta_{\o,N-1}+\ve_1\nu_\o) \EQ(\bx') \rrangle \\
    &+ \llangle \left( \ve_1 D^{(1)} _\o  +\frac{\ve_1 ^2}{2} \n_\o ^2 - \ve_1 a_{[{\o+1}]} \n_\o  \right) \EQ(\bx')  \rrangle \\
    & + \kq_\o (x-x'_{[{\o+1}]}+\ve_2\delta_{\o,N-1}) \llangle (x-m_{\o}^+-m_{{\o}}^-+a_\o-\ve_1\nu_{\o-1})  \EQ(\bx') \rrangle  \\
      &+  {\qe}_\o  \llangle \left( - \ve_1 D^{(1)} _{\o-1} + \frac{\ve_1^2}{2} \n_{\o-1}^2 +(m_{\o}^+ - m_{{\o}}^- -a_\o - \ve_1 )\ve_1 \n_{\o-1}  \right. \\
      & \qquad\qquad\qquad \left. +P_{\o+1}^+ (a_\o)  P_{{\o}}^-(a_\o+\ve_1) \right)\, \EQ(\bx') \  \rrangle 
\end{split}
\end{align}
where 
\begin{align}
\begin{split}    
    & D_{\omega}^{(1)} = \ve_2 k_{\omega} + \sum_{\Box\in {\sK}_{\omega}} \hat{c}_\Box - \sum_{\Box\in {\sK}_{\omega+1}} \hat{c}_\Box = \ve_2 k_\omega + \hat{c}_\omega - \hat{c}_{\omega+1}. \\
    & \hat{c}_\Box = \hat{a}_\alpha + (i-1)\ve_1 + (j-1)\hat\ve_2 ,\qquad \Box_{(i,j)} \in \hat{\l} ^{(\a)}.
\end{split}
\end{align}
At the special values of $x$ in \eqref{def:qq} we get
\begin{multline} \label{eq:fracTQder}
{\CalZ}_\o (x=x'_{[{\o+1}]} - \delta_{\o,N-1}\ve_2,\bx')  
    =  (x_{[{\o+1}]}'-\delta_{\o,N-1}\ve_2-x_\o') P_{[{\o+1}]}^-(x_{[{\o+1}]}+\ve_1) \langle {\EQ} (\bx'+\ve_1 \be_{[{\o+1}]})  \rangle,\\
{\CalZ}_{\o} (x=x'_\o, \bx') 
     = \kq_\o P_{{\o}}^+(x'_\o) (x_\o'-x_{[{\o+1}]}'+\delta_{\o,N-1}\ve_2) \langle {\EQ} (\bx'-\ve_1 {\be}_{\o}) \rangle\ ,
\end{multline}
where 
\beq
{\be}_\o = (\d_{\o,\o'})_{\o'=0}^{N-1} \ .
\eeq 
We now compute, again in two ways, 
\begin{multline} \label{eq:Totemp}
\frac{{\CalZ}_\o (x=x'_\o,\bx') - {\CalZ}_\o (x=x'_{[{\o+1}]} - \delta_{\o,N-1}\ve_2,\bx')}{x_\o'-x_{[{\o+1}]}'+\delta_{\o,N-1}\ve_2} \\
 =^{\rm by \, \eqref{eq:czoxx}} \\
    = \left( x_{[{\o+1}]} ' - a_{[{\o+1}]} + \ve_1 + \ve_1 u_{\o+1} \p_{ u_{\o+1}} + \kq_\o (x_\o ' - m_{\o}^+ - m_{{\o}}^- + a_\o - \ve_1 u_\o \p_{u_\o}) \right) \llangle {\EQ}({\bx}') \rrangle \\
    =^{\rm by \, \eqref{eq:fracTQder}} \\
    = \kq_\o P_{{\o}}^+(x'_\o)  \llangle {\EQ} (\bx'-\ve_1 {\be}_{\o}) \rrangle + P_{[{\o+1}]}^-(x_{[{\o+1}]}+\ve_1) \llangle {\EQ} (\bx'+\ve_1 \be_{[{\o+1}]})  \rrangle
\end{multline}
which we can call the \textit{fractional quantum TQ equations}. 

Again, we normalize ${\bf\Psi}$ by the tree level contribution \eqref{eq:classcal}
\begin{align} \label{eq:classmono}
    \prod_{\o=0}^{N-1} u_\o^{\frac{m_\o^+-a_\o}{\ve_1}},
\end{align}

For the generalized $\bf Q$-observables defined in \eqref{eq:genQspec} the fractional $TQ$ equations simplify to
\begin{align} \label{eq:fracTQnormal}
    {\hat T}_\o (x) \llangle \EQ_\o (x)\rrangle = \llangle \EQ_{\o+1}(x) \rrangle + \qe_\o P_\o (x) \llangle \EQ_{\o-1}(x) \rrangle,\qquad \o=0,1,\cdots, N-1,
\end{align}
with the understanding $\EQ_{\o+N} (x) = \EQ_\o (x+\ve_1)$ and 
\begin{align} \label{eq:txexp}
    {\hat T}_\o (x) =     (1+\kq_\o) x + \ve_1 u_{\o+1} (\p_{u_{\o+1}}-\p_{u_\o}) - m_{\o+1}^+ - \kq_\o m_\o^-. 
\end{align}
as computed from \eqref{eq:Totemp}. 
\subsubsection{$\tilde{\bf Q}$-observable in the presence of the surface defect}
Next we consider the gauge origami setup:
\begin{align}
\begin{split}
    &\hat{N}_{12} = \sum_{{\o'}=0}^{N-1} \left[ e^{a_{\o'}} \CalR_0  +  q_{13}  q_2 ^{\d_{\o',0}} e^{m_{[{\o'-1}]} ^{-} } \CalR_1  + q_3^{-1} e^{m_{\o'}^+}  \CalR_2  \right]  \otimes  {\tilde q}_2^{\o'} \fR_{\o'}, \\
    & \hat{N}_{13} = \sum_{\o'=0}^{N-1} e^{x'_{\o'} + \ve_1} {\tilde q}_2^{\o'} \CalR_0 \otimes \fR_{\o'}, \\
    & \hat{N}_{34} = e^x {\tilde q}_2^\o \CalR_0 \otimes \fR_\o.
\end{split}
\end{align}
The gauged origami partition function is computed to be
\begin{align}
\begin{split}
    {\tilde\CalZ}_{\o} (x, {\bx'}) =  \sum_{\hat{\boldsymbol\lambda}} \prod_{\o=0}^{N-1} \kq_\o^{k_\o} \m_{\ba, \hat{\bl}}^{\BZ_N} \, \tilde{\ET}_\o(x) \star \tilde\EQ(\bx')[\ba, \hat{\boldsymbol\lambda}] 
\end{split}
\end{align}
where OPE of the fractional $qq$-character and the $\tilde{\bf Q}$-observable is given by
\begin{align} \label{eq:dualfracqq}
\begin{split}
    & \tilde{\ET}_\o(x) \star \tilde\EQ(\bx') \\
    & = \prod_{\o'=0}^{N-1} \sum_{\bd} \kq_{\o'}^{\frac{x'_{\o'}}{\ve_1}+d_{\o'}}  \left[ \frac{(x'_\o-x)(x-x'_{[{\o+1}]} + \ve_2\d_{\o,N-1} - d_{[{\o+1}]} \ve_1 )}{x-x'_\o-d_\o\ve_1 } \EY_{[{\o+1}]} (x+\ve_1+\d_{\o,N-1}\ve_2) \right. \\
    & \qquad \qquad \qquad \quad + \left. \kq_\o \frac{(x'_{[{\o+1}]}-x-\ve_2\d_{\o,N-1})(x'_{[{\o-1}]}-x+(d_{[{\o-1}]}+1)\ve_1+\ve_2\d_{\o,0})}{x'_\o-x+(d_\o+1)\ve_1} \frac{P_\o(x)}{\EY_\o(x)} \right] \EZ_\bd(\bx') \tilde\EQ_\bd(\bx'),
\end{split}
\end{align}
where we used the notation of \eqref{eq:dualnot}. Again, the apparent poles at $x = x'_\o + d_\o \ve_1$ actually cancel. See the appendix \ref{sec:abspole} for some details.
The regularity theorem implies $\tilde\CalZ_{\o}(x, {\bx}')$ is a linear function of $x$. 
By expanding the right hand side of \eqref{eq:dualfracqq} in $x$ we relate
$\tilde\CalZ_\o$ to expectation values of $\tilde\EQ_\bd(\bx')$ 
\begin{multline}
{\tilde\CalZ}_{\o}(x, {\bx}') = \llangle\, - x^{2} + x \left( x'_\o +  x'_{[{\o+1}]} - \ve_2\d_{\o,N-1} + (d_{[{\o+1}]} - d_{\o} ) \ve_1 - x'_\o  \right) \tilde\EQ_\bd(\bx')\, \rrangle
\end{multline}

Now, at special values of $x$, $\tilde\CalZ_{\o}$ reduces to vevs of fractional ${\tilde{\bf Q}}$-observables. First we consider the case $x=x'_\o$: 
\beq \label{eq:fracdual1}
\frac{ \tilde{\CalZ}_{\o} (x=x'_\o,\bx')}{(x'_{[{\o+1}]}-x'_\o-\ve_2\d_{\o,N-1})}  
 = \kq_\o P_\o(x'_\o) \, \llangle \tilde\EQ(\bx'-\ve_1 {\be}_\o) \rrangle \ . 
\eeq
Here we used \eqref{eq:EZ-ER conditions}. Next, $x=x'_{[{\o+1}]}-\ve_2\d_{\o,N-1}$ gives
\beq \label{eq:fracdual2}
 \frac{ \tilde{\CalZ}_{\o} (x=x'_{[{\o+1}]}-\ve_2\d_{\o,N-1}, {\bx}')} {(x'_\o-x'_{[{\o+1}]}+\ve_2\d_{\o,N-1})} \\
     = \tilde\EQ(\bx'+\ve_1 {\be}_{[{\o+1}]}) \ , 
\eeq
where we again used the relations \eqref{eq:EZ-ER conditions}.

Let us define a differential operator in fractional couplings, also a degree 1 polynomial in $\bx'$ by 
\beq
    \tilde{T}_\o(\bx')  = x'_{[{\o+1}]} -a_{[{\o+1}]} + \ve_1 + \ve_1 u_{\o+1}\p_{u_{\o+1}} + \ve_2 \d_{\o,N-1} 
     + \kq_\o (x'_\o - m_\o^+-m_\o^- + a_\o + \ve_1 u_\o\p_{u_\o} - \ve_2\d_{\o,0}).
\eeq
Then, by taking the difference between the two equations \eqref{eq:fracdual1} and \eqref{eq:fracdual2} (and changing the notation from $\bx'$ to $\bx$) we arrive at
\begin{align}
    \tilde{T}_{\o}(\bx) \llangle \tilde\EQ(\bx) \rrangle = \llangle \tilde\EQ(\bx+\ve_1 {\be}_{[{\o+1}]} \rrangle + \kq_\o P_{\o}(x_\o) \llangle \tilde\EQ(\bx-\ve_1 e_{\o}) \rrangle,\quad \o=0,1,\cdots, N-1.
\end{align}
These are the fractional quantum TQ equations for the generalized dual $\bf Q$-observables.

By multiplying the classical part
\begin{align}
    u_{0}^{\frac{\ve_2}{\ve_1}} \prod_{\o=0}^{N-1} u_\o^{\frac{m_\o^+-a_\o}{\ve_1}} 
\end{align}
and assign $\bx = \bx_\o := (\underbrace{x,\cdots, x}_{\o+1 \text{\;times}}, \underbrace{x-\ve_1, \cdots, x-\ve_1}_{N-\o-1 \;\text{times}})$, we obtain for $\tilde{\EQ}_\o (x) := \tilde{\EQ}(\bx_\o)$,
\begin{align} \label{eq:fracTQnormal2}
    T_\o (x) \llangle \tilde{\EQ}_\o (x)\rrangle = \llangle \tilde{\EQ}_{\o+1}(x) \rrangle + \qe_\o P_\o (x) \llangle \tilde{\EQ}_{\o-1}(x) \rrangle,\qquad \o=0,1,\cdots, N-1,
\end{align}
where $T_\o (x)$ is precisely the degree $1$ polynomial,
\begin{align}
    T_\o(x) = (1+\kq_\o)x + u_{\o+1}(\p_{u_{\o+1}}-\p_{u_\o}) - m_{\o+1}^+ - \kq_\o m_\o^-, 
\end{align}
that appeared in the fractional quantum TQ equation \eqref{eq:fracTQnormal} for the generalized $\bf Q$-observables. Thus, we have shown that the generalized $\bf Q$-observables and generalized dual $\bf Q$-observables of specific type satisfy the same fractional quantum TQ equation \eqref{eq:fracTQnormal} and \eqref{eq:fracTQnormal2}.

\section{di-Langlands correspondence} \label{sec:hbarlang}
Finally, we present the $\EN=2$ gauge theoretical account of the difference Langlands correspondence. We construct the R-matrices of the Yangian $Y(\fgl(2))$ represented on specific modules from the fractional TQ equations. The consternation of them leads to the universal $\hbar$-oper equation satisfied by the correlation function of the (dual) $\bf Q$-observable and the regular monodromy surface defect. By taking the limit $\ve_2 \to 0$, where the correlation function factorizes by cluster decomposition, we prove that the $Q$-eigenstates constructed as the normalized vacuum expectation values of the regular monodromy surface defect, enumerated by the Coulomb moduli $\ba$, are also common eigenstates of the quantum Hamiltonians of the XXX spin chain.

\subsection{Yangian and universal $\hbar$-oper} \label{sec:yangian}
The Yangian $Y({\fgl}(n))$ of $\fgl(n)$ is an associative algebra, which is reviewed in the appendix \ref{sec:yangian1}. We introduce the universal $\hbar$-oper which generate a maximal commutative subalgebra of $Y(\fgl(n))$. 

It is straightforward to show that, due to the defining relation \eqref{eq:defy}, $T(x) e^{-\hbar \p_x} \in \text{End}(\BC^n) \otimes Y(\fgl(n)) [[x^{-1} , \p_x]]$ is a Manin matrix \cite{Tal:2004,Chervov:2007bb}, i.e. a matrix over a non-commutative ring for which, nevertheless, the determinant can be well-defined by the column expansion \cite{ManinQuantumGroups}. The notion of Manin matrix and its relevant properties are reviewed in the appendix \ref{sec:manin}. The determinant of $T(x) e^{-\hbar \p_x}$ is computed to be
\begin{align}
    \det \left( T(x) e^{-\hbar \p_x} \right) = \text{qdet}\, T(x-(n-1)\hbar)  e^{-n\hbar \p_x}.
\end{align}

A Manin matrix defines another Manin matrix when it is added or multiplied by any constant matrix. Let us be given with a constant matrix $K \in \text{End}(\BC^n)$, and consider a Manin matrix 
\begin{align}
\mathds{1}_n - K T(x) e^{-\hbar \p_x} \in \text{End}(\BC^n) \otimes Y(\fgl(n)) [[x^{-1} , \p_x]].
\end{align}
Note that the constant matrix $K$ introduces a twist in the periodic boundary condition for the XXX spin chain \eqref{eq:XXXmono}. The determinant can be expanded in $e^{-\hbar \p_x}$ as
\begin{align}
\begin{split}
    \det \left(\mathds{1}_n - K T(x) e^{-\hbar \p_x} \right) &= 1+ \sum_{m=1} ^{n} (-1)^m \mathtt{t}_m (x) e^{-m\hbar \p_x} \\
    &= 1 - \mathtt{t}_1 (x) e^{-\hbar \p_x} + \cdots + (-1)^n \mathtt{t}_n (x) e^{-n\hbar \p_x}.
\end{split}
\end{align}
The coefficients $\mathtt{t}_m (x) \in Y(\fgl(n)) [[x^{-1}]]$ are called \textit{universal transfer matrices}.\footnote{The definition of the transfer matrices $\mathtt{t}_m (x)$ obviously depends on the choice of the constant matrix $K$. We will not explicitly indicate this dependence for brevity.} In explicit terms, the transfer matrices are given by
\begin{align} \label{eq:univtrans}
    &\mathtt{t}_m (x) = \sum_{j_1<j_2<\cdots<j_m} \text{qdet}\,((K T(x-(m-1)\hbar))_{j_1 j_2 \cdots j_m}), \quad\quad 1\leq m \leq n,
\end{align}
where $[(A_{j_1 j_2 \cdots j_m})_{ab}]_{a,b=1} ^m = [A_{j_a  j_b}]_{a,b=1}^m$ is a minor of size $m$. In particular, $\mathtt{t}_1 (x) = \text{Tr} (K T(x))$ and $\mathtt{t}_n (x) = \det K\, \text{qdet}\, T(x+n-1)$. It can be shown that the universal transfer matrices mutually commute
\begin{align}
    [\mathtt{t}_m (x) , \mathtt{t}_l (x')] = 0 ,\quad\quad m,l=1,2,\cdots, n.
\end{align}
In fact, if the constant matrix $K$ has a simple spectrum, the coefficients of the universal transfer matrices span a maximal commutative subalgebra of the Yangian, called the Bethe subalgebra $\mathcal{B}(Y(\fgl(n))) \subset Y(\fgl(n))$ \cite{cmp/1104286662,MTV:2006}. Note that $\mathtt{t}_n (x) = \det K \, \text{qdet}\, T(x-(n-1)\hbar)$ generates the center of the Yangian $Z(Y(\fgl(n))) \subset \mathcal{B} (Y(\fgl(n)))$. All the rest of the universal transfer matrices generate non-central elements in the Bethe subalgebra.

Let us define the \textit{quantum} powers\footnote{The procedure developed here is $\hbar$-deformation of the one for $U(\fsl(N) \otimes t^{-1} \BC[[t^{-1}]])$ which was studied in \cite{Chervov:2009ck}. The latter was also used in \cite{Jeong:2023qdr} to realize the universal oper in the $\EN=2$ gauge theory. See also \cite{Chervov:2006xk}.} of the generating matrix $KT(x)$ by
\begin{align}
\begin{split}
    &(KT (x))^{[0]} := \mathds{1}_n \\
    &(KT (x))^{[i]} := KT(x+(i-1)\hbar) KT(x +(i-2)\hbar) \cdots KT(x+\hbar) KT(x), \quad\quad i>0.
\end{split}
\end{align}
Note the recursive relation $(KT (x))^{[i+1]} = (KT (x+\hbar))^{[i]} KT(x)$.

For any $v \in \BC^n$, let us consider $G(x) \in \text{End}(\BC^n) \otimes Y(\fgl(n))[[x^{-1}]]$ defined by $G_{ab} (x) = \sum_{c=1}^n v_c ((KT (x+\hbar))^{[n-a]} )_{cb}$. Then, it can be shown that 
\begin{align} \label{eq:chthm}
\begin{split}
    &G(x -(n-1)\hbar )\left( \mathds{1}_n - KT(x-(n-1)\hbar) e^{-\hbar\p_x} \right) \\
    & = \left[ \mathds{1}_n  - \begin{pmatrix}
         \mathtt{t}_1 (x ) & -\mathtt{t}_2 (x) & \cdots & \cdots & (-1)^{n-2} \mathtt{t}_{n-1} (x)  & (-1)^{n-1} \mathtt{t}_n (x) \\ 1 & 0 & \cdots & \cdots & 0 & 0 \\ 0 & 1 &  \cdots & \cdots & 0 & 0  \\ 
        \cdots & \cdots & \cdots & \cdots & \cdots & \cdots \\ 0 & \cdots &\cdots & 0 &  1 & 0
    \end{pmatrix} e^{-\hbar \p_x} \right] G(x-(n-1)\hbar).
\end{split}
\end{align}
The first line of the identity follows from the Cayley-Hamilton theorem\footnote{$0 = \sum_{m=0} ^n (-1)^m \mathtt{t}_m (x) T^{[n-m]} (x-(n-1)\hbar)$, where we set $\mathtt{t}_0 (x) =1.$} for Manin matrices, proven in \cite{Chervov:2007bb}. All the rest are trivial identities.

Now, let us be given with a solution $\mathbb{Q}(x) = (\cdots,\cdots , \mathbf{Q} (x)) \in \BC^n \otimes \text{End}(V)$, where $V$ is a module over $Y(\fgl(n))$, to the Manin matrix,
\begin{align} \label{eq:diffconn}
    0 = \left.\left( \mathds{1}_n - KT(x) e^{-\hbar\p_x} \right)\right\vert_V \mathbb{Q}(x).
\end{align}
Then by using \eqref{eq:chthm} with $v=(0,\cdots,0,1)$, we obtain the following $\hbar$-difference equation satisfied by $ \mathbf{Q}(x)$, the last component of $\mathbb{Q}(x)$:
\begin{align} \label{eq:TQn}
    0 = \left.\left(1 - \mathtt{t}_1 (x) e^{-\hbar\p_x} + \mathtt{t}_2 (x) e^{-2\hbar\p_x}- \cdots + (-1)^n \mathtt{t}_n (x) e^{-n\hbar\p_x}  \right)\right\vert_{V}  \mathbf{Q} (x ).
\end{align}
We will call this $n$-th order $\hbar$-difference operator the \textit{universal $\hbar$-oper}. It generates the \textit{transfer matrices} $\mathtt{t}_i (x) \vert_V \in \text{End}(V)[[x^{-1}]]$, $i=1,2,\cdots, n$. By construction, the transfer matrices generate a maximal set of mutually commuting operators in $\text{End}(V)$, namely, the quantum Hamiltonians and the central elements. The above $\hbar$-difference equation for the universal $\hbar$-oper represented on the module $V$ is nothing but the operator Baxter's TQ equation where $ \mathbf{Q}(x)$ is a $Q$-operator.

Let us specialize to our main case $Y(\fgl(2))$. Then the above universal $\hbar$-oper equation reads
\begin{align} \label{eq:hopereqyang}
    0 = \left.\left( 1- \mathtt{t}(x) e^{-\hbar\p_x} + \qe P(x) e^{-2\hbar\p_x} \right) \right\vert_{V}  \mathbf{Q}(x+\hbar),
\end{align}
where we set $\det K = \qe$ and $\text{qdet}\, T(x) = P(x)$, so that $\mathtt{t}_2 (x) = \det K \, \text{qdet} T(x-\hbar) = \qe P(x-\hbar)$, and $\mathtt{t}(x) := \mathtt{t}_1 (x-\hbar)$. 

We will construct the universal $\hbar$-oper represented on the space $\CalH$ (see section \ref{subsubsec:basis}) in our $\EN=2$ gauge theoretical setup. It is crucial that the $\EN=2$ theory provides not only the universal ${\hbar}$-oper itself, but also its solutions, namely, the $Q$-operators, by ${\bf Q}/\tilde{\bf Q}$-observables. 

\subsection{R-matrices from fractional TQ equations} \label{subsubsec:rmatgauge}
We construct the R-matrices \eqref{eq:rmat1} of the Yangian of $\fgl(2)$ from the fractional quantum TQ equations \eqref{eq:fracTQnormal} and \eqref{eq:fracTQnormal2}. Let us rewrite the fractional TQ equation \eqref{eq:fracTQnormal} and \eqref{eq:fracTQnormal2} in a matrix form as
\begin{align} \label{eq:matform1}
     \Xi_{\o+1} (x) =  \begin{pmatrix}
        T_\o (x) & -\qe_\o P^- _\o (x) \\ P^+ _{\o+1}(x) & 0 
    \end{pmatrix} \Xi_\o (x) =: L_\o (x) \Xi_\o (x),\quad \text{the same for }\; \tilde{\Xi}_\o (x),
\end{align}
for all $\o=0,1,\cdots, N-1$, where we repackaged the generalized (dual) $\bf Q$-observables into columns
\begin{align}
    \Xi_\o (x) :=  \begin{pmatrix}
        \llangle {\EQ_{\o} (x)} \rrangle \\ P^+ _\o (x) \llangle \EQ_{\o-1} (x) \rrangle
    \end{pmatrix}, \qquad  \tilde{\Xi}_\o (x) :=  \begin{pmatrix}
        \llangle {\tilde{\EQ}_{\o} (x)} \rrangle \\ P^+ _\o (x) \llangle \tilde{\EQ}_{\o-1} (x) \rrangle
    \end{pmatrix}.
\end{align}
Satisfying exactly the same equations, the procedure that follows will be identical for $\Xi_\o (x)$ and $\tilde{\Xi}_\o (x)$. We will only explicitly write $\Xi_\o (x)$. 

We make a transformation $\Theta_\o (x) = g_\o (x) \Xi_\o (x)$ with
\begin{align} \label{eq:ggauge}
    g_\o (x) = -\frac{1}{\ve_1 u_{[{\o}]}} \begin{pmatrix}
        1 & - 1 - \frac{\ve_1 u_\o \p_{u_\o}}{P^+ _\o (x)}  \\ 0 & - \frac{\ve_1 u_{[{\o}]}}{P^+ _\o (x)}
    \end{pmatrix}, \qquad g_\o (x) ^{-1} = \begin{pmatrix}
        -\ve_1 u_{[{\o}]} & P^+ _\o (x) +\ve_1  u_\o \p_{u_\o}   \\ 0 & P^+ _\o (x)
    \end{pmatrix},
\end{align}
which gives
\begin{align}
    \Theta_\o (x) = \begin{pmatrix}
        \frac{1}{\ve_1 u_{[{\o}]}} (-\llangle \EQ_\o (x) \rrangle + P^+_\o (x) \llangle \EQ_{\o-1} (x) \rrangle) +\p_{u_\o} \llangle \EQ_{\o-1}(x) \rrangle \\ \llangle \EQ_{\o-1}(x) \rrangle
    \end{pmatrix}.
\end{align}
Then, the matrix form \eqref{eq:matform1} of the fractional TQ equation, for $\o=0,1,\cdots, N-2$, becomes
\begin{align} \label{eq:Rfromgauge1}
\begin{split}
    \Theta_{\o+1} (x) &= g_{\o+1} (x) \begin{pmatrix}
        T_\o (x) & -\qe_\o P^- _\o (x) \\ P^+ _{\o+1}(x) & 0 
    \end{pmatrix} g_\o  (x) ^{-1}  \Theta_\o (x) \\
    & =  \begin{pmatrix}
         x-\ve_1 u_\o\p_{u_\o}-m^-_\o-\ve_1 & -(m^+_\o-m_\o^--\ve_1) \p_{u_{\o}} + \ve_1 u_\o \p^2_{u_\o} \\
        -\ve_1 u_\o & x + \ve_1 u_\o \p_{u_\o} - m_{\o}^+
    \end{pmatrix} \Theta_\o (x) \\
    &= \left[ x  - \th_\o - \ve_1 \begin{pmatrix}
        \bs_\o^0  & \bs_\o^- \\ \bs_\o^+ & -\bs_\o^0 
    \end{pmatrix}  \right] \Theta_\o (x) \\
    & =: R_\o (x-\th_\o) \Theta_\o (x).
\end{split}
\end{align}
The last component $\o=N-1$ of the fractional TQ equation requires more care, as $[N] =0$, giving
\begin{align} \label{eq:Rfromgauge2}
\begin{split}
    \Theta_N (x ) &= \begin{pmatrix}
        \qe & 0 \\ 0 & 1
    \end{pmatrix} \left[ x -\th_{N-1} - \ve_1 \begin{pmatrix}
        \bs_{N-1} ^0  & \bs_{N-1} ^- \\ \bs_{N-1} ^+ & -\bs_{N-1} ^0 
    \end{pmatrix}  \right] \Theta_{N-1} (x) \\
    & =: K R_{N-1} (x-\th_{N-1})\Theta_{N-1} (x).
\end{split}
\end{align}
where we defined the twist matrix $K = \text{diag} (\qe,1)$ parameterized by the gauge coupling $\qe$.

Thus, we discover the fractional TQ equations give rise to the R-matrix of the Yangian $Y(\fgl(2))$, 
\begin{align} \label{eq:rmatgauge}
    R_\o(x) := x - \ve_1 \begin{pmatrix}
        \bs_\o^0  & \bs_\o^- \\ \bs_\o^+ & -\bs_\o^0 
    \end{pmatrix} \in \text{End}(\BC^2) \otimes \text{End}(\CalH_\o),
\end{align}
that we introduced in \eqref{eq:rmat1} with the identification $\ve_1 = \hbar$ between the Yangian deformation parameter and the $\O$-background parameter, up to a trivial redefinition of multiplication by $x$. Indeed, $\mathbf{s}^{\pm,0} _\o$ denote the generators of $\fsl(2) \subset \fgl(2)$ represented as differential operators in the monodromy defect parameter $u_\o$,
\begin{align} \label{eq:sl2gen}
\begin{split}
        &[\bs^0 _\o ,\bs^\pm _\o] = \pm \bs^\pm _\o,\qquad [\bs^+ _\o, \bs^- _\o] = 2 \bs^0 _\o,\\
    &\bs_\o^0 = u_\o \p_{u_\o} - s_\o, \quad \bs_\o^+ = u_\o, \quad \bs_\o^- = 2s_\o \p_{u_\o} - u_\o \p_{u_\o}^2, 
\end{split} \qquad \o=0,1,\cdots, N-1,
\end{align}
acting on the space of Laurent polynomials $\CalH_\o = u_\o ^{\frac{m^+ _\o -a_\o}{\ve_1}} \BC(u_\o)$. The spin $s_\o$ is given in terms of the mass parameters by
\begin{align} \label{eq:spin}
   s_\o = \frac{m^+_\o - m_\o^- -\ve_1}{2\ve_1}, \qquad \o=0,1,\cdots, N-1.
\end{align}
The trace part acts trivially. Finally, the evaluation parameter $\th_\o$ is encoded in the rest of the degrees of freedom of the mass parameters by
\begin{align} \label{eq:evalpa}
    \th_\o = \frac{m^+ _\o + m^- _\o+\ve_1 }{2 }, \qquad \o=0,1,\cdots, N-1.
\end{align}
Note that the $\fgl(2)$-module $\CalH_\o$ is irreducible for generic values of gauge theory parameters. Moreover, it is a bi-infinite module, possessing neither a highest-weight state nor a lowest-weight state.

Note that the quantum determinant \eqref{eq:qdetdef} of the R-matrix is computed to be a simple combination of the mass parameters,
\begin{align} \label{eq:qdetr}
    \text{qdet}(R_\o(x-\th_\o)) = (x-m^+ _\o)(x-m^- _\o) = P_\o (x),\qquad \o=0,1,\cdots, N-1.
\end{align}

To present the $\fgl(2)$-module $\CalH_\o$ more explicitly as a $\ve_1$-difference module, we introduce variables $(w_\o)_{\o=0}^{N-1}$ constrained by $\sum_{\o=0} ^{N-1} w_\o =0$, which are Fourier dual of the monodromy defect parameters $(u_\o)_{\o=0}^{N-1}$, namely, $u_\o = e^{-\ve_1 \p_{w_\o}}$. Then we can express $\CalH_\o$ as the vector space spanned by $\ve_1$-difference operators, 
\begin{align}
    \CalH_\o =\bigoplus_{l \in \BZ} \BC \,e^{(a_\o- m ^+_\o  - l \ve_1) \p_{w_\o}} , \qquad \o=0,1,\cdots, N-1,
\end{align}
on which the $\fsl(2) \subset \fgl(2)$ generators act now by $\ve_1$-difference operators,
\begin{align}
  &\bs_\o^0 = \frac{w_\o}{\ve_1} - s_\o, \quad \bs_\o^+ = e^{-\ve_1 \p_{w_{\o}}}, \quad \bs_\o^- = \left( \frac{w_\o}{\ve_1}-1 \right) \left( 2s_\o  - \frac{w_\o }{\ve_1 }\right) e^{\ve_1 \p_{w_\o}} .
\end{align}
In principle, we would be able to construct this $\fgl(2)$-module as a (twisted) $\hbar$-difference module on the moduli space $\text{Bun}_{GL(2)} (\BP^1;D)$ of parabolic $GL(2)$-bundles over $\BP^1$ with a framing at $\infty \in \BP^1$ and parabolic structures at $N$ marked points $D \subset \BP^1 \setminus\{\infty\}= \BC$, by a $\hbar$-deformed version of the Beilinson-Bernstein localization \cite{BB}. In particular, $(w_\o)_{\o=0}^{N-1}$ would be identified with holomorphic coordinates on an open patch of $\text{Bun}_{GL(2)} (\BP^1;D)$. In this sense, the vacuum expectation value $\langle \Psi \rangle_\ba$ of the regular monodromy surface defect should be viewed as a section of this (twisted) $\hbar$-difference module on $\text{Bun}_{GL(2)} (\BP^1;D)$, enumerated by the Coulomb moduli $\ba$. More precisely, recall that we split each Coulomb parameter $a_\o$ into $\frac{a_\o}{\ve_1}\, \text{mod}\, \BZ$ and its $\ve_1$-integral shifts, and treat the former to be fixed. The aforementioned sections of the (twisted) $\hbar$-difference module over $\text{Bun}_{GL(2)} (\BP^1;D)$ are enumerated by the latter, the $\ve_1$-integral shifts in the Coulomb moduli $\ba$. In other words, the vacuum expectation value of the regular monodromy surface defect provides a distinguished basis of the (twisted) $\hbar$-difference module on $\text{Bun}_{GL(2)} (\BP^1;D)$ enumerated by the $\ve_1$-integral shifts of the Coulomb moduli $\ba$.

This is not the approach taken in the present work, however, and we only give a concrete $Y(\fgl(2))$-module arising from our $\EN=2$ gauge theory setup.

\subsection{Monodromy matrix and universal $\hbar$-oper}
Note that the column vectors $\Theta_\o (x)$ of vacuum expectation values of generalized $\bf Q$-observables are valued in $\text{End}(\BC^2) \otimes \tilde{\CalH}$, where $\tilde{\CalH}$ is the space of Laurent polynomials in monodromy defect parameters $(u_\o)_{\o=0} ^{N-1}$, 
\begin{align} \label{eq:states}
    \tilde{\CalH}=\bigotimes_{\o=0} ^{N-1} \CalH_\o,\qquad \CalH_\o = u_\o^{\frac{m^+_\o -a_\o}{\ve_1}} \BC(u_\o).
\end{align}
 We view $\tilde{\CalH}$ as a evaluation module over the Yangian $Y(\fgl(2))$ (defined in \eqref{eq:sl2gen} and \eqref{eq:evalpa}, cf. section \ref{subsec:rmat}). 
In terms of the associated XXX spin chain with $N$ sites, $\tilde{\CalH}$ is precisely the space of states. By concatenating all the $N$ R-matrices \eqref{eq:Rfromgauge1} and \eqref{eq:Rfromgauge2}, we arrive at a difference equation,
\begin{align}\label{eq:diffeqmat}
\begin{split}
    \Theta_N (x) &= K R_{N-1}(x-\th_{N-1}) R_{N-2}(x-\th_{N-2}) \cdots R_{0} (x-\th_0)  \Theta_{0}(x) \\
    & =K T(x) \vert_{\tilde{\CalH}} \, e^{-\ve_1 \p_x} \Theta_N (x),
\end{split}
\end{align}
where we used the periodicity $\Theta_{\o+N} (x) = \Theta_\o (x+\ve_1) = e^{\ve_1 \p_x} \Theta_\o (x)$ in the second equality. The product of R-matrices appearing on the right hand side is exactly the monodromy matrix,
\begin{align} \label{eq:monogauge}
    T(x)\vert_{\tilde{\CalH}} = R_{N-1} (x-\th_{N-1}) R_{N-2} (x-\th_{N-2})\cdots R_0 (x-\th_0) \in \text{End}(\BC^2)\otimes \text{End}(\tilde{\CalH} ),
\end{align}
for the XXX spin chain with $N$ sites. In particular, it agrees with the monodromy matrix introduced in \eqref{eq:XXXmono} with the relevant Yangian module chosen to be $\tilde{\CalH}$ \eqref{eq:states}. We remind that the monodromy matrix is nothing but the generating matrix $T(x) \in \text{End}(\BC^2)\otimes Y(\fgl(2))[[x^{-1}]]$ for the Yangian represented on $\tilde{\CalH}$, and equivalently the universal R-matrix represented on the tensor product, $ T(x)\vert_{\tilde{\CalH}} = (\r_{\BC^2} \otimes \r_{\CalH} ) (\CalR(x))$. 

Note also that the quantum determinant of the monodromy matrix is a simple degree $2N$ polynomial whose coefficients are the mass parameters,
\begin{align}
    \text{qdet}(T(x))\vert_{\tilde{\CalH}} = \prod_{\o=0} ^{N-1} \text{qdet}(R_\o (x-\th_\o)) =  \prod_{\o=0} ^{N-1}  \prod_\pm (x-m^\pm _\o) = P(x),
\end{align}
by the factorization property \eqref{eq:qdetfac} and the quantum determinants \eqref{eq:qdetr} of individual R-matrices.

The difference equation \eqref{eq:diffeqmat} is reorganized into the form of \eqref{eq:diffconn}, a Manin matrix represented on the Yangian module $\tilde{\CalH}$ annihilating the column vector $\Theta_N (x)$,
\begin{align}
    0 = \left.\left[ \mathds{1}_2- K T(x) e^{-\ve_1 \p_x}\right]\right\vert_{\tilde{\CalH}} \Theta_N (x).
\end{align}
Note that the bottom component of the column vector $\Theta_N (x)$ is precisely $\llangle \EQ_{N-1} (x) \rrangle = \langle Q(x) \Psi(\mathbf{u}) \rangle_\ba$, namely, the correlation function of the $\bf Q$-observable and the regular monodromy surface defect without any $0$-observable on their interface. As we have seen in \eqref{eq:Qoper}, the $\bf Q$-observable can be viewed as the $Q$-operator $\mathbf{Q}(x) \in \text{End} \left(\CalH \right)$ since it defines an action on the monodromy surface defect by its insertion in the correlation function. It follows that the equation \eqref{eq:TQn} satisfied by the bottom component of $\Theta_N (x)$ can be written as an operator equation valued in $\text{End}\left(\CalH \right)$,
\begin{align} \label{eq:qunivhoper}
    0 = \left[1 - \hat{t} (x) e^{-\ve_1 \p_x} +  \qe P(x) e^{-2\ve_1 \p_x} \right] \mathbf{Q}(x+\ve_1),
\end{align}
where we used $\det K = \qe$ and $\text{qdet} (T(x))\vert_{\CalH} = P(x)$. This is precisely the universal $\hbar$-oper equation represented on $\CalH$ (equivalently, the operator Baxter TQ equation). Here, we defined the \textit{transfer matrix} $\hat{t}(x)\in \text{End}\left(\CalH \right)$ by the trace,
\begin{align} \label{eq:transfergauge}
    \hat{t}(x) := \text{Tr}_{\BC^2} K T(x+\ve_1)\vert_{\CalH },
\end{align}
which is none other than the universal transfer matrix $\mathtt{t}_1 (x)$ \eqref{eq:univtrans} represented on the module $\tilde{\CalH}$ (recall that we restrict to the degree-zero subspace $\CalH$ by fixing the total momentum).\footnote{Comparing with the universal transfer matrix $\mathtt{t}_1 (x) = \text{Tr}_{\BC^2} (KT(x)) $ that we defined earlier in section \ref{sec:yangian}, there is an additive shift of the argument by $\ve_1$. We make such a redefinition merely for a conventional purpose, which only amounts to rearranging its coefficients, namely, the quantum Hamiltonians. \label{fn:shift}} Since the universal transfer matrix generates the non-central elements of the Bethe subalgebra $\CalB(Y(\fgl(2))) \subset Y(\fgl(2))$, the coefficients of the transfer matrix $\hat{t}(x)$ span a maximal set of mutually commuting operators acting on the space of states $\CalH$. They are precisely the quantum Hamiltonians of the XXX spin chain system with $N$ spin sites.

Among $N+1$ coefficients of degree $N$ polynomial $\hat{t}(x)$, the leading coefficient of $x^N$ is simply $1+\qe$. The next-to-leading coefficient of $x^{N-1}$ involves the total momentum, $\sum_{\o=0} ^{N-1} \ve_1 u_\o \p_{u_\o}$, which becomes a number when restricted to $\CalH$. Therefore, we obtain $N-1$ mutually commuting quantum Hamiltonians in total, which we collectively denote by $\hat{H}_k \in \text{End}\left(\CalH  \right)$, $k=2,3,\cdots, N$.

We emphasize that the XXX spin chain that we constructed above is defined upon bi-infinite modules $\CalH_\o = u_\o ^{\frac{m^+_\o -a_\o}{\ve_1}} \BC(u_\o)$ \eqref{eq:sl2gen}. Recall that we regard the values of $\frac{a_\a}{\ve_1}$ to be fixed mod $\BZ$. If we now specialize this parameter $\frac{a_\o}{\ve_1}  \mod  \BZ $ to the locus $\frac{m^+ _\o - a_\o}{\ve_1} \in \BZ$ (namely, $t_\o = 1$), the bi-infinite module under consideration becomes reducible, containing a lowest-weight submodule. Note that the loci $\frac{m^+ _\o - a_\o}{\ve_1} \in \BZ$ are precisely where a higgsing from four-dimensional $\EN=2$ theory to a two-dimensional $\EN=(2,2)$ theory is triggered \cite{Dorey:2011pa}. After going through the higgsing, we recover the standard Bethe/gauge correspondence \cite{Nekrasov:2009uh,Nekrasov:2009ui} where the algebraic Bethe ansatz equation is identified with the vacuum equation of the emergent two-dimensional $\EN=(2,2)$ theory. See \cite{Jeong:2021bbh} also for more elaboration on the higgsing.

\subsection{Spectral equations for XXX spin chain} \label{subsubsec:hlanglands}
Finally, we show that the eigenstates for the $Q$-operators are also common eigenstates of the quantum Hamiltonians of the $\fgl(2)$ XXX spin chain with $N$ sites. The first few spectral equations were derived in \cite{Lee:2020hfu} without using the $Q$-operators constructed by the ${\bf Q}/{\tilde{\mathbf{Q}}}$-observables. Here, we verify the all $N-1$ spectral equations simultaneously, for arbitrary $N$.

Let us remind that the normalized vacuum expectation value $\psi(\ba;\mathbf{u})$ of the regular monodromy surface defect in the limit $\ve_2 \to 0$ provides distinguished basis of the space $\CalH$ of states which diagonalizes the action of the $Q$-operators. Such an eigenvalue property can be understood as the factorization of the correlation function of the two surface defects in the limit $\ve_2 \to 0$,
\begin{align}
    \lim_{\ve_2 \to 0} \Big\langle Q(x) \Psi (\mathbf{u}) \Big \rangle = e^{\frac{\widetilde{\EuScript{W}} (\ba;\qe)}{\ve_2}} Q(\ba;x) \psi(\ba;\mathbf{u};\qe).
\end{align}

Now, we apply the limit $\ve_2\to 0$ to the universal $\hbar$-oper equation \eqref{eq:qunivhoper} acting on the distinguished basis element $\psi(\ba;\mathbf{u}) \in \CalH$. Then the $Q$-operator is converted into its eigenvalue $Q(\ba;x)$, so that we have
\begin{align} \label{eq:univoperres}
    0 = \left[1 - \hat{t} (x) e^{-\ve_1 \p_x} +  \qe P(x) e^{-2\ve_1 \p_x} \right] Q(\ba;x+\ve_1) \psi(\ba;\mathbf{u};\qe).
\end{align}
Here, it should be stressed that the difference operator $e^{-\ve_1 \p_x}$ acts only on the normalized vacuum expectation value $Q(\ba;x+\ve_1)$ of the $\bf Q$-observable, while the operator $\hat{t}(x) \in \text{End}\left(\CalH\right)$ acts only on the normalized vacuum expectation value $\psi(\ba;\mathbf{u};\qe)$ of the monodromy surface defect.

Meanwhile, it was independently shown in 
\eqref{eq:qhopereq} that the $\ve_2 \to 0$ limit of the normalized vacuum expectation value $Q(\ba;x)$ of the $\bf Q$-observable by itself satisfies the $\hbar$-oper equation, which reads
\begin{align} \label{eq:operres}
    0 = \left[1 - {t} (\ba;x) e^{-\ve_1 \p_x} +  \qe P(x) e^{-2\ve_1 \p_x} \right] Q(\ba;x+\ve_1).
\end{align}

The two equations \eqref{eq:univoperres} and \eqref{eq:operres} look almost identical, but actually there is a crucial difference: the coefficients of $\hat{t} (x)$ are operators in $\text{End}(\CalH)$, while the coefficients of $t(\ba;x)$ are numbers given by the normalized vacuum expectation values of the local chiral observables. By multiplying $\psi(\ba;\mathbf{u};\qe)$ to the second equation and subtracting the two, we get
\begin{align}
    0 = \begin{pmatrix}
        \left(\hat{t}(x) - t(\ba;x) \right) \psi(\ba;\mathbf{u}) & 0 \\ 0 & 0 
    \end{pmatrix} \begin{pmatrix}
       Q (\ba;x )  &  \tilde{Q} (\ba;x )\\ Q (\ba;x-\ve_1) & \tilde{Q} (\ba;x-\ve_1)
    \end{pmatrix},
\end{align}
where we organized $\hbar$-jets of two solutions, the $\bf Q$-observable and the dual $\bf Q$-observable, into a $2 \times 2$ matrix. The determinant of this matrix is precisely the $\hbar$-Wronskian for the $\hbar$-oper difference equation, which was computed in \eqref{eq:hwrons}. We remind that it is an entire function in $x \in \BC$ with simple zeros only at discrete loci, so that we can invert the $2\times 2$ matrix at generic $x$ to get
\begin{align}
     0= \left(\hat{t}(x) - t(\ba;x) \right) \psi(\ba;\mathbf{u}).
\end{align}
Since the equation holds for generic $x \in \BC$, each coefficient of the degree $N$ polynomial vanishes individually. The first two equations are trivially satisfied, and the rest of $N-1$ coefficients finally give
\begin{align}
    0 = \left( \hat{H}_k - E_k (\ba) \right) \psi(\ba;\mathbf{u};\qe), \qquad k=2,3,\cdots, N,
\end{align}
where $\hat{H}_k \in \text{End}(\CalH)$ is the $k$-th quantum Hamiltonian of the $\fgl(2)$ XXX spin chain system with $N$ sites obtained in \eqref{eq:transfergauge}, while $E_k (\ba)$ is a combination of normalized vacuum expectation values of local observables obtained in \eqref{eq:qhopereq}. Thus, we conclude that the eigenstates $\psi(\ba;\mathbf{u})$ of the $Q$-operators, constructed as the normalized vacuum expectation value of the regular monodromy surface defect, are also common eigenstates of the quantum Hamiltonians $\hat{H}_k$ of the associated XXX spin chain with $N$ sites. The corresponding eigenvalues $E_k (\ba)$ are normalized vacuum expectation values of the local chiral observables, which parameterize the space of $\hbar$-opers by our construction of the $\hbar$-oper and its solutions in section \ref{subsec:tqoper}.

\section{Discussion} \label{sec:discussion}

We have clarified how the $\hbar$-Langlands correspondence can be formulated in the four-dimensional $\EN=2$ gauge theory with the help of two kinds of surface defects $-$ regular monodromy surface defect and the $\mathbf{Q}$-observable. We show that the vacuum expectation value of the former gives a distinguished basis $\psi(\ba)$ for the degree-zero subspace $\CalH$ of an evaluation module over the Yangian $Y(\fgl(2))$, while the insertion of the $\mathbf{Q}$-observable on top of it gives the action of a $Q$-operator. The action was shown to be diagonal in the limit $\ve_2 \to 0$ due to the cluster decomposition of the surface defects, with the \textit{eigenvalue} being the $Q$-function. Using this construction, we showed that the $Q$-eigenstate constructed by a regular monodromy surface defect is also a common eigenstate of the quantum Hamiltonians of the associated XXX spin chain. The result can be regarded as a $\hbar$-deformation of the geometric Langlands correspondence \cite{BD1,BD2}, realized in the $\EN=2$ gauge theory setting.

While this study sheds light on crucial aspects, the subject merits further exploration in various layers.

\paragraph{$q$-Langlands correspondence and 5d $\EN=1$ uplift}
It would be desirable to uplift our formulation to the five-dimensional $\EN=1$ field theories compactified on a circle. In the twisted M-theory setting, this amounts to replacing the holomorphic surface $\BC \times \BC^\times$ by $\BC^\times \times \BC^\times$. Doing so, we would be able to study (quantum) $q$-Langlands correspondence with ramifications \cite{frenkel1996,Aganagic:2017smx}, where the surface defects are replaced by codimension-two defects wrapping the circle. While the monodromy surface defect will still be defined by singular behavior of fields along a codimension-two surface, interestingly, both $\mathbf{Q}$-observable and canonical surface defect will uplift to \textit{multiplicative} $\mathbf{Q}$-observables, defined by coupling the five-dimensional theory to a three-dimensional $\EN=2$ gauge theory either in the Coulomb phase or in the Higgs phase. The bispectral duality between two XXZ spin chains is expected from the isomorphism between two moduli spaces of multiplicative Higgs bundles on $\BC^\times$ (on $\BP^1$ with a framing at $0,\infty \in \BP^1$), with the rank of the bundle and the number of regular singularities swapped \cite{Mukhin2006,Mironov:2012uh,Bulycheva:2012ct,Gaiotto:2013bwa}. In particular, this construction will allow incorporating bi-infinite (i.e., non-highest-weight) modules of quantum affine algebra $U_q (\widehat{\fsl}(N))$ on the automorphic side of the $q$-Langlands correspondence with ramifications. See \cite{Aganagic:2017gsx,Koroteev:2018jht, Frenkel:2020iqq, Haouzi:2023doo} for more previous studies of the $q$-Langlands correspondence. 

\paragraph{Affinization and circular quiver theories}
In the present work, we considered the linear quiver $\EN=2$ gauge theory. It would be interesting to apply our methodology to the circular quiver theories. In the twisted M-theory setting, this amounts to replacing the holomorphic surface $\BC \times \BC^\times$ by $\BC \times T^2$. The moduli space of Higgs bundles on $T^2$ and the associated geometric Langlands correspondence can be analyzed in the usual way as here. On the other hand, in the bispectral point of view, the moduli space of multiplicative Higgs bundles on $\BC$ should be modified with a further affinization since the transverse holomorphic surface is now $T^2$ rather than $\BC^\times$. See \cite{Chen:2019vvt,Grekov:2023fek} for previous studies in the $\EN=2$ gauge theory side.

\paragraph{Yangian double and quantum vertex algebra}
There should be a vertex algebra approach to the $\hbar$-Langlands correspondence parallel to the one for the ordinary Langlands correspondence. Let us briefly recall the ordinary case. Let $\fg$ be a finite-dimensional simple Lie algebra over $\BC$. The vacuum module $V_k (\hfg) = \text{Ind}^{\hfg} _{\hfg_+ \oplus \BC K} \BC v_k$ of the affine Lie algebra at any level $k\in \BC$ is endowed with a vertex algebra structure, whose center $Z(\hfg_k)$ is a commutative associative algebra. The center is given by the $\hfg_+$-invariant subspace, $Z(\hfg_k) = V_{k} (\hfg) ^{\hfg_+}$, which is trivial at all $k \in \BC$ except at the critical level $k=-h^\vee$ where $h^\vee$ is the dual Coxeter number of $\fg$ \cite{Feigin:1991wy}. As vector spaces, the vacuum module $V_k (\hfg)$ is isomorphic to the universal enveloping algebra $ U(\hfg_-)$. The induced injective map $Z(\hfg_k) \hookrightarrow U(\hfg_-)$ turns out to be an algebra morphism. Hence, the non-trivial center $Z(\hfg_{-h^\vee})$ can be viewed as a maximally commutative subalgebra of $U(\hfg_-)$, which we call the \textit{universal Gaudin algebra}.\footnote{It is said to be universal since the coproduct of $U(\hfg_-)$ and the evaluation homomorphism $U(\hfg_-) \to U(\fg)$ give an algebra morphism $U(\hfg_-) \to U(\fg)^{\otimes(n+2)}$ for any $n$, under which the universal Gaudin algebra maps to a maximally commutative subalgebra in $U(\fg)^{\otimes(n+2)}$, called the Gaudin algebra}. Then it was shown that it is isomorphic to the classical $\CalW$-algebra associated to ${}^L \fg$ as a Poisson algebra, by the Miura transformation of the latter.

The story parallels after the $\hbar$-deformation. The Yangian double of $\fgl(n)$ with a central extension, which we denote by $DY (\fgl(n))$, was defined in \cite{KenjiIohara_1996}. It can be viewed as a $\hbar$-deformation of the (completed) universal enveloping algebra of the affine Kac-Moody algebra $\widehat{\fgl}(n)$. The vacuum module $\CalV_{c} (DY(\fgl(n)))$ of the Yangian double is endowed with a \textit{quantum vertex algebra} structure \cite{Jing:2018exs}, whose concept was originally defined in \cite{Frenkel:1996nz, Etingof2000}.\footnote{The term \textit{quantum} sometimes indicates deviating from the critical level. In this sense, calling it \textit{quantum} vertex algebra may cause a confusion. We could have referred to it as \textit{$\hbar$-vertex algebra}.} The Yangian double has a non-trivial center at the critical level $c=-n$, $Z(DY(\fgl(n))_{-n})$ \cite{Fan2022CenterOT}. As a vector space, the vacuum module is isomorphic to the Yangian $Y(\fgl(n))$, and the induced map $Z(DY(\fgl(n))_{-n}) \hookrightarrow Y(\fgl(n))$ is an injective algebra morphism. Thus, the non-trivial center can be viewed as a maximally commutative subalgebra of $Y(\fgl(n))$, i.e., the Bethe subalgebra. It should also be isomorphic to the classical $\hbar$-$\mathcal{W}$-algebra \cite{Hou:1996fx,ding1998} associated to $\fgl(n)$ as a Poisson algebra, due to the $\hbar$-Miura transformation of the latter. The statement is expected to follow from a limit of \cite{frenkel1996}.

In this context, it would be nice to express the vacuum expectation value of the regular monodromy surface defect explicitly as a $\hbar$-conformal block of the quantum vertex algebra for the Yangian double. Such a presentation would naturally provide its geometric construction as a section of a $\hbar$-difference module over $\text{Bun}_{GL(n)} (\BP^1;D)$. We also anticipate that the $Q$-operator would be given by the insertion of a degenerate vertex operator for a module over the Yangian double $DY(\fgl(n))$ which is a $\hbar$-deformation of the spectral flow module for $\widehat{\fgl}(n)$.

\paragraph{Analytic Langlands correspondence from $\EN=2$ gauge theory}
In recent studies \cite{Etingof:2019pni,Etingof:2021eub,Etingof:2021eeu,Etingof:2023drx}, the analytic version of the Langlands correspondence was formulated. In particular, the Hecke operator that we obtained as a contour integral of the correlation function of parallel surface defects in \cite{Jeong:2023qdr} was reconstructed as the \textit{chiral} Hecke operator \cite{Etingof:2023drx}. It would be interesting to incorporate the analytic Langlands correspondence in our $\EN=2$ gauge theoretical framework by replacing a part of the four-dimensional worldvolume by a compact $\BP^1$. The analytic version of the $\hbar$-(and $q$-)Langlands correspondence is also a subject to be developed. See \cite{Gaiotto:2021tsq} for the account of the analytic Langlands correspondence in the GL-twisted $\EN=4$ gauge theory setup. 

The quantum analytic Langlands correspondence realized as an one-parametric deformation, established recently in \cite{Gaiotto:2024tpl}, is expected to arise by turning on both $\O$-background parameters, where $\ve_1$ is associated to the isometry of the compact $\BP^1$ while $\ve_2$ is associated to the isometry of the non-compact $\BR^2$ as usual.

\paragraph{R-matrices for bi-infinite modules from stable envelopes}
In this work, we constructed the R-matrices of the Yangian $Y(\fgl(2))$ on the tensor product of a bi-infinite (i.e., non-highest-weight) module and $\BC^2$. Meanwhile, the R-matrices on highest-weight modules admit a geometric construction from stable envelopes \cite{Maulik:2012wi,Aganagic:2016jmx}. The stable envelope was realized in the context of the Bethe/gauge correspondence by the Janus interface interpolating different values of real masses \cite{Bullimore:2017lwu,Dedushenko:2021mds,Bullimore:2021rnr,Ishtiaque:2023acr}. It would be interesting to extend this gauge theoretical construction of the stable envelopes so that the R-matrices on bi-infinite modules would be incorporated, perhaps by regarding the $\BC_{\ve_2}$-plane as a cigar and compactifying along its circle fiber.

\appendix

\section{Yangian of $\fgl(n)$ and Manin matrices} \label{sec:appA}

\subsection{Yangian and universal R-matrix} \label{sec:yangian1}
We review the definition of Yangian, as well as its coproduct, quantum determinant, and the universal R-matrix.

\subsubsection{Definition}
Let us recall the definition of the Yangian of $\mathfrak{gl}(n)$, which we denote by $Y(\mathfrak{gl}(n))$.\footnote{See \cite{Molev:2002ghc, Loebbert:2016cdm} for comprehensive reviews.} The Yangian of $\mathfrak{gl}(n)$ is a unital associative algebra with generators $T_{ab} [s]$, $a,b=1,2,\cdots, n$ and $s \in \BZ_{> 0}$. We can write the generating series as
\begin{align}
    T_{ab} (x) = \d_{ab} -\hbar \sum_{s=1} ^\infty T_{ab} [s] x^{-s} \in Y(\mathfrak{gl}(n)) [[x^{-1}]], \quad\quad a,b=1,\cdots, n.
\end{align}
The defining relation of $Y(\mathfrak{gl}(n))$ is
\begin{align} \label{eq:defy}
    (x-x') [T_{ab} (x), T_{cd} (x')] = \hbar \left(T_{cb} (x) T_{ad} (x') - T_{cb} (x')T_{ad} (x) \right), \quad\quad 
 a,b,c,d=1,\cdots, n.
\end{align}

Let $E_{ab}$, $a,b=1,\cdots, n$, be the standard basis of $\text{End}(\BC^n)$. We can combine the generating series into a matrix $T(x) = \sum_{a,b=1} ^n E_{ab} \otimes T_{ab} (x) \in \text{End}(\BC^n) \otimes Y(\mathfrak{gl}(n))[[x^{-1}]]$. The rational $R$-matrix $R(x) \in \text{End} ( \BC^n \otimes \BC^n)$ defined by 
\begin{align} \label{eq:rrmat}
\begin{split}
R(x) &= \mathds{1}_n \otimes \mathds{1}_n - \frac{\hbar}{x} P \\
&=\mathds{1}_n \otimes \mathds{1}_n -\frac{\hbar}{x}  \sum_{a,b=1} ^n E_{ab} \otimes E_{ba} ,
\end{split}
\end{align}
where $P(v \otimes w) = w \otimes v$ is the permutation operator. The defining commutation relation \eqref{eq:defy} for $Y(\fgl(n))$ is equivalent to the RTT-relation
\begin{align} \label{eq:RTT}
    R^{(12)} (x-x') T^{(13)} (x) T^{(23)} (x') = T^{(23)} (x') T^{(13)} (x) R^{(12)} (x-x'),
\end{align}
valued in $\text{End}(\BC^n) \otimes \text{End}(\BC^n) \otimes Y(\mathfrak{gl}(n))[[x^{-1}]]$. Here, the superscripts indicate where $R$ and $T$ are valued among the tensor product. For instance, $R^{(12)} (x) = R(x) \otimes \text{id} \in \text{End}(\BC^n) \otimes \text{End}(\BC^n) \otimes Y(\mathfrak{gl}(n))[[x^{-1}]]$.

\subsubsection{Coproduct and quantum determinant}
The Yangian $Y(\mathfrak{gl}(n))$ is a Hopf algebra. In particular, it is endowed with a coproduct $\Delta: Y(\mathfrak{gl}(n)) \to Y(\mathfrak{gl}(n)) \otimes Y(\mathfrak{gl}(n))$ defined by
\begin{align}
    \Delta(T_{ab}(x)) = \sum_{c=1} ^n T_{cb} (x) \otimes T_{ac} (x), \quad a,b=1,\cdots, n.
\end{align}
This relation can be also expressed as
\begin{align}
    (\text{id} \otimes \Delta) T(x) = T^{(13)} (x) T^{(12)} (x),
\end{align}
as an element of $\text{End}(\BC^n) \otimes Y(\mathfrak{gl}(n)) [[x^{-1}]] \otimes Y(\mathfrak{gl}(n)) [[x^{-1}]]$.

The quantum determinant $\text{qdet}: \text{End}(\BC^n )\otimes Y(\mathfrak{gl}(n)) [[x^{-1}]] \to Y(\mathfrak{gl}(n))[[x^{-1}]]$ is defined by
\begin{align} \label{eq:qdetdef}
\begin{split}
    \text{qdet}\, T(x) &= \sum_{\s \in S_n} \text{sgn}(\s) T_{1\s(1)}(x) \cdots T_{n\s(n)} (x + (n-1)\hbar)\\
    &= \sum_{\s \in S_n} \text{sgn}(\s) T_{\s(1)1}(x+ (n-1)\hbar) \cdots T_{\s(n)n} (x).
\end{split}
\end{align}
Under the coproduct the quantum determinant factorizes
\begin{align} \label{eq:qdetfac}
    \D (\text{qdet} \, T(x)) = \text{qdet}\, T(x) \otimes \text{qdet}\, T(x).
\end{align}
The quantum determinant can be expanded as a formal series as
\begin{align}
    \text{qdet}\, T(x) = 1 + d_1 x^{-1} + d_2 x^{-2} + \cdots \in Y(\fgl(n)) [[x^{-1}]].
\end{align}
The coefficients $(d_i)_{i \geq 1}$ generate the center of the Yangian $Z (Y(\fgl(n))) \subset Y(\fgl(n))$.

We may generalize the notion of quantum determinant to the minors of the generating matrix $T(x)$. Let $m \leq n$ and consider the $m\times m$ minor $T^{i_1 i_2 \cdots i_m} _{j_1 j_2 \cdots j_m} (x)$ with elements $(T^{i_1 i_2\cdots i_m} _{j_1 j_2 \cdots j_m})_{ab} (x) = T _{i_a j_b} (x)$. Then, we define its quantum determinant by
\begin{align}
\begin{split}
    \text{qdet}\, T^{i_1 i_2 \cdots i_m} _{j_1 j_2 \cdots j_m} (x) &= \sum_{s \in S_m} \text{sgn}(\s) T_{i_1 j_{s(1)}} (x)  T_{i_2 j_{s(2)}} (x+\hbar)  \cdots T_{i_m j_{s(m)}} (x+ (m-1)\hbar) \\
    & = \sum_{s \in S_m} \text{sgn}(\s) T_{i_{s(1)} j_1 } (x+ (m-1)\hbar)  T_{i_{s(2)} j_2 } (x+(m-2)\hbar)  \cdots T_{i_{s(m)} j_m } (x) .
\end{split}
\end{align}

\subsubsection{Universal R-matrix}
There exists a universal R-matrix $\CalR(x) \in \left(Y(\fgl(n)) \otimes Y(\fgl(n)) \right)[[x^{-1}]] $ satisfying the Yang-Baxter equation
\begin{align}  \label{eq:YBgen}
    \CalR^{(12)} (x-x') \CalR^{(13)}(x) \CalR^{(23)}(x') = \CalR^{(23)} (x') \CalR^{(13)} (x) \CalR^{(12)} (x-x'),
\end{align}
valued in $\left( Y(\fgl(n)) \right)^{\otimes 3} [[x^{-1},x'^{-1}]]$ \cite{Drinfeld:466366}. For any representation $U,V,W$ ($\r_U : Y(\fgl(n)) \to \text{End}(U)$, etc) of the Yangian, we obtain the R-matrices
\begin{align}
    R_{U,V} (x) = (\r_U \otimes \r_V) (\CalR(x)),
\end{align}
which satisfy the Yang-Baxter equation
\begin{align}
    R_{U,V} (x-x') R_{U,W} (x) R_{V,W} (x') = R_{V,W} (x') R_{U,W} (x) R_{U,V} (x-x'),
\end{align}
valued in $\text{End}(U)\otimes\text{End}(V)\otimes\text{End}(W)$.

\subsection{R-matrices and monodromy matrix} \label{subsec:rmat}
We start to introduce representations for the Yangian. We construct R-matrices, and then the monodromy matrix by their concatenation. We construct the XXX spin chain with $N$ sites.

\subsubsection{Evaluation module}
It is clear that the Yangian $Y(\mathfrak{gl}(n))$ contains the universal enveloping algebra $U(\mathfrak{gl}(n))$ as a Hopf subalgebra. The embedding $\iota:U(\mathfrak{gl}(n)) \hookrightarrow Y(\mathfrak{gl}(n))$ is given by $e_{ab} \mapsto T_{ba} [1]$, where $e_{ab}$ are the standard generators of $\mathfrak{gl}(n)$, $[e_{ab},e_{cd}] = \d_{bc} e_{ad} - \d_{ad} e_{cb}$. Moreover, there is an evaluation homomorphism $\epsilon : Y(\mathfrak{gl}(n)) \to U(\mathfrak{gl}(n))$ defined by $\epsilon : T_{ab} [1] \mapsto e_{ba}$ for $a,b=1,\cdots, n$, and $\epsilon:T_{ab} [s] \mapsto 0$ for $s>1$ and all $a,b=1,\cdots, n$. Note that $\epsilon \circ \iota =  \text{id}$.

There is an one-parameter $\th \in \BC$ family of automorphisms $s_\th : Y(\mathfrak{gl}(n)) \to Y(\mathfrak{gl}(n))$ defined by $s_\th (T(x)) = T(x-\th)$. Namely, we get the action of $s_\th$ on the generators by expanding the right hand side in $x^{-1}$. Note that $\epsilon\circ s_\theta\circ \iota= \text{id}$. A $\mathfrak{gl}(n)$-module $V$ can be promoted to a $Y(\mathfrak{gl}(n))$-module $V(\th)$ with the help of of the map $\epsilon \circ s_\th : Y(\fgl(n)) \to U(\fgl(n))$. This is called \textit{evaluation module} with the evaluation parameter $\th$.  

\subsubsection{R-matrices and monodromy matrix}
Now, consider the standard $n$-dimensional representation $\BC^n$ of $U(\fgl(n))$ and the associated evaluation Yangian module $\r_{\BC^n} : Y(\fgl(n)) \to \text{End}(\BC^n)$. Then, we can recover the rational R-matrix \eqref{eq:rrmat},
\begin{align} 
\begin{split}
    &R(x) = (\r_{\BC^n} \otimes \r_{\BC^n}) (\CalR(x))  \in \text{End}(\BC^n)\otimes \text{End}(\BC^n),
\end{split}
\end{align}
and the generating matrix for the Yangian $Y(\fgl(n))$,
\begin{align} \label{eq:generatingT}
    T(x) = (\r_{\BC^n} \otimes \text{id}) (\CalR(x)) \in \text{End}(\BC^n)\otimes Y(\fgl(n)),
\end{align}
by applying this representation to the universal R-matrix. The Yang Baxter equation satisfied by the rational R-matrix and the RTT relation \eqref{eq:RTT} for the $Y(\fgl(n))$ are consequences of \eqref{eq:YBgen}.

For representation $V$ of $\fgl(n)$, consider the associated evaluation Yangian module,  and the evaluation homomorphism $\r_{V} : Y(\fgl(n))\to \text{End}(V)$. Then, the generating matrix represented on $V$ is computed to be
\begin{align} \label{eq:rmat1}
\begin{split}
   T(x) \vert _V  &= (\text{id} \otimes \r_{V}) (T(x)) = \mathds{1}_n - \frac{\hbar \sum_{a,b=1} ^n E_{ab} \otimes e_{ba}\vert_{V}}{x} \in \text{End} (\BC^n) \otimes \text{End}(V) \\
   & =: R_V(x),
\end{split}
\end{align}
which, by \eqref{eq:generatingT}, can also be viewed as the universal R-matrix represented on the tensor product, $R_V (x) = (\r_{\BC^n} \otimes \r_{V}) (\CalR(x))$. For this reason, we call $R_V (x) $ an R-matrix.

Let us given with $N$ $\mathfrak{gl}(n)$-modules $\left(\CalH_\o \right)_{\o=0} ^{N-1}$ and $N$ evaluation parameters $(\th_\o)_{\o=0} ^{N-1} \in \BC^{N}$. Then we construct a $Y(\mathfrak{gl}(n))$-module 
\begin{align}
 \CalH  := \bigotimes_{\o=0} ^{N-1} \CalH_\o  = \CalH_0  \otimes \cdots \otimes \CalH_{N-1} 
\end{align}
with the help of $ \epsilon^{\otimes N}  \circ (s_{\th_0} \otimes \cdots  \otimes s_{\th_{N-1}} ) \circ \Delta^{N-1}$. Then, the generating matrix $T(x)$ represented on the module $\CalH$ is computed to be
\begin{align} \label{eq:XXXmono}
\begin{split}
    T(x)\vert_{\CalH} &= \left( \mathds{1}_n  - \frac{\hbar \sum_{a,b=1} ^n E_{ab} \otimes e_{ba} ^{(N-1)}}{x-\th_0}  \right) \cdots  \left( \mathds{1}_n - \frac{\hbar \sum_{a,b=1} ^n E_{ab} \otimes e_{ba} ^{(0)}}{x-\th_{N-1}}  \right) \\
    &\in \text{End}(\BC^n) \otimes \text{End}(\CalH),
\end{split}
\end{align}
where $e_{ab} ^{(\o)} = 1 \otimes \cdots \otimes 1 \otimes \underbrace{e_{ab}\vert_{\CalH_\o}}_{\o\text{-th}} \otimes 1 \otimes \cdots \otimes 1 $. In terms of the R-matrices \eqref{eq:rmat1}, it is written as
\begin{align}
    T(x)\vert_{\CalH} = R_{N-1}(x-\th_{N-1}) R_{N-2} (x-\th_{N-2}) \cdots R_0 (x-\th_0),
\end{align}
where we denote $R_\o (x) := R_{\CalH_\o} (x)$. This is precisely the known expression for the monodromy matrix for the periodic XXX spin chain with $N$ sites. Each site is labelled by the $\fgl(n)$-module $\CalH_\o$ and the evaluation parameter $\th_\o$.

\subsection{Manin matrices} \label{sec:manin}
We give a brief review of the definition and properties of Manin matrices \cite{ManinQuantumGroups} relevant to this work. See \cite{Chervov:2007bb} for a comprehensive review. 

Let $\CalR$ be a non-commutative associative ring. Let $M \in \text{Hom}(\BC^n, \BC^m) \otimes \CalR$ be a $m \times n$ matrix with its entries in $\CalR$. We call $M$ a Manin matrix if
\begin{itemize}
    \item Elements in the same column commute with themselves.
    \item Commutators of cross terms of any $2 \times 2$ minor of $M$ are equal; namely, \begin{align}
        [M_{ij},M_{kl}] = [M_{kj},M_{il}] \quad \text{for all} \quad i,j,k,l.
    \end{align}
\end{itemize}

Let $M \in \text{End}(\BC^n)\otimes \CalR$ be a Manin matrix. Then we define the determinant of $M$ by the column expansion,
\begin{align}
    \det M := \sum_{\s \in S_n} (-1)^\s \prod_{i=1,2,\cdots, n} ^\curvearrowright M_{\s(i),i},
\end{align}
where $S_n$ is the permutation group and the product is taken in the order of column that the entries lie in. It can be shown that the determinant is not dependent of the order of the columns, and thus it is well-defined as determinant.

Some relevant properties of Manin matrices are:
\begin{itemize}
    \item Any minor of a Manin matrix is a Manin matrix.
    \item If $A$ and $B$ are Manin matrices and $[A_{ij},B_{kl}] = 0$ for all $i,j,k,l$, then $A+B$ is a Manin matrix.
    \item If $A$ and $B$ are Manin matrices and $[A_{ij},B_{kl}] = 0$ for all $i,j,k,l$, then $AB$ is a Manin matrix and $\det (AB) = \det A \det B$.
\end{itemize}

\section{Absence of poles in $\tilde{\ET}_\o(x,\bx')$} \label{sec:abspole}
The residue of the right hand side of \eqref{eq:dualfracqq} at $x=x'_\o +d_\o \ve_1$ is proportional to
\begin{align}
\begin{split}
    & d_\o \ve_1 (x'_{\o}-x'_{[{\o+1}]} + d_\o\ve_1 - d_{[{\o+1}]} \ve_1 + \ve_2\d_{\o,N-1}  ) \EY_{[{\o+1}]}(x'_\o+d_\o\ve_1) \\
    & \prod_{\o'=0}^{N-1} \prod_{j=1}^{d_{\o'}} \frac{ x'_{[{\o'-1}]} - x'_{\o'}+ d_{[{\o'-1}]}\ve_1-(j-1)\ve_1 + \ve_2\d_{\o',0} }{j\ve_1} \frac{Q_{\o'}(x_{\o'}') M_{\o'}(x'_{\o'}+d_{\o'}\ve_1)}{Q_{\o'}(x'_{\o'}+d_{\o'}\ve_1) Q_{[{\o'+1}]}(x'_{\o'}+(d_{\o'}+1)\ve_1 + \ve_2 \d_{\o',N-1}) } \\
    & - (x'_\o - x'_{[{\o+1}]}+d_\o\ve_1-\ve_2\d_{\o,N-1})(x'_{[{\o-1}]}-x'_\o + (d_{[{\o-1}]}-d_\o+1)\ve_1 + \ve_2 \d_{\o,0} ) \frac{P_\o(x'_\o+d_\o\ve_1)}{\EY_\o(x'_\o+d_\o\ve_1)} \\
    & \quad \times  \prod_{\o'=0}^{N-1} \prod_{j=1}^{d_{\o'}-\d_{\o',\o}} \frac{ x'_{[{\o'-1}]} - x'_{\o'}+ (d_{[{\o'-1}]}-\d_{[{\o+1}],\o'})\ve_1-(j-1)\ve_1 + \ve_2\d_{\o',0} }{j\ve_1} \\
    & \quad \times \frac{Q_{\o'}(x_{\o'}') M_{\o'}(x'_{\o'}+(d_{\o'}-\d_{\o',\o})\ve_1)}{Q_{\o'}(x'_{\o'}+(d_{\o'}-\d_{\o',\o})\ve_1) Q_{[{\o'+1}]}(x'_{\o'}+(d_{\o'}+1-\d_{\o',\o})\ve_1 + \ve_2 \d_{\o',N-1}) } \\
    = & \prod_{\o'\neq\o,\o+1} \prod_{j=1}^{d_{\o'}-\d_{\o,\o'}} \frac{ x'_{[{\o'-1}]} - x'_{\o'}+ d_{[{\o'-1}]}\ve_1-(j-1)\ve_1 + \ve_2\d_{\o,0} }{j\ve_1} \\
    & \left[ \frac{ \prod_{j=1}^{d_\o} x'_{[{\o-1}]} - x'_{\o}+ d_{[{\o-1}]}\ve_1-(j-1)\ve_1 + \ve_2\d_{\o,0}  }{\prod_{j=1}^{d_\o-1} j\ve_1 } \frac{\prod_{j=1}^{d_{[{\o+1}]}+1} x'_\o - x'_{[{\o+1}]} + d_\o\ve_1 - (j-1)\ve_1 + \ve_2\d_{\o,0} }{ \prod_{j=1}^{d_{[{\o+1}]}} j\ve_1 } \right. \\
    & \quad - \left. \frac{ \prod_{j=1}^{d_\o} x'_{[{\o-1}]} - x'_{\o}+ d_{[{\o-1}]}\ve_1-(j-1)\ve_1 + \ve_2\d_{\o,0}  }{\prod_{j=1}^{d_\o-1} j\ve_1 } \frac{\prod_{j=1}^{d_{[{\o+1}]}+1} x'_\o - x'_{[{\o+1}]} + d_\o\ve_1 - (j-1)\ve_1 + \ve_2\d_{\o,0} }{ \prod_{j=1}^{d_{[{\o+1}]}} j\ve_1 } \right] \\
    & \times \frac{Q_{\o'}(x_{\o'}') M_{\o'}(x'_{\o'}+d_{\o'}\ve_1)}{Q_{\o'}(x'_{\o'}+d_{\o'}\ve_1) Q_{[{\o'+1}]}(x'_{\o'}+(d_{\o'}+1-\d_{\o,\o'})\ve_1 + \ve_2 \d_{\o',N-1}) } \\
    = & \ 0
\end{split}
\end{align}

\bibliographystyle{utphys}

\bibliography{reference}
\end{document}